\documentclass[aps,prb,reprint,superscriptaddress,nofootinbib]{revtex4-2}
\usepackage{amsmath}
\usepackage{graphicx}
\usepackage{amsfonts}
\usepackage{color}
\usepackage{bm}
\usepackage{subfigure}
\usepackage{lipsum}
\usepackage{comment}
\usepackage{dsfont}
\usepackage{bm}
\usepackage{amsmath}
\usepackage{amssymb}
\usepackage{graphicx} 
\usepackage{braket}
\usepackage{mathtools}
\usepackage{times}
\usepackage{enumerate}

\newcommand{\bea}{\begin{eqnarray}}
\newcommand{\eea}{\end{eqnarray}}
\usepackage[colorlinks,linkcolor=blue,urlcolor=blue,citecolor=blue]{hyperref}

\begin{document}

\title{Thermal state preparation by repeated interactions at and beyond the Lindblad limit}



\author{Carlos Ramon-Escandell}
\email{}
\affiliation{Department of Physics and Centre for Quantum Information and Quantum Control, University of Toronto, 60 Saint George St., Toronto, Ontario, M5S 1A7, Canada}

\author{Alessandro Prositto}
\email{}
\affiliation{Department of Physics and Centre for Quantum Information and Quantum Control, University of Toronto, 60 Saint George St., Toronto, Ontario, M5S 1A7, Canada}

\author{Dvira Segal}
\email{dvira.segal@utoronto.ca}
\affiliation{Department of Chemistry, University of Toronto, 80 Saint George St., Toronto, Ontario, M5S 3H6, Canada}
\affiliation{Department of Physics and Centre for Quantum Information and Quantum Control, University of Toronto, 60 Saint George St., Toronto, Ontario, M5S 1A7, Canada}

\date{\today}

\begin{abstract}
We study the nature of thermalization dynamics and the associated preparation (simulation) time under the repeated interaction protocol uncovering  a generic anomalous, Mpemba-like trend. As a case study, we focus on a three-level system and analyze its dynamics in two complementary regimes, where the system-ancilla interaction strength is either large or small. Focusing on the estimation of the simulation time, we derive closed-form expressions for the minimum number of collisions, or minimal simulation time, required to achieve a thermal state, which is within $\epsilon$  distance to the target thermal state. 
At zero temperature, we analytically identify a set of points  
(interaction strength $\times$ their duration) that minimize the simulation time. 
At nonzero temperature, we observe a Mpemba-like effect: Starting from a maximally mixed state, thermalization to an intermediate-temperature state takes longer than to a lower-temperature one. We provide an accurate analytical approximation for this phenomenon and demonstrate its occurrence in larger systems and under randomized interaction strengths.
The prevalence of the Mpemba effect in thermal state preparation presents a significant challenge for preparing states in large systems, an open problem calling for new strategies.
\end{abstract}

\maketitle

\section{Introduction}
\label{sec:Introduction}

The repeated interaction (RI) protocol is receiving growing attention as a versatile approach for the study of open quantum systems \cite{RevRI, Martinez2016, Ciccarello2017, Cusumano2022} and as a quantum algorithm framework \cite{Pocrnic25, Zambrini, Goold2021,Poletti2023,Donadi2024,Vedral2024}. In the most basic RI approach, the environment is represented as a sequence of subsystems (often called ``ancillas") that sequentially and independently interact with the system for a finite duration before being discarded. Each RI step involves a unitary evolution governed by a joint system-ancilla Hamiltonian, after which the ancilla is traced out, leaving the system in a new, non-unitary state \cite{RevRI, Cusumano2022}. 

The RI framework provides a physically-intuitive and mathematically tractable way to bridge microscopic unitary interactions and coarse-grained open quantum system dynamics \cite{RevRI}, and as such it lends itself for different applications: 
It provides a foundation for deriving the Lindblad quantum master equation and other Markovian and non-Markovian equations of motion \cite{Martinez2016, Merkli14, Attal05, Lorenzo17, Giovannetti12, Karevski09, Luchnikov17, Plenio,Merkli25}. It can be used to study equilibration processes \cite{Rau,Gisin2002, Campbell2020,Landi2021,David2023, Ghosh2024,Segal2024,Landi25}, thermalization dynamics \cite{Buzek02,Parrondo2022, David2024,Landi24,Campbell2020}, and thermodynamic processes \cite{Barra2015, Kosloff2019, Xia2022,Haack1,Haack2,Landi19c,Landi24,Juzar}. It has also been used to explore information-theoretic properties in open quantum dynamics \cite{Paternostro2015,Hanson2017, Beever2024,Strasberg17,Strasberg19,Juzar}. 
As for physical realizations, a linear optics implementation described in Ref. \citenum{expAllP} was realized in Ref. \citenum{expAll} capturing dynamics from perfectly Markovian up to strongly non-Markovian cases. 

More recently, the RI framework has gained attention not only as a theoretical tool but also as a practical paradigm for implementing open quantum system dynamics and preparing thermal states on quantum computing platforms \cite{Gemma, Pocrnic25, Zambrini, Goold2021}. Its inherently modular structure, where a system interacts sequentially with ancillary units, naturally aligns with gate-based operations of digital quantum computers. Quantum algorithms that build on the RI protocol were tested on near-term devices and further developed for fault-tolerant quantum computing machines \cite{Poletti2023,Donadi2024,Vedral2024}.

Focusing on the Markovian implementation of the RI scheme, with each ancilla independently interacting with the system, and only once, the Lindblad limit of the repeated interaction protocol is a well-established result \cite{Attal05, Zambrini, Merkli14, Lorenzo17, Luchnikov17, Giovannetti12, Karevski09, Pocrnic25}. It is derived by assuming a {\it stroboscopic evolution}: Each ancilla interacts with the system for a short time $\tau$, and the interaction strength $J$ is scaled such that the product, $J^2\tau$, remains a constant as $\tau\to 0$. Under these conditions, the discrete sequence of system-ancilla unitaries, followed by ancilla resets, converges to a continuous-time Markovian quantum dynamical semigroup described by a Lindblad master equation \cite{BookOQS}.
However, we reiterate that this convergence is based on the stroboscopic and strong-coupling assumptions. This motivates the first research question we address in this work:
(i) {\it What kinds of open quantum dynamics can be realized by a Markovian repeated interaction protocol, when we move beyond the stroboscopic-Lindblad (SL) limit?}

As the system evolves towards its steady state, which in this work is thermal, and matching the temperature of the ancillas, an important consideration in quantum algorithms is resource estimation for thermal state preparation \cite{Hagan,Wang2022}, in particular an estimate for simulation time \cite{Hagan}. This brings us to the second question that we address in this study:
(ii) {\it  What is the simulation time (or number of RI steps) required to reach the thermal state (within $\epsilon$ distance) under the RI protocol? How does this time depend on the parameters of the system?}

\begin{figure}[tbp]
    \centering
\includegraphics[width=\linewidth]{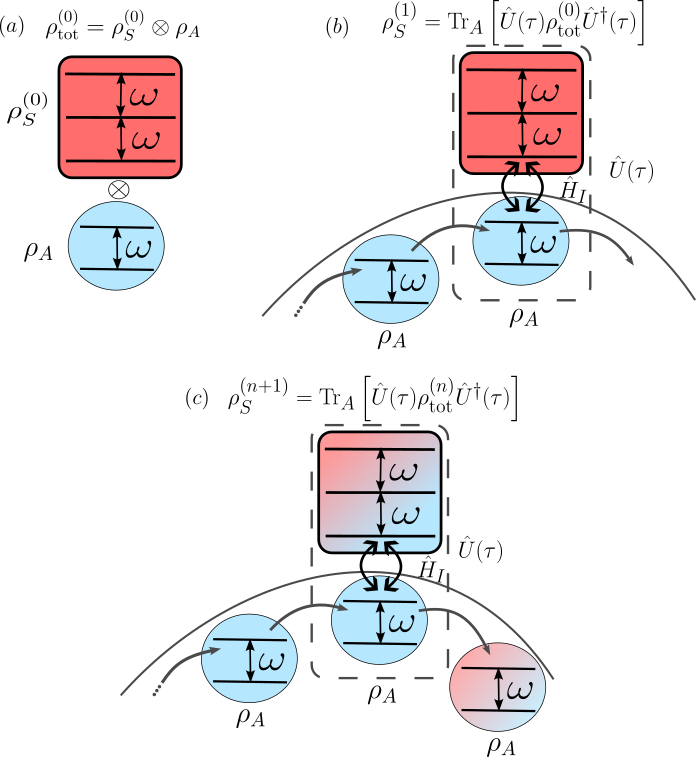} 
\caption{An illustration of thermal state preparation via the repeated interaction protocol, representing the cooling of a hot (red) system towards the state of the cold (blue) bath of ancilla qubits.
(a) The RI process starts by preparing a product state of the system and an ancilla.
(b) After one collision, the system evolves into $\rho_S^{(1)}$, while the first ancilla is discarded.
(c) After $n$ collisions, the system is cooler.  Discarded ancilla heats up to some extent due to heat exchange with the system.
}
    \label{fig:scheme}
\end{figure}

To address these two questions, we focus on a three-level system as a case study, with a flip-flop type 
interaction with the ancillas, see Fig. \ref{fig:scheme}. 
We solve its repeated interaction dynamics exactly and find that the population and coherence dynamics are decoupled, and that coherences evolve in a {\it nonmonotonic} manner.
This exact result enables us to
focus on two distinct dynamical regimes. The first is the Stroboscopic-Lindblad limit, where the interaction time is short but the interaction energy is large enough, so that $J^2\tau$ is a constant. 
The second complementary regime, which we refer to as the ``$J\tau1$ regime", assumes a weak interaction strength $J$, but a sufficiently long interaction time $\tau$ such that the product, $J\tau$, is of order 1.

Our main results on estimation of simulation time for thermalization, defined as reaching a state within $\epsilon$ trace distance of the ancilla's thermal state, are as follows:

(i) We solve the RI dynamics exactly at {\it zero temperature} and derive a lower bound on the total simulation time, $T_{\text{sim}}$, required for cooling.
In the $J\tau1$ regime, we identify a set of optimal interaction parameters under which zero-temperature cooling of an $d$-level system can occur in as few as $d-1$ interaction steps.

(ii) At {\it nonzero temperature}, we observe and explain analytically a surprising Mpemba-like effect: Starting from a maximally mixed state, it can take {\it longer} to reach a thermal state at high temperature than a thermal state at low temperature \cite{Raz17,Klich19,quantumM,FelixM,Goold24}. We explain this observation in both the Stroboscopic-Lindblad and the $J\tau1$ regimes. 

(iii) Extending our analysis to a system with $d$ levels, we show that the Mpemba effect arises generically whenever $d>2$, when starting from a maximally mixed initial state.

(iv) Generalizing the $J\tau1$ regime to arbitrary random Hermitian system-ancilla interaction Hamiltonians, we show numerically that thermalization occurs, unlike in the SL limit, and that the Mpemba effect persists.

The plan of the paper is as follows.
In Sec. \ref{sec:qnqb} we introduce the three-level system model and outline the basic principles of repeated interaction dynamics, focusing on energy-conserving interactions that lead to thermalization. In Sec. \ref{sec:EOM}, we derive exact equations of motion for the system under the RI scheme and analyze thermalization dynamics in two regimes: the Stroboscopic-Lindblad limit and the $J\tau1$ regime. 
Sec. \ref{sec:time} is dedicated to estimating the simulation time required for thermal state preparation when starting with a completely mixed state, providing lower bounds at zero temperatures. 
In Sec. \ref{sec:Mpemba}, we discuss the emergence of Mpemba-like behavior in thermalization and provide an approximate, yet accurate analytical result that follows
its occurrence. We further present numerical simulations that demonstrate thermalization and Mpemba dynamics in randomized RI models, and for other multi-level models.
We conclude in Sec. \ref{sec:summ} with a summary and an outlook.


\vspace{-0.35cm}
\section{REPEATED INTERACTION MODEL with Energy conserving interactions} \label{sec:qnqb}

In theoretical studies of quantum phenomena, there is always the need of --- and tension between --- two schools: one aims at working with models as general as possible, even at the cost of relying on approximations, while the other focuses on solving concrete models exactly, to extract insights that can later be generalized.

With the objective to develop a general quantum algorithm for thermal state preparation using the RI scheme, Ref. \citenum{Hagan} adopted a randomized interaction Hamiltonian. Using that approach, they showed that a thermal state is an approximate fixed point for arbitrary interaction Hamiltonians, and they bounded the runtime. 

Here, we take the complementary approach. We focus on a three-level system as a case study, solve it exactly, and derive first concrete then general insights into the simulation time required for thermal state preparation using the RI scheme. Specifically, we are able to derive and explain the nonmonotonic behavior of the required RI collisions for thermal state preparation, a curious observation \cite{Hagan} that was lacking a physical explanation.

We focus on systems described by the
following Hamiltonians,
\begin{equation}
    \label{eq: free-qudit-systemHamiltonian}
    \hat{H}_S = -\omega_S \hat{S}_z^{(s)},
\end{equation}
where $s = \frac{1}{2},1,\frac{3}{2}...$ represents the spin value. For example, the spin-1 case would correspond to a three-level system (qutrit) with equidistant energy levels. A good choice of basis for working with spin operators is the spin eigenbasis, where, given a spin quantum number $s$, there are $2s+1$ projections labeled by the magnetic quantum number $m_s$. This basis is spanned by the set $\left\{\ket{s,m_s=+s},\ket{s,m_s=s-1},\ldots,\ket{s,m_s=-s}\right\}$.
The ground state for this system Hamiltonian corresponds to the highest projection state $\ket{s,m_s=+s}$.

Each ancilla is a qubit system ($s= 1/2$), described by the Hamiltonian
\begin{equation}
    \label{eq: free-ancilla-Hamiltonian}
    \hat{H}_A = -\omega_A \hat{S}_z^{(\frac{1}{2})} = -\frac{\omega_A}{2}\hat{\sigma}_z,
\end{equation}
with $\hat \sigma_z$ a Pauli-z operator.
We consider that the ancillas are always initialized in the corresponding canonical thermal state at inverse temperature $\beta$,
\begin{equation}
\label{eq: ancilla_density_matrix}
    \rho_A(\hat{H}_A) = \frac{e^{-\beta \hat{H}_A}}{\mathcal{Z}_A} = \begin{pmatrix}
        p_A&0\\
        0&1-p_A
    \end{pmatrix},
\end{equation}
where $\mathcal{Z}_A = \text{Tr}[\exp(-\beta\hat{H}_A)]$ is the partition function and $p_A = 1/(1+e^{-\beta\omega_A})$ is the thermal population of the ground state of the ancilla.

To model the system-ancilla interaction, we focus on energy-conserving flip-flop transitions between adjacent pairs of states. Working in the computational basis, we establish the following mapping to the spin basis, $\left\{\ket{0}\xrightarrow[]{}\ket{s,m_s=+s},\ket{1}\xrightarrow[]{}\ket{s,m_s =s-1}\right.$$,\ldots, $ $\ket{2s}$$\left.\xrightarrow[]{}\ket{s,m_s =-s}\right\}$, which allows us to express the interaction Hamiltonian as
\begin{equation}
\label{eq:interaction-Hamiltonian-qudit-qubit}
    \hat{H}_I = \sum_{k=0}^{d-2}J_{k+1,k}\left(\ket{k+1,\downarrow}\bra{k,\uparrow}+H.c. \right).
\end{equation}
Here $d = 2s +1$ is the system dimension, $\ket{\uparrow}, \ket{\downarrow}$ are the spin-up/down states of the ancilla, with the spin-down state as the ground state of the ancilla.
$H.c.$ denotes the Hermitian conjugate. This choice of interaction Hamiltonian differs from previously-proposed models \cite{Giovannetti12, Scrani19}, which typically involved the tensor product of the system and ancilla spin operators explicitly. The issue with those interaction Hamiltonians is that employing explicit spin operators introduces state-dependent prefactors that differ across spin dimensions, artificially causing certain transitions to gain stronger coupling strengths. Such prefactors complicate a fair and consistent comparison between different spin-dimensional models and can lead to artificially accelerated thermalization processes. 

The total Hamiltonian of each system-ancilla collision is thus
\bea
\label{eq: total-hamiltonian}
    \hat{H}_{\text{tot}} &=& \hat{H}_S \otimes\mathbb{I}_A +\mathbb{I}_S\otimes\hat{H}_A+\hat{H}_I 
    \nonumber\\
    &=& \hat{H}_0+\hat{H}_I.
\eea
At each step of the RI scheme, the evolution of the density matrix of the system is described by the following completely positive trace-preserving (CPTP) map
\begin{equation}
\label{eq: CPTPmap}
    \rho_S^{(n+1)} = \text{Tr}_A\left[\hat{U}(\tau)\left(\rho_S^{(n)}\otimes\rho_A(\hat{H}_A)\right)\hat{U}^{\dagger}(\tau)\right],
\end{equation}
where $\text{Tr}_A[\cdot]$ refers to trace over the degrees of freedom of the ancilla, and $\hat{U}(\tau)$ is the total collision unitary,
\begin{equation}
    \label{eq: total-collision-unitary}
    \hat{U}(\tau) = e^{-i(\hat{H}_0+\hat{H}_I)\tau},
\end{equation}
with $\tau$ indicating the duration of each collision. Throughout this study, we work in units where $\hbar=1$. 

We assume that the system is initially prepared in a general state, $\rho_S^{(0)}$. In certain cases, we will consider $\rho_S^{(0)}$ to be diagonal in the energy eigenbasis of $\hat{H}_S$, such as the maximally mixed state or a thermal state at a certain temperature. The full initial state of the joint system-ancilla pair can be expressed as $\rho_{\text{tot}}^{(0)}=\rho_S^{(0)}\otimes\rho_A(\hat{H}_A)$ and it satisfies $(\rho^{(0)}_{\text{tot}})^{\dagger} = \rho_{\text{tot}}^{(0)} \geq 0$, and $\text{Tr}\left[\rho_{\text{tot}}^{(0)}\right] = 1$. 
As for the steady state (fixed point) of the system, we denote it by $\rho_S^*$.

While in some other collision models, the authors considered more generic frameworks where each collision has a different duration \cite{Gennaro09}, or incorporated free evolution of the system between collisions \cite{Zambrini}, we consider a simplified protocol where each collision has a fixed and identical duration $\tau$ and the system immediately interacts with a freshly prepared ancilla, without allowing the system to have free evolution between collisions. 

Moreover, as previously mentioned, in this work, we focus on interactions that conserve energy, i.e.,
\begin{equation}
\label{eq: energy-preserving-condition}
    \left[\hat{U}(\tau),\hat{H}_0\right]=0.
\end{equation}
This condition ensures that the canonical thermal state of the system $\rho_S(\hat{H}_S) = e^{-\beta\hat{H}_S}/\mathcal{Z}_S$ is a fixed point of the dynamics, placing our model within the class of \textit{thermal maps} \cite{Barra17}.
To satisfy this, we need to impose that the system and ancilla are on resonance, which requires a matching energy splitting $\omega_S = \omega_A$ \cite{Arisoy2019}. Although energy conservation does not strictly impose any constraint on the individual values of the coefficients $J_{k+1,k}$, the total Hamiltonian should remain Hermitian. As a result, the couplings must satisfy $J_{k+1,k}=J_{k,k+1}$, where we assumed real-valued interactions. In principle, these couplings could vary across transitions and we would still satisfy the energy-conservation condition. However, for simplicity and to allow for a clear analytical treatment, unless otherwise stated, we restrict ourselves to the case of isotropic interactions, where all couplings have the same value, $J=J_{k+1,k}$, which we assume to be positive without loss of generality. 

Consequently, the dynamics of our collision model is constrained by three parameters: The interaction coupling $J$, the collision duration $\tau$, and the system and ancilla frequencies $\omega$. In the next section, we derive exact equations of motion under this setup and analyze the thermalization behavior in the SL and the $J\tau1$ regimes, using a three-level system as a representative example.



\vspace{-0.15cm}
\section{EQUATIONS OF MOTION}
\label{sec:EOM}

For a three-level system $(s=1)$, the total Hamiltonian in Eq. (\ref{eq: total-hamiltonian}) is expressed in matrix form as
\begin{equation}
    \hat{H}_{tot}=
\begin{pmatrix}
-\frac{3\omega}{2} & 0 & 0 & 0 & 0 & 0 \\
0 & -\frac{\omega}{2} & J & 0 & 0 & 0 \\
0 & J & -\frac{\omega}{2} & 0 & 0 & 0 \\
0 & 0 & 0 & \frac{\omega}{2} &  J & 0 \\
0 & 0 & 0 & J & \frac{\omega}{2} & 0 \\
0 & 0 & 0 & 0 & 0 & \frac{3\omega}{2}
\end{pmatrix}.
\end{equation}
\vspace{3mm}
The corresponding collision unitary describing the dynamics during each collision is
\begin{widetext}
    \begin{equation}
\label{eq: collision-unitary-qutrit-qubit}
\hat{U}(\tau) =
\begin{pmatrix}
e^{\frac{3 i \tau \omega}{2}} & 0 & 0 & 0 & 0 & 0 \\
0 & e^{\frac{i \tau \omega}{2}} \cos(J \tau) & -i e^{\frac{i \tau \omega}{2}} \sin( J \tau) & 0 & 0 & 0 \\
0 & -i e^{\frac{i \tau \omega}{2}} \sin(J \tau) & e^{\frac{i \tau \omega}{2}} \cos(J \tau) & 0 & 0 & 0 \\
0 & 0 & 0 & e^{-\frac{i \tau \omega}{2}} \cos(J \tau) & -i e^{-\frac{i \tau \omega}{2}} \sin(J \tau) & 0 \\
0 & 0 & 0 & -i e^{-\frac{i \tau \omega}{2}} \sin(J \tau) & e^{-\frac{i \tau \omega}{2}} \cos(J \tau) & 0 \\
0 & 0 & 0 & 0 & 0 & e^{-\frac{3 i \tau \omega}{2}}
\end{pmatrix}.
\end{equation}
\end{widetext}

From the unitary operator, we compute the reduced density matrix of the system after each collision step using Eq. (\ref{eq: CPTPmap}). At the $n$-th step, the density matrix of the system is written generically as
\begin{equation}
    \rho_{S}^{(n)}=\begin{pmatrix}
        p_1^{(n)}&c_{12}^{(n)}&c_{13}^{(n)}\\
        [c_{12}^{(n)}]^*&p_2^{(n)}&c_{23}^{(n)}\\
        [c_{13}^{(n)}]^*&[c_{23}^{(n)}]^*&1-p_1^{(n)}-p_2^{(n)}
    \end{pmatrix},
\end{equation}
where $p_1$ is the population of the ground state; $p_2$ and $p_3$ are the populations of the intermediate and highest level in the manifold, respectively; the off-diagonal terms $c_{12}$, $c_{13}$, and $c_{23}$ correspond to the coherences between the levels 1-2, 1-3, and 2-3, respectively.
In the next subsection, we write down the explicit recursive equations for the elements of this density matrix.
\vspace{-0.2cm}
\subsection{Exact expressions}

Using the RI dynamics, Eq. (\ref{eq: CPTPmap}),
we obtain the following recursive expressions for the level populations, 
\begin{equation}
\label{eq:GSpopulationqutrit-qubit}
\begin{aligned}
    p_1^{(n+1)}= \frac{1}{2}&\Bigl[
  p_{1}^{(n)}\bigl((1 + p_{A}) + (1 - p_{A})\cos\bigl(2J\tau\bigr)\bigr)\\
  &+
  p_{2}^{(n)}p_{A}\bigl(1 - \cos\bigl(2J\tau\bigr)\bigr)
\Bigr],
\end{aligned}
\end{equation}
\begin{equation}
\label{eq:p2populationqutrit-qubit}
\begin{aligned}
    &p_2^{(n+1)}=\frac{1}{2}\Bigl[
  p_{1}^{(n)} + p_{2}^{(n)} + p_{A}
  -(2p_{1}^{(n)} + p_{2}^{(n)})p_{A}\\
  & +\bigl(-p_{1}^{(n)} + p_{2}^{(n)} +(-1 + 2p_{1}^{(n)}+p_{2}^{(n)})p_{A}\bigr)
    \cos\bigl(2J\tau\bigr)\Bigr],
\end{aligned}
\end{equation}
\begin{equation}
\label{eq:p3populationqutrit-qubit}
    \begin{aligned}
        p_3^{(n+1)} = \frac{1}{2}&\Bigg\{ p_2^{(n)}\Big[(1-p_A)(1-\cos(2J\tau))\Big]\\
       &- p_3^{(n)}\Big[ p_A(1-\cos(2J\tau))-2\Big]\Bigg\}.
    \end{aligned}
\end{equation}
Interestingly, the population dynamics described by Eqs. (\ref{eq:GSpopulationqutrit-qubit}), (\ref{eq:p2populationqutrit-qubit}) and (\ref{eq:p3populationqutrit-qubit}) is decoupled from the coherences terms. Furthermore, these equations can be expressed more compactly by considering that under energy-conserving interaction conditions, the system must reach thermal equilibrium at the temperature of the ancilla \cite{Barra17}, 
\bea
\label{eq:thermal_populations}
    p_1^* = \frac{p_A^2}{1-p_A+p_A^2},\ 
    p_2^* = \frac{p_A(1-p_A)}{1-p_A+p_A^2},\ 
    p_3^* = \frac{(1-p_A)^2}{1-p_A+p_A^2},
    \nonumber
\eea
where the ground state population of the ancilla is $p_A=\frac{1}{1+e^{-\beta\omega}}$.
Subtracting these steady-state values from the population equations simplifies the recursive relations to
\bea
\label{eq:populations-with-ss_grouping}
        p_1^{(n+1)}-p_1^*&=&\eta_{11}(p_1^{(n)}-p_1^*)+\eta_{12}(p_2^{(n)}-p_2^*),
        \nonumber\\
         p_2^{(n+1)} -p_2^* &=& \eta_{21}(p_1^{(n)}-p_1^*)+\eta_{22}(p_2^{(n)}-p_2^*)
        \nonumber\\
        &+&\eta_{23}(p_3^{(n)}-p_3^*),\nonumber\\
        p_3^{(n+1)}-p_3^* &=& \eta_{32}(p_2^{(n)}-p_2^*)+\eta_{33}(p_3^{(n)}-p_3^*).
\eea
Here, we have defined the following dimensionless rates, $0\leq\eta\leq 1$,
\bea
\label{eq:qutrit-qubit-eta-values}
    \eta_{11} &=& \frac{1}{2}[(1+p_A)+(1-p_A)\cos(2J\tau)],
    \nonumber\\
    \eta_{12} &=& \eta_{23} =\frac{1}{2}p_A(1-\cos(2J\tau)),
\nonumber\\
    \eta_{21} &=& \eta_{32}= \frac{1}{2}(1-p_A)(1-\cos(2J\tau)),
    \nonumber\\
    \eta_{22} &=& \frac{1}{2}(1+\cos(2J\tau)),
   \nonumber\\
    \eta_{33} &=& -\frac{1}{2}\left[p_A(1-\cos(2J\tau))-2\right].
\eea
\begin{figure*}[tbp]
    \centering
    \includegraphics[width=\linewidth]{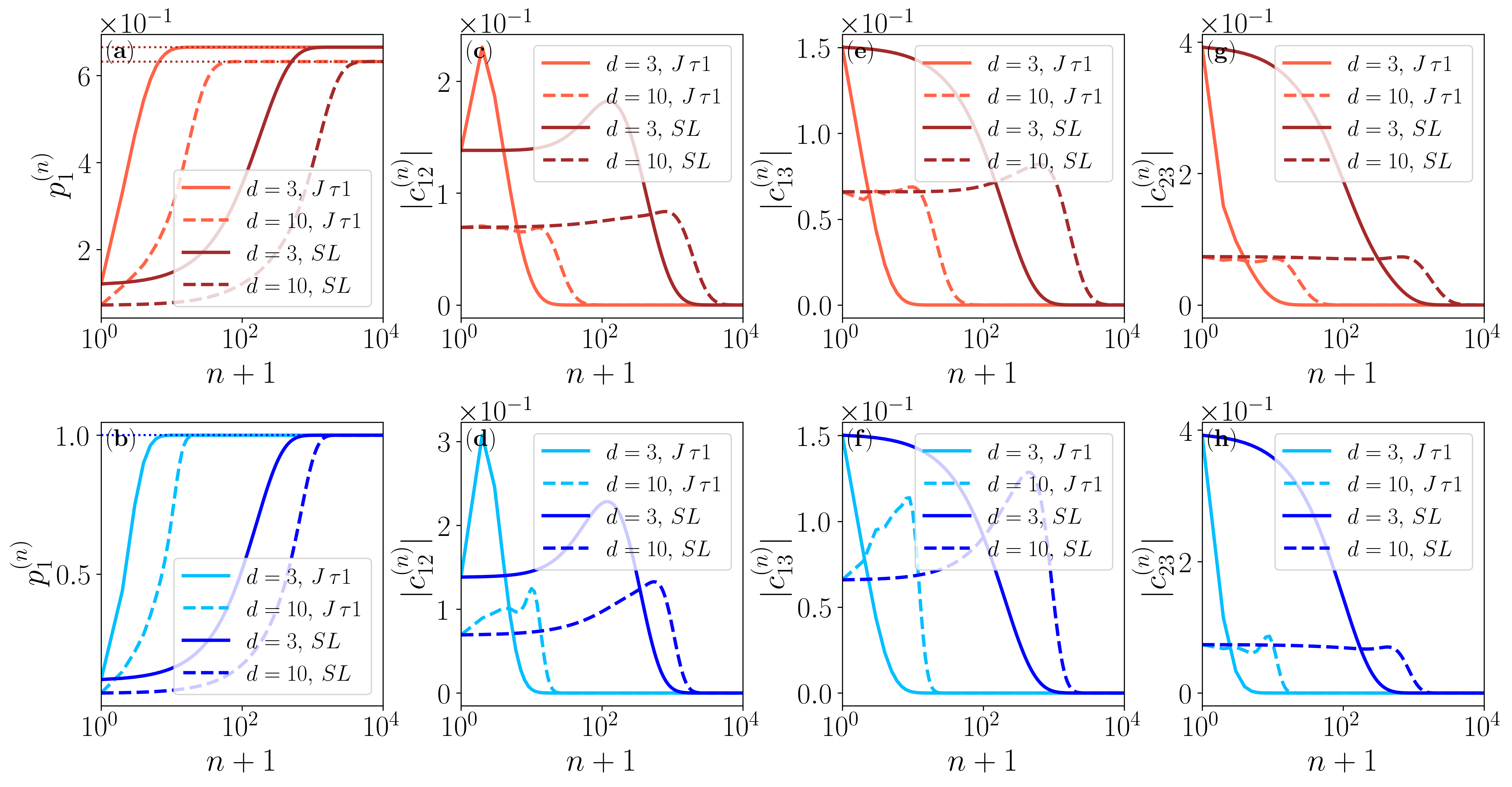} 
    \caption{Dynamics of the ground-state population and selected coherences for spin-$s$ systems interacting with thermal qubit ancillas at two different temperatures, $\beta=1$ (top panels) and $\beta=10$ (bottom panels). We compare two system sizes, $d=3$ (spin-$1$, solid lines) and $d=10$ (spin-$9/2$, dashed lines), and two interaction regimes: the $J\tau1$ limit (red for the top panel and light blue for the bottom panel) and stroboscopic-Lindblad limit (brown for the top panel and blue for the bottom panel). Panels (a) and (b) show the evolution of the ground-state population of the system, $p_1$, with dotted lines indicating the corresponding thermal state of the system. Panels (c)-(h) show the decay of coherences $|c_{12}|$, $|c_{13}|$ and $|c_{23}|$. 
    All simulations were performed under the following conditions: $\omega=1$, $J\tau=1$ with $J = 10^{-3}$ and $\tau = 10^3$ for the $J\tau1$ limit and $J = 10$ with $\tau = 10^{-2}$ for the stroboscopic-Lindblad limit. The initial state for both systems is given by a random initial state $\rho_S^{(0)}$. 
    }
    \label{fig:multiplot-p1-and-coherences}
\end{figure*}
%
%
The dynamics corresponding to the ground-state population $p_1$ is shown in Fig. \ref{fig:multiplot-p1-and-coherences} (a) and (b) at two different inverse temperatures, $\beta=1$ and $\beta=10$, respectively. In each plot, the stroboscopic-Lindblad limits (in darker colors) and the $J\tau1$ limits (in lighter colors) are compared for two different systems of dimensions $s=1$ $(d=3)$ in solid lines and $s=9/2$ $(d=10)$ in dashed lines. 
We observe the following: (i)
The $J\tau1$ limit requires fewer collision steps to bring the system to thermal equilibrium than the Stroboscopic-Lindblad case. This holds at different temperatures, as well as for systems of different dimensions.
(ii) In the low temperature limit, where $p_A \simeq 1$, the system reaches the thermal state in {\it fewer} iterations than for a higher temperature; compare panel (a) (high target temperature) to panel (b) (low target temperature) in Fig. \ref{fig:multiplot-p1-and-coherences}. This Mpemba-like effect is discussed in Sec. \ref{sec:Mpemba}.

We similarly derive the recursive relations for coherences. Starting from the explicit RI expressions, we find that these are decoupled from the populations, obeying
%
%
%
\begin{equation}
    \begin{aligned}
        \label{eq: equation for the coherences}     &c_{12}^{(n+1)}=e^{i\tau\omega}\left[c_{12}^{(n)}\psi_{11}+c_{23}^{(n)}\psi_{13}\right],\\
        &c_{13}^{(n+1)}=e^{2i\tau\omega}c_{13}^{(n)}\psi_{22},\\
        &c_{23}^{(n+1)}=e^{i\tau\omega}\left[c_{12}^{(n)}\psi_{31}+c_{23}^{(n)}\psi_{33}\right],
    \end{aligned}
\end{equation}
where the coefficients $\psi_{ij}$ are
\bea
\label{eq:psi11andpsi13first}
    \psi_{11} &=& \frac{1}{2}\left[1-p_A+2p_A\cos(J\tau)+(1-p_A)\cos(2J\tau)\right],
    \nonumber\\
     \psi_{13} &=& \frac{p_A}{2}\left[1-\cos(2J\tau)\right], \,\,\,
    \psi_{22} = \cos(J\tau),
    \nonumber\\
        \psi_{31}&=&\frac{1}{2}\left[1 - p_A -(1-p_A) \cos(2 J \tau)\right],\nonumber\\
        \psi_{33} &=& \frac{1}{2}\left[p_A+2(1-p_A)\cos(J\tau)+p_A\cos(2J\tau)\right].
\eea
From Eq. (\ref{eq: equation for the coherences}), we note that $c_{13}$ evolves independently through a simple 
decay form, modulated by $\cos(J\tau)$. This behavior can be observed in panels (e) and (f) of Fig. \ref{fig:multiplot-p1-and-coherences}, where $|c_{13}|$ decays monotonically to zero for both temperature regimes (see $d=3$ full lines). In contrast, the dynamics of $c_{12}$ and $c_{23}$ is more complicated, since these terms are dynamically coupled through the coefficients $\psi_{13}$ and $\psi_{31}$. For $\beta=1$, the term $c_{23}$ is coupled to $c_{12}$, while for large $\beta$, $\psi_{31}$ essentially vanishes, resulting in a decoupled decay for $c_{23}$. We do not explicitly observe this coupling in panel (e) because even at inverse temperature $\beta=1$, the term $\psi_{31}$ is rather small, only becoming influential at higher temperatures. However, $c_{12}$ remains dynamically coupled to $c_{23}$ even in this low-temperature limit due to the $\psi_{13}$ term, sustaining a transient feed-in from the decaying of $c_{23}$, which gives rise to the nonmonotonic dynamics present at panels (c) and (d) of Fig. \ref{fig:multiplot-p1-and-coherences}.

With the general expressions at hand, we now proceed to solve the dynamics of the three-level system at zero temperature, first in the general case, without constraining $\tau$ and $J$, see Sec. \ref{subsec:EOM_Jtau1}, 
then at the stroboscopic-Lindblad regime, Sec. \ref{subsec:EOM_SL_limit}.

\subsection{Dynamics at zero temperature}
\label{subsec:EOM_Jtau1}

In this subsection, we solve for the dynamics of the three-level system assuming zero temperatures with the ancilla being fully polarized in its ground state, $p_A=1$. In addition, we define $\lambda =\cos(2J\tau)$ and introduce the shorthand notation $\lambda_+ =\frac{1}{2}\left(1+\lambda\right )$ and $\lambda_- =\frac{1}{2}\left(1-\lambda\right )$. Note that $\lambda_++\lambda_-=1$ and that 
$\lambda_{\pm}\leq 1$.
Under this zero temperature setting, the coefficients from Eqs. (\ref{eq:qutrit-qubit-eta-values}) simplify to
\bea
        \eta_{11} &=& 1, \,\,\,\eta_{21}=\eta_{32}=0,
        \nonumber\\
        \eta_{12}&=&\eta_{23} =\lambda_-, \,\,\,\,
        \eta_{22}=\eta_{33} =\lambda_+.
\eea
Substituting these expressions into the recursive relations, Eq. (\ref{eq:populations-with-ss_grouping}), we obtain
\begin{equation}
\label{eq:populations-for-pA1-with-lambdaM}
    \begin{aligned}
        & p_1^{(n+1)}=p_1^{(n)}+\lambda_- p_2^{(n)},\\
        & p_2^{(n+1)} =\lambda_+p_2^{(n)}+\lambda_-p_3^{(n)},\\
        & p_3^{(n+1)}=\lambda_+p_3^{(n)}.
    \end{aligned}
\end{equation}
These equations convey a clear intuition:
The population of the ground state after $n+1$ collisions is given by its population after $n$ collisions, plus an additional contribution, $\lambda_-p_2^{(n)}$ due to decay of population from the level above it (2). Similarly, the population of level 2 after $n+1$ collisions is given by what remains in that state,  $(1-\lambda_-)p_2^{(n)}$, and an added population due to decay from the level above (3). The population of level 3 after each collision is reduced (or left unchanged) by the factor $(1-\lambda_-)p_3^{(n)}$.

We solve equations
(\ref{eq:populations-for-pA1-with-lambdaM})
in Appendix \ref{AppendixT0}, arriving at
%
\bea
 \label{eq:p1-pA1-important-resultM}
   p_1^{(n)} &=& 1-(\lambda_+)^n \left[p_2^{(0)}+p_3^{(0)}\left(1+n\frac{\lambda_-}{\lambda_+}\right)\right],
   \nonumber\\
    p_2^{(n)} &=& (\lambda_+)^{(n-1)} \left[\lambda_+p_2^{(0)}+n\lambda_-p_3^{(0)}\right],
    \nonumber\\
   1&=& p_1^{(n)}+p_2^{(n)}+p_3^{(n)}.
\eea
Since $\lambda_{\pm}\leq 1$, 
the system eventually reaches the zero temperature limit, $p_1^{*}\to 1$, $p_{2,3}^{* }\to 0$ in the steady state, besides at special points when the inequality is saturated.

We turn to the evolution of the coherences under the same limit of zero temperature for the ancilla, $p_A=1$. Substituting this condition into Eq. (\ref{eq:psi11andpsi13first}), we find that the coefficients simplify to
\bea
        \label{eq: constants-of-the-coherences}
        \psi_{11} &=& \psi_{22} = \mu,
        \nonumber\\
        \psi_{13} &=&\lambda_-, \,\,\
         \psi_{33} = \lambda_+,\,\,\,
        \psi_{31} = 0,
\eea
where we define $\mu = \cos(J\tau)$. Substituting these into the recurrence relation, Eq. (\ref{eq: equation for the coherences}), yields 
%
%
%
\bea
    \label{eq:c13-markovian-map}
    c_{12}^{(n)}&=&e^{i\tau\omega n} \bigg[\left(c_{12}^{(0)}-\frac{\lambda_-c_{23}^{(0)}}{\lambda_+-\mu}\right)\mu^n
        + \frac{\lambda_-c_{23}^{(0)}}{\lambda_+-\mu}(\lambda_+)^n
        \Bigg],
            \nonumber\\
    c_{13}^{(n)} &=& e^{2i\tau\omega n} \mu^n c_{13}^{(0)},
    \nonumber\\
    c_{23}^{(n)}&=&e^{i\tau\omega n} (\lambda_+)^nc_{23}^{(0)}.
\eea
For details, see Appendix \ref{AppendixT0}.
Since $|\mu|\leq 1$ and $\lambda_{\pm}\leq 1$, coherences generically decay to zero, besides special points that saturate these inequalities, where the dynamics is frozen.

We emphasize that so far we made no assumptions on the strength of the interaction $J$ and correspondingly the collision time, $\tau$.

%

\subsection{Dynamics in the Stroboscopic-Lindblad limit}
\label{subsec:EOM_SL_limit}

We now derive equations of motion that govern the dynamics in the stroboscopic Lindblad limit, which corresponds to $J\tau \xrightarrow[]{}0$ while keeping $J^2\tau$ constant. Specifically, we choose $J^2\tau =\Gamma$, where $\Gamma$ defines an effective rate constant. We start our derivation for the case of general temperature of the ancilla, then solve the dynamics at zero temperature. 

To proceed, we expand the coefficients from Eqs. (\ref{eq:qutrit-qubit-eta-values})
to second order in $J\tau$. 
This yields, e.g., that $\eta_{11} \simeq 1- J^2\tau^2(1-p_A)$ while $\eta_{12} \simeq p_AJ^2\tau^2$. Continuing in this manner for all terms, 
%
and substituting these results into the population recursion relations, Eq. (\ref{eq:populations-with-ss_grouping}), we obtain 
\begin{equation}
\label{eq: populations-before-Lindblad-limit}
    \begin{aligned}
        &\frac{\Delta p_1}{\tau} = J^2\tau\left\{ -p_1^{(n)}(1-p_A)+p_2^{(n)}p_A\right\},\\
        & \frac{\Delta p_2}{\tau}=J^2 \tau\left\{ p_1^{(n)}(1-p_A)-p_2^{(n)}+p_Ap_3^{(n)}\right\},\\
        &\frac{\Delta p_3}{\tau}=J^2\tau\left\{p_2^{(n)}(1-p_A)-p_3^{(n)}p_A\right\}.
    \end{aligned}
\end{equation}
To convert this discrete map to a continuous-time description, we must take the limit mentioned above. This leads to a set of coupled differential equations for the populations,
\begin{equation}
\label{eq: populations-qutrit-qubit-Lindblad}
    \begin{aligned}
        & \dot{p_1}(t) = \Gamma \left[-p_1(t)(1-p_A)+p_2(t)p_A \right],\\
        &\dot{p}_2(t)=\Gamma\left[ p_1(t)(1-p_A)-p_2(t)+p_3(t)p_A\right],\\
        & \dot{p}_3(t) = \Gamma\left[p_2(t)(1-p_A)-p_3(t)p_A\right].
    \end{aligned}
\end{equation}
Similarly, we derive equations of motion for the dynamics of coherences by first expanding the $\psi_{ij}$ coefficients from Eqs. (\ref{eq:psi11andpsi13first}) to second order in $J\tau$.
To derive the corresponding differential equations, we additionally assume that $\tau\omega\xrightarrow[]{}0$ and $\omega^2\tau\xrightarrow[]{}0$, which corresponds to a regime where $J\gg\omega$. Under these conditions, the recursive relations for the coherences translate into the following set of differential equations,
\begin{equation}
\label{eq:diff-eq-for-coherences-qutrit-qubit}
    \begin{aligned}
        &\dot{c}_{12}(t)=\Gamma\left[-\frac{1}{2}(2-p_A)c_{12}(t)+p_Ac_{23}(t)\right],\\
        &\dot{c}_{13}(t)=-\frac{\Gamma}{2}c_{13}(t),\\
        &\dot{c}_{23}(t)=-\Gamma\left[\frac{1}{2}(1+p_A)c_{23}(t)+(1-p_A)c_{12}(t)\right].
    \end{aligned}
\end{equation}
We observe that $c_{13}$ decays exponentially at a rate constant proportional to $\Gamma$, while $c_{12}$ and $c_{23}$ are coupled with mutual feed-in terms modulated by $p_A$, as we have discussed in the previous section.

Analyzing the low-temperature limit corresponding to $p_A = 1$, Eq. (\ref{eq: populations-qutrit-qubit-Lindblad}) simplifies to
\begin{equation}
    \label{eq:populations-qutrit-qubit-Lindblad-pA1}
    \begin{aligned}
        & \dot{p_1}(t) = \Gamma p_2(t),\\
        &\dot{p}_2(t)=\Gamma\left[-p_2(t)+p_3(t)\right],\\
        & \dot{p}_3(t) = -\Gamma p_3(t).
    \end{aligned}
\end{equation}
This set of equations shows clear hierarchical relaxation processes, where the population of the highest excited state $p_3(t)$ decays exponentially to zero at a rate constant $\Gamma$, feeding the population of the intermediate state $p_2(t)$, which also decays to the ground state $p_1(t)$. In the long time limit, all the population decays to the ground state.
%
%
%
Solving these equations, we obtain
\begin{equation}
    \begin{aligned}
        &p_1(t) = 1-e^{-\Gamma t}\left[p_2(0)+p_3(0)(1+\Gamma t)\right],\\
        &p_2(t) = e^{-\Gamma t}\left[p_2(0)+\Gamma tp_3(0)\right],\\
        & p_3(t) = e^{-\Gamma t}p_3(0).
    \end{aligned}
    \label{eq:popSLT0}
\end{equation}

For the coherences, Eq. (\ref{eq:diff-eq-for-coherences-qutrit-qubit}) becomes
\begin{equation}
\label{eq:diff-eq-for-coherences-qutrit-qubit-pA1}
    \begin{aligned}
        &\dot{c}_{12}(t)=\Gamma\left[-\frac{1}{2}c_{12}(t)+c_{23}(t)\right],\\
        &\dot{c}_{13}(t)=-\frac{\Gamma}{2}c_{13}(t),\\
        &\dot{c}_{23}(t)=-{\Gamma}c_{23}(t).
    \end{aligned}
\end{equation}
In this zero-temperature regime, $c_{23}(t)$ is decoupled from $c_{12}(t)$ and it decays exponentially at a rate two times larger than that of $c_{13}(t)$. This can be observed in Fig. \ref{fig:multiplot-p1-and-coherences}: a careful comparison shows that panel (f) displays a decay two times slower than in panel (h) for the $d=3$ case. Meanwhile, the coherence $c_{12}(t)$ is dynamically coupled to $c_{23}(t)$, receiving transient contributions from the decaying upper coherence before itself decaying to zero,
\begin{equation}
    \begin{aligned}
        &c_{12}(t) = 
        e^{-\Gamma t/2 }c_{12}(0) + 2
        \left(e^{-\Gamma t/2 }-e^{-\Gamma t}\right)c_{23}(0),\\
        &c_{13}(t) = e^{-\Gamma t/2 }c_{13}(0),\\
        & c_{23}(t) = e^{-\Gamma t}c_{23}(0).
    \end{aligned}
\end{equation}
%


We conclude section \ref{sec:EOM} with a few comments.

First, we note that solving the recursive equations (\ref{eq:GSpopulationqutrit-qubit})–(\ref{eq:p3populationqutrit-qubit}) exactly at arbitrary temperature is a nontrivial task; here, we provide the exact solution in the zero-temperature limit.
Moreover, in Sec. \ref{sec:Mpemba}, we derive an accurate approximate expression for the simulation time required for thermal state preparation {\it at an arbitrary temperature}. There, we show that to estimate the simulation time effectively, it suffices to analyze the slowest dynamical mode, specifically, its decay rate and the projection of the initial condition onto this mode.

Second, we note that in the SL regime, the dynamics (\ref{eq: populations-qutrit-qubit-Lindblad})
is quite simple. Once again we solve it here exactly only at zero temperature, and we left the analysis at arbitrary temperature to Sec. \ref{sec:Mpemba} where we focus on the slowest evolving mode as a key to the estimation of simulation time for the preparation of a thermal state.

\begin{figure*}[t]
\centering
\includegraphics[width=\linewidth]{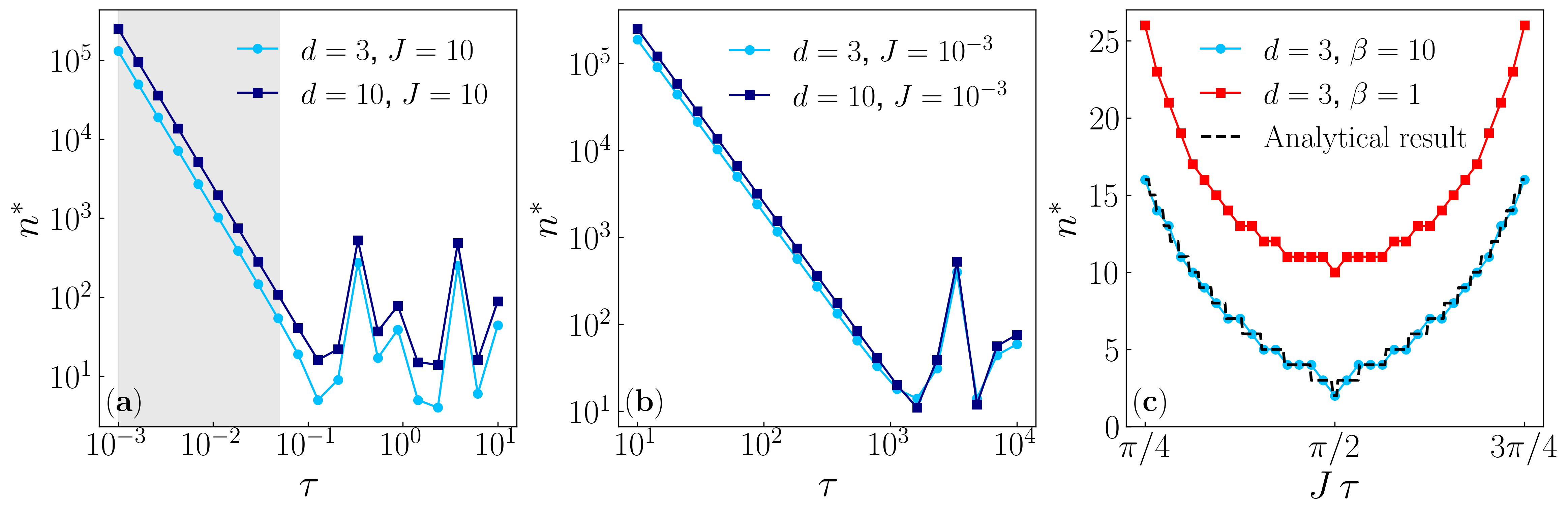} 
    \caption{
Minimum number of collisions required for the thermalization of a system starting from a completely mixed state and  targeting a low temperature state ($\beta=10$) within trace distance $\epsilon$, using (a) $J=10$ and (b) $J=10^{-3}$.
We present results for system with $d=3$ and $d=10$ states; the minimum number of collisions required to thermalize either systems follows the same pattern.
In panel (a), the shaded region corresponds to the SL limit where $J\tau \xrightarrow[]{}0$ but $J^2\tau = \Gamma$ with $\Gamma \sim 1$. 
In panel (b), the zone around the first minimum corresponds to the $J\tau1$ limit. 
(c) A zoom-in on the region around  $J\tau \sim1$, showing that the minimum occurs precisely at $J\tau = \pi/2$. 
We present results at low (light blue) and intermediate (red) ancilla (target thermalization) temperatures; in both cases, the minimum takes place at $J\tau = \pi/2$.
We use the following settings: 
The initial state of the system is maximally mixed, $\omega=1$, and $\epsilon=10^{-4}$. }
\label{fig:total-sim-time-plot1}
\end{figure*}

\section{
Ground state preparation}
\label{sec:time}

The task considered in this paper is to prepare a thermal state, typically from an infinite-temperature completely mixed state.  
The total simulation time, $T_{\text{sim}}$, quantifies the duration needed for the system to reach a specific thermal equilibrium state from a certain initial condition. In the continuous-time limit, this corresponds to the time required for the state of the system to approach the target thermal state $\rho_S^*$ within a specified trace distance $\epsilon$. Formally, we define this convergence condition as
\begin{equation}
\label{eq: trace-distance-calculation}
D(\rho_S(T_{\text{sim}}),\rho_S^{*}) =\frac{1}{2}\text{Tr}|\rho_S(T_{\text{sim}})-\rho_S^{*}|\leq \epsilon,
\end{equation}
where $D(\cdot \ , \ \cdot)$ corresponds to the trace distance between the two states \cite{NielsenChuang}.

In the discrete-time case, e.g., corresponding to the $J\tau1$ limit, the total simulation time is expressed as
\begin{equation}
    T_{\text{sim}} = n^*\tau,
\end{equation}
where $n^*$ is the minimum number of collisions required for the system to satisfy the condition 
$D(\rho_S^{(n^*)},\rho_S^{*}) \leq \epsilon$, and $\tau$ is the finite duration of collisions.

\subsection{Key observations}
Focusing on the three-level system and fixing the coupling value $J$, we study which dynamical regime, and at what parameters, gives us the shortest simulation time. As we have seen in Fig. \ref{fig:multiplot-p1-and-coherences}, the system thermalizes in fewer iterations for the $J\tau1$ limit, with the convergence further accelerated at the low-temperature limit. We now examine aspects of this behavior.

In Fig. \ref{fig:total-sim-time-plot1}, we inquire which interaction regime, SL or $J\tau1$, achieves thermalization faster. We
present the minimum number of steps required to satisfy thermalization within the $\epsilon$ distance as a function of the interaction time, testing thermalization for both 3- and 10-level systems. 

The Stroboscopic-Lindblad regime is analyzed in Fig. \ref{fig:total-sim-time-plot1}(a), presented in the shaded region.
Using $J=10$, we find that the first minimum in $n^*$ appears {\it outside the SL regime}, showing up around $J\tau = 1$. This result holds for both 3-level and 10-level systems. 
This indicates that for a system with more energy levels, the dynamics is similar. Importantly, the SL limit lies outside the optimal regime for thermal state preparation. 
Next, in Fig. \ref{fig:total-sim-time-plot1}(b), we turn to the weak $J$ case, and repeat the analysis with a fixed coupling $J=10^{-3}$. We once again verify that the first local minimum of $n^*$ occurs near $J\tau = 1$, corresponding to the $J\tau1$ operational regime. 

Based on Fig. \ref{fig:total-sim-time-plot1}(a)-(b), we conclude that the optimal regime of operation, leading to smallest $n^*$, shows once $J\tau\approx1$, different from the common SL regime.
To better understand the behavior around this point, in Fig. \ref{fig:total-sim-time-plot1} (c), we plot $n^*$ as a function of $J\tau$ for the qutrit system at two different temperatures, $\beta = 1$ (red) and $\beta = 10$ (blue). For both temperatures, we observe a minimum in simulation time at $J\tau = \pi/2$. As $J\tau$ departs from this point, the number of collisions starts to increase, diverging at the limiting values $J\tau = 0$ and $J\tau = \pi$. This explains why in panels (a) and (b), on the left part of the plot, which corresponds to values where $J\tau \xrightarrow[]{}0$ we obtain a larger value for $n^*$. Furthermore, after reaching the first minimum, we start observing oscillations in $n^*$. In these oscillatory dynamics, the minima correspond to values $J\tau=(2k+1)\pi/2$, and the maxima correspond to $J\tau = k\pi$ with $k\in \mathbb{N}$.

The total simulation time can be analytically derived at zero temperature; we achieve
this result in Eq.
(\ref{eq:nstar-value}). 
We also present this zero-temperature analytical result
in Fig. \ref{fig:total-sim-time-plot1}(c)
as a black dashed line. The apparent step-like structure of the analytical curve comes from rounding the theoretical value of $n^*$ to the nearest integer during the evaluation.

\subsection{Estimation of simulation time in the $J\tau1$ limit}
\label{subsec:Tsim_Jtau1_limit}

We derive here an analytical expression for $T_{\text{sim}}$ in the $J\tau1$ limit and assuming a low (zero) temperature for the ancillas, which is the target temperature for the qutrit.
Assuming that the initial state of the qutrit system is diagonal, Eq. (\ref{eq: trace-distance-calculation}) simplifies to
\begin{align}
    D(\rho_S^{(n^*)},\rho_S^*) &= \frac{1}{2}\left\{\left|p_1^{(n^*)}-1\right|+\left|p_2^{(n^*)}\right|+\left|p_3^{(n^*)}\right|\right\} \leq \epsilon \nonumber \\
    &= \frac{1}{2}\left\{1 - p_1^{(n^*)} + p_2^{(n^*)} + p_3^{(n^*)} \right\} \nonumber \\
    &= 1 - p_1^{(n^*)} \leq \epsilon
    \label{eq:trace-distance-simplification}
\end{align}

Substituting the solution for the ground state population of the system from Eq. (\ref{eq:p1-pA1-important-resultM}), and saturating the inequality, we solve for $n^*$,
\begin{equation}
    \label{eq: condition-nstar-qutrit-qubit}
    (\lambda_+)^{n^*}\left[p_2^{(0)}+p_3^{(0)}\left(1+n^*\frac{\lambda_-}{\lambda_+}\right)\right] = \epsilon.
\end{equation}
Recall that 
$   \lambda_+ =\cos^2(J\tau)$, $\lambda_-=\sin^2(J\tau)$.
This expression can be rewritten as
\begin{equation}
\label{eq: condition-nstar-qutrit-qubit-rewritten}
p_2^{(0)}+p_3^{(0)}+p_3^{(0)}\frac{\lambda_-}{\lambda_+}n^* = \epsilon e^{-n^*\ln(\lambda_+)}.
\end{equation}
%
The formal solution of this equation in terms of $n^*$ is
\begin{widetext}
\begin{equation}
\label{eq:nstar-value}
n^* = -\frac{\lambda_+}{\lambda_-}\left(\frac{p_2^{(0)}+p_3^{(0)}}{p_3^{(0)}}\right)
+\frac{1}{\ln(\lambda_+)} \, W_{-1}\left[
(\ln \lambda_+)\frac{\lambda_+ \epsilon}{\lambda_- p_3^{(0)}}
\cdot (\lambda_+)^{\frac{\lambda_+\left(p_2^{(0)}+p_3^{(0)}\right)}{\lambda_- p_3^{(0)}}}
\right],
\end{equation}
\end{widetext}
where $W_{-1}(z)$ corresponds to the lower $k=-1$ branch of the Lambert function, $w=W_{-1}(z)$ solving
$we^{w}=z$. 
Details of arriving at this solution, 
the justification for using this specific branch, and the conditions that ensure that this result gives a real value are discussed in Appendix \ref{sec: Lambert function}.

We now study scenarios in which the condition in Eq. (\ref{eq: condition-for-epsilon-lambert}), $z\geq -1/e$, is violated, thus making  Eq.~(\ref{eq:nstar-value}) either divergent or invalid. These special points are instrumental for optimizing the RI scheme for thermal state preparation tasks.
A singularity in Eq. (\ref{eq:nstar-value}) arises when $\lambda_+ = 1$, 
which occurs when
\begin{equation}
    \label{eq: condition-Jtau-to-freeze-dynamics}
    J\tau = \pi k, \quad k \in \mathbb{N}.
\end{equation}
Recalling Eq. (\ref{eq:populations-for-pA1-with-lambdaM}), at these points the dynamics of the populations are frozen, 
    $p_{i}^{(n+1)} = p_{i}^{(n)}$ with $i=1,2,3$.

\begin{figure}[h]
    \centering
    \includegraphics[width=1\linewidth]{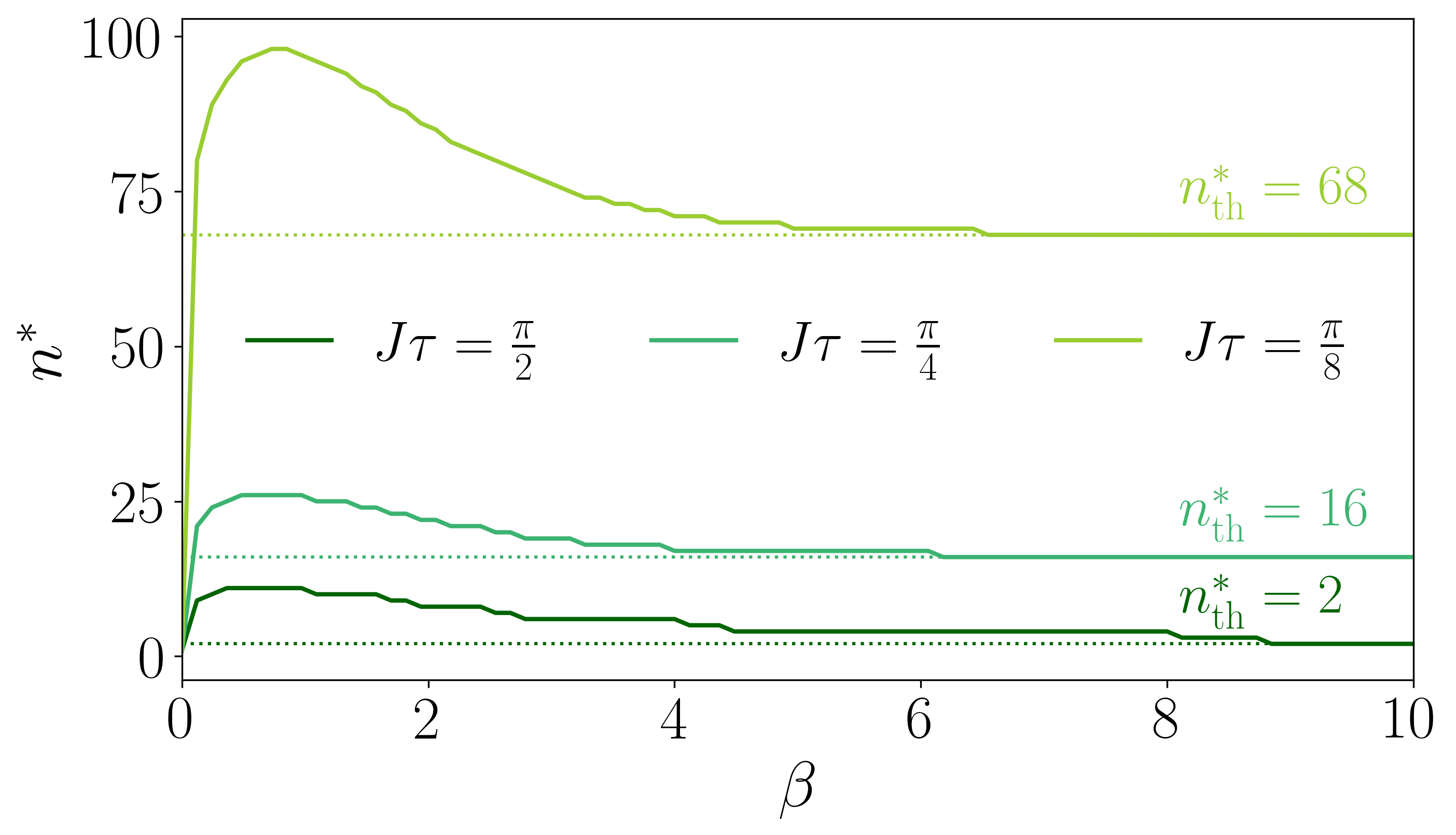} 
    \caption{$J\tau1$ limit. Minimum number of collisions, $n^*$, required for a three-level system to thermalize within $\epsilon$ to temperature $\beta^{-1}$ from a completely mixed state. Different curves correspond to $J\tau = \pi/2$ (dark green), $J\tau = \pi/4$ (medium green) $J\tau = \pi/8$ (light green). The solid lines correspond to  numerical simulations;  dotted lines ($n^*_{\text{th}}$) depict the analytical result at zero temperature predicted by Eq. (\ref{eq:nstar-value}). 
    Energy splitting, identical for the qutrit system and the ancilla is given by $\omega = 1$; the precision is set to $\epsilon = 10^{-4}$. 
    }
    \label{fig:nstar_vs_beta_Jtau1}
\end{figure}

Another scenario that requires special attention corresponds to $\lambda_+ = 0$ and 
$\lambda_- = 1$, which implies that
\begin{equation}
    J\tau = \frac{\pi}{2}(2k+1),
    \quad k \in \mathbb{N}.
\end{equation} 
%
In this limit we get after the first collision, Eq. (\ref{eq:populations-for-pA1-with-lambdaM}),
\begin{equation}
    \label{eq:populations-particular-caseJtaupihalf}
    \begin{aligned}
        &p_1^{(1)} = p_1^{(0)}+p_2^{(0)},\\
        & p_2^{(1)}=p_3^{(0)},\\
        &p_3^{(1)} = 0.
    \end{aligned}
\end{equation}
After the second collision, we have
\begin{equation}
    \label{eq:populations-particular-caseJtaupihalf}
    \begin{aligned}
        &p_1^{(2)} = p_1^{(1)}+p_2^{(1)}=p_1^{(0)}+p_2^{(0)}+p_3^{(0)}=1,\\
        & p_2^{(2)}=p_3^{(1)}=0,\\
        &p_3^{(1)} = 0.
    \end{aligned}
\end{equation}
One can similarly verify that coherences follow similar trends, vanishing after two collisions. 
Therefore, the system thermalizes at $n^* = 2$, which corresponds to two collisions. We observe this phenomenon in Fig. \ref{fig:nstar_vs_beta_Jtau1}, plotted in dark green.  There, as we increase $\beta$ we find that $n^*$ approaches 2 for $J\tau=\pi/2$ when $\beta\to \infty$.
The nonmonotnic behavior of $n^*$
with $\beta$ is explained in Sec. \ref{sec:Mpemba}.

\subsection{Estimation of $T_{\text{sim}}$ in the stroboscopic-Lindblad limit}
\label{subsec:Tsim_SL_limit}

So far, we have not made any assumptions about the strength of the interaction and its duration. We now focus on the SL regime and we compute the total simulation time in the zero-temperature limit.
Based on the solution for the dynamics, Eq. (\ref{eq:popSLT0}), the trace distance condition, Eq. (\ref{eq: trace-distance-calculation}) becomes
\begin{equation}
\label{eq: trace-distance-qt-qb-fulleq}
    D(\rho_S(T_{\text{sim}}) = e^{-\Gamma T_\text{{sim}}} \left[p_2(0)+p_3(0)+p_3(0)\Gamma T_\text{{sim}} \right]\leq \epsilon.
\end{equation}
To simplify notation, we define
$\alpha = p_2(0)+p_3(0)$, $ \beta = \Gamma p_3(0)$.
Thus, Eq. (\ref{eq: trace-distance-qt-qb-fulleq}) becomes
\begin{equation}
    \alpha+\beta T_\text{{sim}} = e^{\Gamma T_\text{{sim}}} \epsilon,
\end{equation}
where we have saturated the bound. Rearranging, we define $y = \alpha + \beta T_\text{{sim}} \xrightarrow[]{} T_\text{{sim}}=\frac{y-\alpha}{\beta}$ and get
\begin{equation}
    \begin{aligned}
  -\frac{\Gamma}{\beta}ye^{-\frac{\Gamma}{\beta}y} = -\frac{\Gamma}{\beta}\epsilon e^{-\frac{\Gamma}{\beta}\alpha}.
    \end{aligned}
\end{equation}
This equation matches the standard Lambert function definition,
$    Y e^Y = X$
where we identify
$Y = -\frac{\Gamma}{\beta} y$, $X = -\frac{\Gamma}{\beta} e^{-\frac{\Gamma}{\beta} \alpha} \epsilon$.
Thus, the total simulation time is
\begin{equation}
\label{eq: Total_sim_time-qutrit-qubit}
    T_{\text{sim}} = -\frac{1}{\Gamma}\left\{ 1+\frac{p_2(0)}{p_3(0)}+W_{-1}\left[-\frac{\epsilon}{p_3(0)} \;e^{-\left(1+\frac{p_2(0)}{p_3(0)}\right)} \right]\right\},
\end{equation}
where once again $W_{-1}(z)$ corresponds to the lower $k=-1$ branch of the Lambert function, $w=W_{-1}(z)$.
The justification for the selection of the lower branch is discussed in Appendix \ref{sec: Lambert function}. Furthermore, in the SL limit, we observe that the Lambert function depends only on the initial conditions and the precision parameter, $\epsilon$. This should be contrasted with the general solution in Eq. (\ref{eq:nstar-value}) that depends explicitly on the values of the interaction parameters $J$ and $\tau$.

The simulation time $T_{\rm sim}$ in the SL limit is  presented in Fig. \ref{fig:TsimLindblad_epsilon_eigenvalues}(a),
obtained from direct numerical simulations, while
Eq. (\ref{eq: Total_sim_time-qutrit-qubit})
is depicted as the asymptotic zero temperature limit.
This figure once again demonstrates a counterintuitive relaxation process, similar to
Fig. \ref{fig:nstar_vs_beta_Jtau1}, which was created in the $J\tau1$ regime. For a maximally mixed initial state (corresponding to an infinite-temperature initial state), it takes longer to cool to intermediate temperatures than to lower temperatures. This phenomenon, which appears in both the SL and the $J\tau1$ limit,
is analogous to the Mpemba effect \cite{Raz17,Klich19, quantumM, FelixM, Goold24}, and we address it in the next section.
As a final note, with respect to Eq. (\ref{eq: Total_sim_time-qutrit-qubit}), in Fig. \ref{fig:TsimLindblad_epsilon_eigenvalues}(b), we observe that $T_{\text{sim}}$ increases logarithmically as the precision parameter $\epsilon$ is reduced. 

\section{Thermal state preparation to nonzero temperature: The Mpemba effect}
\label{sec:Mpemba}

The Mpemba effect is an anomaly in relaxation dynamics where a system initially at a higher temperature can cool faster than the same system when starting from a lower temperature. 
Consider a temperature hierarchy,
$T_h>T_c>T_b$, with $T_b$ the temperature of the environment. In the Mpemba effect, a system initialized at $T_h$ reaches $T_b$ more quickly than a system prepared at $T_c$, despite starting further away. Initially observed in water \cite{MpembaW}, this phenomenon has since been reported in a range of physical systems \cite{MpembaRevB}. Within the framework of nonequilibrium thermodynamics, the Markovian Mpemba effect has been analyzed in minimal models such as three-level systems governed by memoryless stochastic dynamics \cite{Raz17}. In such cases, the effect is understood through the spectral properties of the system's Liouvillian generator: specifically, by examining the projection of the initial state onto the {\it slowest decaying} eigenmode. A strong Markovian Mpemba effect occurs when this projection vanishes for the hot initial condition but remains nonzero for the colder one, leading to fast thermalization dynamics under $T_h$. A weaker version of the Mpemba effect arises when the projection is smaller for the hot state, still resulting in it anomalously more quickly approaching equilibrium at $T_b$ \cite{Klich19}.
Extensions of the classical Mpemba effects to quantum systems were discussed in Refs. \cite{quantumM, FelixM, Goold24}.

We emphasize that the Markovian Mpemba effect involves preparing the system in two distinct initial states, each corresponding to a different temperature, and comparing their relaxation times with a common background temperature, $T_b$.

\begin{figure*}[t!]
    \centering
    \includegraphics[width=1\linewidth]{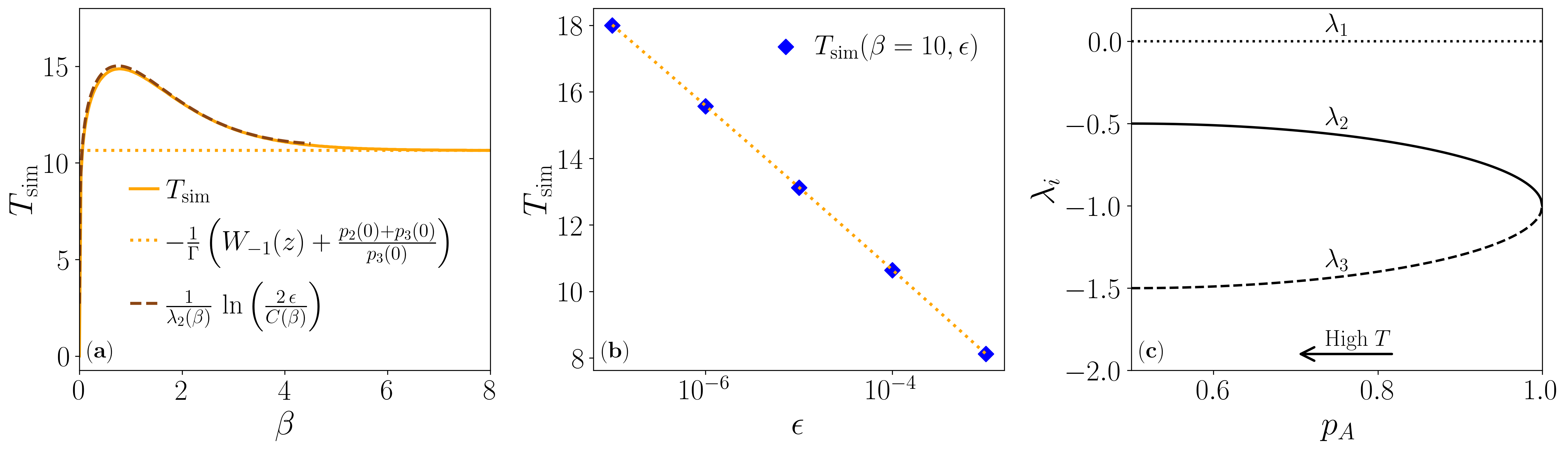} 
    \caption{(a) Total simulation time $T_{\text{sim}}$ in the stroboscopic-Lindblad limit as a function of the inverse temperature $\beta$. The solid line corresponds to the numerical results, the dotted line represents the asymptotic expression Eq. (\ref{eq:nstar-value}), and the dashed brown curve corresponds to the analytic result in Eq. (\ref{eq:Mpemba_Tsim_lambda2}), correctly describing the behavior for intermediate temperatures. We fixed the energy splitting to $\omega=1$, the precision to $\epsilon=10^{-4}$, and $J= 10$ with  $\tau=10^{-2}$ to efficiently capture the SL limit for $\Gamma=1$. (b) Blue diamonds correspond to the numerical results for $\beta = 10$ with varying $\epsilon$. As the required precision increases (i.e., smaller $\epsilon$), the simulation time $T_{\text{sim}}$ grows logarithmically. The dotted lines represent the analytical result given by Eq. (\ref{eq:Mpemba_Tsim_lambda2}). (c) Eigenvalues of the Liouvillian at finite temperature, with $\lambda_1 = 0$, and $\lambda_2 \geq \lambda_3$.
    }
\label{fig:TsimLindblad_epsilon_eigenvalues}
\end{figure*}

In Figures \ref{fig:nstar_vs_beta_Jtau1} and \ref{fig:TsimLindblad_epsilon_eigenvalues}(a), we demonstrate a complementary, Mpemba-like effect taking place in thermal state preparation with the RI protocol. Here, the system is initialized in a maximally mixed state, representative of a high-temperature condition, while the temperature of the environment (ancilla) is varied. Counterintuitively, we observe that over a broad range of ancilla temperatures, the time required to reach a colder stationary state ($\beta\gg1$), of a larger distance from the initial condition, is {\it shorter} than the time needed to reach a hotter one ($\beta\gtrsim1$), which is closer to the initial condition. This anomalous relaxation behavior, characteristic of the Mpemba effect, is illustrated in both the $J\tau1$ regime and the stroboscopic-Lindblad regime, as shown in Figures \ref{fig:nstar_vs_beta_Jtau1} and \ref{fig:TsimLindblad_epsilon_eigenvalues}(a), respectively.

In what follows, we first focus on the stroboscopic-Lindblad limit to uncover the underlying mechanism responsible for this behavior, and then extend our analysis to the general discrete case, including the $J\tau1$ limit to investigate the occurrence of Mpemba dynamics in that regime as well.
We further show that the effect is robust:
It exists, and in fact is further emphasized, at multi-level ($d>3$) systems, and it survives under randomized interactions.  

\subsection{Mpemba effect in the SL limit}
\label{subsec:Mepmba_inSLlimit}

In the SL limit, the evolution of the populations at general {\it non-zero} temperatures is given by the coupled differential equations
(\ref{eq: populations-qutrit-qubit-Lindblad}), written here in matrix form,
\begin{equation}
\label{eq:diff_eq_populations_nonzeroT}
    \frac{d}{dt} \begin{pmatrix}
        p_1(t)\\
        p_2(t)\\
        p_3(t)
    \end{pmatrix} = 
    \Gamma
    \begin{pmatrix}
        -(1-p_A) & p_A & 0\\
        1-p_A & -1 & p_A\\
        0 & 1-p_A & -p_A
    \end{pmatrix}
    \begin{pmatrix}
        p_1(t)\\
        p_2(t)\\
        p_3(t)
    \end{pmatrix}.
\end{equation}
This can be expressed more compactly in a vector form as $\dot{\textbf{p}}(t) = \mathcal{L}\textbf{p}(t)$,
where the superoperator $\mathcal{L}$, known as the Liouvillian, generally includes both unitary and dissipative contributions.
Since $\mathcal{L}$ is non-Hermitian, we must consider its biorthogonal decomposition
\begin{equation}
\label{eq:biorthogonal_equations_Liouvillian}
\mathcal{L} \mathbf{v}_i = \lambda_i \mathbf{v}_i, \quad \mathcal{L}^T \mathbf{u}_i = \lambda_i \mathbf{u}_i, \quad
\end{equation}
where $\{\mathbf{v}_i\}$ and $\{\mathbf{u}_i\}$ form sets of right and left eigenvectors, respectively, satisfying the biorthogonality condition $\mathbf{u}_i^T \cdot \mathbf{v}_j = \delta_{ij}$. The corresponding eigenvalues are 
\begin{equation}
\label{eq:eigenvalues_Liouvillian}
    \lambda_1 = 0, \quad \lambda_2 = -\Gamma(1-\theta), \quad \lambda_3 = -\Gamma(1+\theta),
\end{equation}
with $\theta = \sqrt{p_A(1 - p_A)} \in [0,1/2]$. We illustrate the three eigenvalues of the Liouvillian in Fig. \ref{fig:TsimLindblad_epsilon_eigenvalues}(c). 

As for the eigenvectors,
a convenient choice for the right eigenvector $\{\mathbf{v_1}\}$ is based on the equilibrium solution,
\begin{equation}
    \mathbf{v}_1 = (p_1^*,p_2^*,p_3^*)^T,
\end{equation}
where the normalization is fixed by choosing $v_1^{(1)} = p_1^*$. To construct $\mathbf{v}_2$, we set $v_2^{(3)} = 1$, which gives
\begin{equation}
    \mathbf{v}_2 = \left(-\frac{\theta}{1-p_A},\frac{\theta-1+p_A}{1-p_A},1\right)^T.
\end{equation}
Similarly, for $\mathbf{v}_3$ we set $v_3^{(3)} = 1$, yielding
\begin{equation}
    \mathbf{v}_3 = \left(\frac{\theta}{1-p_A},\frac{-\theta-1+p_A}{1-p_A},1\right)^T.
\end{equation}
For the set of eigenvectors $\{ \mathbf{u_i}\}$, as it will be clear later, we just need to compute the eigenvector $\mathbf{\{u_2\}}$, which can be obtained by solving the second eigenvalue problem. It is properly chosen to satisfy $\mathbf{u}_2^T \cdot \mathbf{v}_j = \delta_{2j}$,
\begin{equation}
\label{eq:u2_vector}
    \mathbf{u}_2 = \frac{1}{2(1-\theta)}\left( -\frac{(1 - p_A)\theta}{p_A}, -1 + p_A + \theta, p_A \right)^T.
\end{equation}
Next, to avoid directly solving Eq. (\ref{eq:diff_eq_populations_nonzeroT}), we propose the following ansatz based on the spectral decomposition,
\begin{equation}
\label{eq:general_ansatz}
    p_i(t) = p^*_i + \sum_{k=2}^dc_{k,i} e^{\lambda_k t}.
\end{equation}
Here, $\lambda_n$ are the eigenvalues of $\mathcal{L}$ ordered as $0= \lambda_1 >\lambda_2\geq\lambda_3\geq\ldots\geq\lambda_N$ and $p^*=(p_1^*,p_2^*,p_3^*)^T$ is the stationary state. For our qutrit system, $d=3$, the eigenvalues are given by Eq. (\ref{eq:eigenvalues_Liouvillian}).

At long times, the slowest decaying eigenmode $\lambda_2$ dominates the dynamics of the system. Therefore, we can approximate Eq. (\ref{eq:general_ansatz}), with $d=3$, by
\begin{equation}
\label{eq: populations_for_long_time}
    p_i(t>\lambda_3^{-1}) \simeq p_i^*+c_{2,i}e^{\lambda_2 t}.
\end{equation}
The total simulated time can be computed as
\begin{equation}
\begin{aligned}
    &\frac{1}{2}\left\{ \left|p_1(T_{\text{sim}})-p_1^*\right|+\left|p_2(T_{\text{sim}})-p_2^*\right|+\left|p_3(T_{\text{sim}})-p_3^*\right|\right\} \leq \epsilon.\\
\end{aligned}
\end{equation}
Saturating the inequality, we write
\begin{equation}
    \begin{aligned}
 & \frac{1}{2}\left\{ \left(|c_{2,1}|+|c_{2,2}|+|c_{2,3}|\right)e^{\lambda_2T_{\text{sim}}}\right\} = \epsilon,\\
\end{aligned}
\end{equation}
where we used Eq. (\ref{eq: populations_for_long_time}). Defining now $C=|c_{2,1}|+|c_{2,2}|+|c_{2,3}|$ we solve for $T_{\text{sim}}$,
\begin{equation}
\label{eq:Mpemba_Tsim_lambda2}
    T_{\text{sim}} \approx \frac{1}{\lambda_2}\ln\left(\frac{2\epsilon}{C}\right) = -\frac{1}{\Gamma(1-\sqrt{p_A(1-p_A)})}\ln\left(\frac{2\epsilon}{C}\right).
\end{equation}
To fully capture the behavior of $T_{\text{sim}}$, we need to compute the value of $C$, which reflects the contribution of the slowest decaying mode to the initial condition. 
We expand the initial deviation from equilibrium, $\Delta p_i(0) = p_i(0)-p_i^*$, as
\begin{equation}
\label{eq:ansatz_initial_conditions}
    \Delta\mathbf{p}(0) = \alpha_2\mathbf{v}_2+\alpha_3 \mathbf{v}_3,
\end{equation}
where the condition $\mathbf{1}^T \cdot \Delta\mathbf{p}(0) = 0$ implies $\alpha_1 = 0$. The time evolution of deviation from equilibrium is governed by
\begin{equation}
    \Delta p_i(t) = \alpha_2 v_i^{(2)} e^{\lambda_2 t} + \alpha_3 v_i^{(3)} e^{\lambda_3 t},
\end{equation}
from which we identify $ c_{2,i} = \alpha_2 v_i^{(2)}$.
To compute $\alpha_2$, we project Eq.~(\ref{eq:ansatz_initial_conditions}) onto the left eigenvector $\mathbf{u}_2^T$, given in Eq. (\ref{eq:u2_vector}), which fulfills the biorthogonality condition $\mathbf{u}_j^T\cdot \mathbf{v}_i = \delta_{ji}$,
\begin{equation}
    \label{eq: coefficient alpha2}
    \alpha_2=\mathbf{u}_2^T\cdot\Delta \mathbf{p}(0).
\end{equation}
Explicitly, we find that
\begin{equation}
    \begin{aligned}
    \alpha_2 &= \frac{1}{2(1-\theta)}\bigg[ 
    \frac{(-1 + p_A)\theta}{p_A} \cdot \Delta p_1(0)\\
    &-(1 - p_A - \theta) \cdot \Delta p_2(0) 
    + p_A \cdot \Delta p_3(0) 
    \bigg].
    \label{eq:alpha}
\end{aligned}
\end{equation}
Finally, the coefficient $C$ becomes
\begin{equation}
    \begin{aligned}
        C &=|\alpha_2|\left(|p_2^*|+\left|\frac{\theta-1+p_A}{1-p_A}\right|+\left|\frac{p_A-1-\theta}{1-p_A}\right| \right)\\
        &=|\alpha_2|\left(p_2^*+\frac{2\theta}{1-p_A} \right),
        \label{eq:C}
    \end{aligned}
\end{equation}
which contains the contribution of the initial state to the long-time dynamics. Recall that $\theta = \sqrt{p_A(1 - p_A)}$.

The analytical expression of $T_{\text{sim}}$ given by Eq. (\ref{eq:Mpemba_Tsim_lambda2}) is one of our main results. It describes the simulation time in the SL limit from a maximally mixed state to a state at temperature $\beta^{-1}$, and it uncovers the Mpemba-like effect physics:
The expression reproduces the nonmonotonic behavior for intermediate temperatures as one can observe in Fig. \ref{fig:TsimLindblad_epsilon_eigenvalues}(a) in the dashed brown line, correctly describing the Mpemba-like effect observed for a maximally mixed initial state.

We further use Eq. (\ref{eq:Mpemba_Tsim_lambda2}) to gain physical insight into a key feature observed in Fig. \ref{fig:TsimLindblad_epsilon_eigenvalues}(a):
 the nonmonotonic (turnover) behavior of
$T_{\rm sim}$  as the ancilla's temperature decreases from $\beta\to0$ to $\beta\gg1$, and the corresponding Mpemba-like effect.
Three factors govern the behavior of $T_{\rm sim}$:
(i) The proximity of the system's initial state to that of the ancilla, quantified by the coefficient $\alpha_2$, see Eqs. (\ref{eq:alpha})-(\ref{eq:C}), 
(ii) the projection of the initial state to the slow eigenmode, captured by $C$ being large enough, and (iii) the slowest relaxation timescale, characterized by $\lambda_2$.

When the system's initial state is within a small distance, on the order of $\epsilon$ (or less), from the ancilla, $|\alpha_2| \to 0$,  $T_{\rm sim}$ is small.
Furthermore, the simulation time continues to decrease (to zero) as the ancilla's temperature increases and approaches the completely mixed state, similar to the initial condition of the system. In Figure \ref{fig:TsimLindblad_epsilon_eigenvalues}(a), this regime corresponds to the high-temperature limit of the ancilla, near $\beta\to 0$.

In contrast, when $C$ is not small, the dominant contribution to Eq. (\ref{eq:Mpemba_Tsim_lambda2}) arises from the slowest relaxing mode, governed by $\lambda_2$; for generic initial conditions, all nonzero eigenvectors contribute.
As shown in Figure \ref{fig:TsimLindblad_epsilon_eigenvalues}(c), the magnitude of  $\lambda_2$ 
{\it decreases} monotonically with increasing temperature. Consequently, relaxation to a low-temperature ancilla state occurs more rapidly than to a high-temperature one, which leads to the counterintuitive trend observed in Fig. \ref{fig:TsimLindblad_epsilon_eigenvalues}(a), where the total simulation time $T_{\rm sim}$
is shorter when approaching colder final states. This region appears on the right-hand side of the turnover.

\begin{figure}[h]
    \centering
\includegraphics[width=1\linewidth]{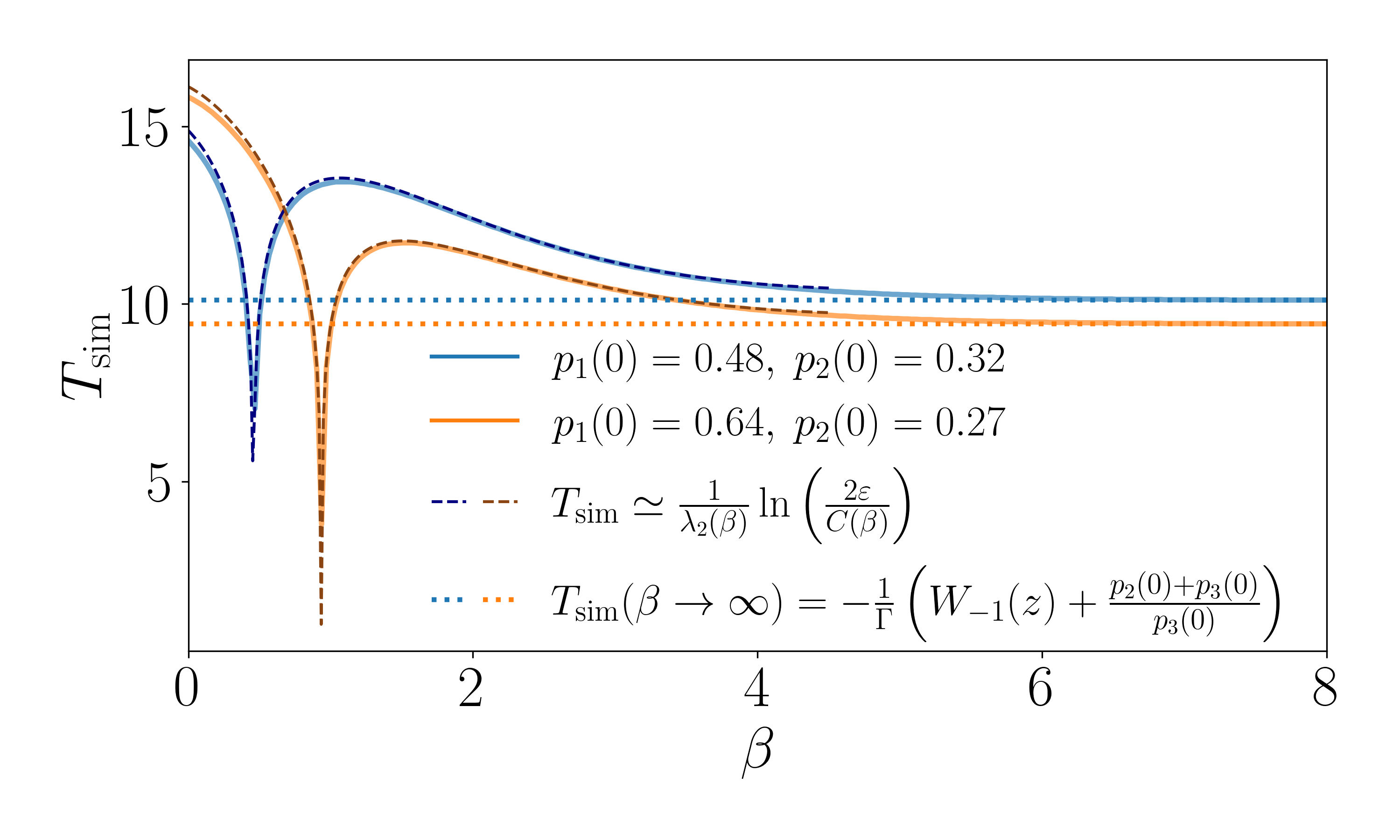} 
 \caption{Simulation time $T_{\text{sim}}$ in the Stroboscopic-Lindblad limit for selected diagonal initial states. The dip corresponds to initial conditions for which $\alpha_2 \simeq 0$, causing the dynamics to be dominated by $\lambda_3$. We use $\omega=1$ and the precision is set to $\epsilon=10^{-4}$. $J=10$, $\tau = 10^{-2}$, capturing the SL limit. 
}
    \label{fig:Tsimwithdip}
\end{figure}

Additionally, Eq. (\ref{eq:Mpemba_Tsim_lambda2})  can be tested to explore the behavior of $T_{\text{sim}}$ for any other initial diagonal state for the system. For example, in Fig. \ref{fig:Tsimwithdip}, we observe a dip in $T_{\text{sim}}$ at intermediate temperatures. This characteristic corresponds to the value where $\alpha_2 \rightarrow 0$, in which case the contribution of the slow mode vanishes and the decay is governed by the faster mode, $\lambda_3$. 
To the right side of this dip, once again the Mpemba effect shows due to the suppression of the slow eigenvalue $\lambda_2$ with temperature.

These factors, the closeness of the initial condition of the system to the target (ancilla) state, and the dependence of the slow eigenvalue of the Markovian process on temperature, yield an analogous Mpemba-like effect in the $J\tau1$ limit, as we observe in  Figs. \ref{fig:nstar_vs_beta_Jtau1} and \ref{fig:eigenvalues_Jtau1limit}, and discuss next in mathematical language.


\subsection{Mpemba effect in the $J\tau1$ regime}

\begin{figure*}[t!]
    \centering
    \includegraphics[width=1\linewidth]{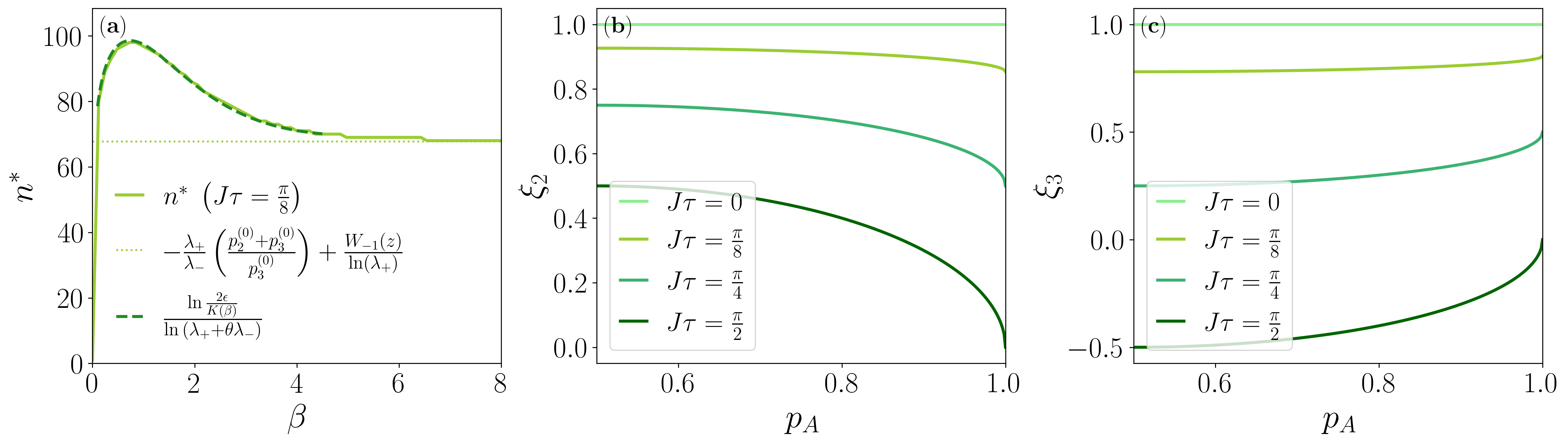} 
    \caption{
    Mpemba effect in the $J\tau1$ limit.
    (a) Total number of collisions $n^*$ as a function of $\beta$ for $J\tau = \frac{\pi}{8}$. The dotted line corresponds to the zero temperature solution from Eq. (\ref{eq:nstar-value}); the dashed green line corresponds to Eq. (\ref{eq:TsimMpembaJtau1}). (b)-(c) Eigenvalues of the stochastic rate matrix, $\xi_2$ and $\xi_3$ respectively, shown as a function of $p_A$ for different collision times, where $p_A\to 1$ corresponds to zero temperature and $p_A \to 0.5$ corresponds to the high temperature limit. }
\label{fig:eigenvalues_Jtau1limit}
\end{figure*}

We demonstrate in Fig. \ref{fig:eigenvalues_Jtau1limit}(a) the existence of the Mpemba-like effect in the $J\tau 1$ limit, complementing Fig. \ref{fig:nstar_vs_beta_Jtau1}.
In fact, we make no approximations as to the respective values of $J$ and $\tau$, and our results are completely general in that respect. However, when $J\tau \to 0$, 
The simulation time scales as $n^* \propto 1/(J\tau)^2$, leading to high cost, contrasting, e.g., the case $J\tau=\pi/2$, representative of the $J\tau1$ regime, 
for which only a few 
preparation steps are required, as demonstrated in Fig. \ref{fig:total-sim-time-plot1}(c).

Attempting to explain the effect, we rewrite Eq. (\ref{eq:populations-with-ss_grouping}) as
\begin{equation}
    \Delta\mathbf{p}^{(n+1)} = \begin{pmatrix}
        \eta_{11} & \eta_{12} &0\\
        \eta_{21} & \eta_{22} & \eta_{12}\\
        0& \eta_{21} & \eta_{33}
    \end{pmatrix}
    \Delta \mathbf{p}^{(n)} \equiv{\Lambda} \,\Delta \mathbf{p}^{(n)},
\end{equation}
where $\Lambda$ corresponds to a stochastic matrix, with coefficients defined in Eq. (\ref{eq:qutrit-qubit-eta-values}). 
The eigenvalues of the matrix are 
\bea
\label{eq:eigenvalues_stochastic_matrix}
    \xi_1 &=& 1, 
    \nonumber\\
     \xi_2 &=& \lambda_+ + \theta\lambda_-
     \nonumber\\
     &=& \cos^2(J\tau) + \sqrt{p_A(1-p_A)} \sin^2(J\tau),
     \nonumber\\
      \quad \xi_3 &=& \lambda_+-\theta\lambda_-
      \nonumber\\
       &=& \cos^2(J\tau) - \sqrt{p_A(1-p_A)} \sin^2(J\tau),
       \nonumber\\
\eea
which satisfy $1=\xi_1>\xi_2 \geq \xi_3 $, where
$0<\xi_2<1$ and $-1/2\leq\xi_3\leq 1$.
These inequalities ensure that the system reaches a fixed point, unless $J\tau$ is given by integer multiples of $\pi$, in which case there is no evolution.
It should be remembered that a small magnitude for $\xi$, as close as possible to zero, ensures a fast approach to the fixed point.

In Fig. \ref{fig:eigenvalues_Jtau1limit}(b) we observe that when $J\tau = \frac{\pi}{2}$, 
$\xi_2$ approaches zero as $p_A \xrightarrow[]{}1$. This explains the fast decay we found in Fig. \ref{fig:nstar_vs_beta_Jtau1} for this value at zero temperature. In contrast, for $J\tau \to 0$, the three eigenvalues approach $1$, which explains why we do not observe evolution at this point, that is, $p_i^{(n+1)} = p_i^{(n)}$. For intermediate values, $0<J\tau<\pi/2$, the eigenvalues become closer to one from below as $J\tau \to 0$. This explains the increase
of $n^*$ in this direction, observed in 
Fig. \ref{fig:total-sim-time-plot1}(c). 

Fig. \ref{fig:eigenvalues_Jtau1limit}(b)-(c) reveals the underlying mechanism of the Mpemba effect in the general RI case. On the one hand, at a very high temperature, the target state is similar to the initial mixed state of the system, resulting in $n^*\to 0$. 
The number of RI steps should start growing as we reduce the temperature, which we observe in this figure.
The turnover however takes place because of the relaxation dynamics:
Since $1\geq|\xi_2|\geq|\xi_3|$, $\xi_2$ dictates the long-term evolution.
However, according to Fig. \ref{fig:eigenvalues_Jtau1limit}(b), $\xi_2$ gets smaller when reducing temperature ($p_A\to 1)$. As a result, the relaxation process is made {\it faster} as we go to lower temperatures, compared to an intermediate temperature. 

We now provide an estimate for $n^* $. 
Since $\Lambda$ is non-symmetric, we  consider its biorthogonal decomposition. We refer to $\{ \mathbf{r}_i\}$ and $\{\mathbf{s}_i\}$ as the sets of right and left eigenvectors, respectively. A convenient choice for the right eigenvectors $\{ \mathbf{r}_i\}$ is the following,
\begin{equation}
    \mathbf{r}_1 = \left(p_1^*, p_2^*, p_3^* \right)^T,
\end{equation}
Setting $r_2^{(3)} = 1$ we obtain
\begin{equation}
    \mathbf{r}_2 = \left(- \frac{p_A}{\theta}, -1+\frac{p_A}{\theta},1\right)^T,
\end{equation}
and similarly for the third eigenvector,
\begin{equation}
    \mathbf{r}_3 = \left(\frac{p_A}{\theta}, -1-\frac{p_A}{\theta},1 \right)^T.
\end{equation}
With this choice of eigenvectors,
the biorthogonal eigenvector $\mathbf{s}_2$ takes the same expression as $\mathbf{u}_2$ from Eq. (\ref{eq:u2_vector}), and satisfies $\mathbf{s}_2^T \cdot \mathbf{r}_j  =\delta_{2j}$.
%
%

We now propose the following ansatz to describe the RI evolution towards the fixed-point thermal state,
\begin{equation}
    p_i^{(n)} = p_i ^* + \sum_{k = 2}^d \Tilde{c}_{k,i} \xi_k^n.
\end{equation}
Here,  $\Tilde{c}_{k,i}$ are coefficients that depend on the initial conditions, which we can compute, and $\xi_k$ are the eigenvalues of the rate matrix. 
The total simulated time is computed from
\begin{equation}
\begin{aligned}
    &\frac{1}{2}\left\{ \left|p_1^{(n^*)}-p_1^*\right|+\left|p_2^{(n^*)}-p_2^*\right|+\left|p_3^{(n^*)}-p_3^*\right|\right\} \leq \epsilon.\\
\end{aligned}
\end{equation}
Assuming that $\xi_2$ dominates the RI dynamics after enough steps, 
we find an approximation for the total number of collisions, 
\begin{equation}
\label{eq:TsimMpembaJtau1}
    n^* \approx \frac{\ln\left( \frac{2\epsilon}{K}\right)}{\ln(\xi_2)} = \frac{\ln\left( \frac{2\epsilon}{K}\right)}{\ln\left[\cos^2(J\tau)+\sqrt{p_A(1-p_A)}\sin^2(J\tau)\right]},
\end{equation}
where $K = |\Tilde{c}_{2,1}|+|\Tilde{c}_{2,2}|+|\Tilde{c}_{2,3}|$.

Proceeding in the same way as we did in Sec. \ref{subsec:Mepmba_inSLlimit}, we obtain 
\begin{equation}
    K = |\alpha_2| \left(\left|p_2^*\right| + \left|-1+\frac{p_A}{\theta}\right| + \left|-1-\frac{p_A}{\theta}\right|\right),
\end{equation}
where $\alpha_2 = \mathbf{s}_2^T\cdot\Delta\mathbf{p}(0)$ and the result is given by Eq. (\ref{eq: coefficient alpha2}),
$\Delta \mathbf{p}= \mathbf{p}(0)-\mathbf{p}^*$ is the deviation of the initial condition from the fixed-point thermal state, and $\theta=\sqrt{p_A(1-p_A)}$.

Equation (\ref{eq:TsimMpembaJtau1}) is one of our main results. It describes the minimal number of RI steps required for thermal state preparation at temperature $\beta^{-1}$, starting from a maximally mixed state. It holds for any $J$ and $\tau$, and as such it generalizes Eq. (\ref{eq:Mpemba_Tsim_lambda2}) which holds in the SL limit.
In Fig. \ref{fig:eigenvalues_Jtau1limit}(a), we show that Eq. (\ref{eq:TsimMpembaJtau1}) provides an excellent approximation to the required number of RI steps for intermediate temperatures. Importantly, this approximate expression successfully describes the observed Mpemba effect.


\subsection{Mpemba effect in larger systems}

\begin{figure*}[t]
    \centering
\includegraphics[width=1\linewidth]{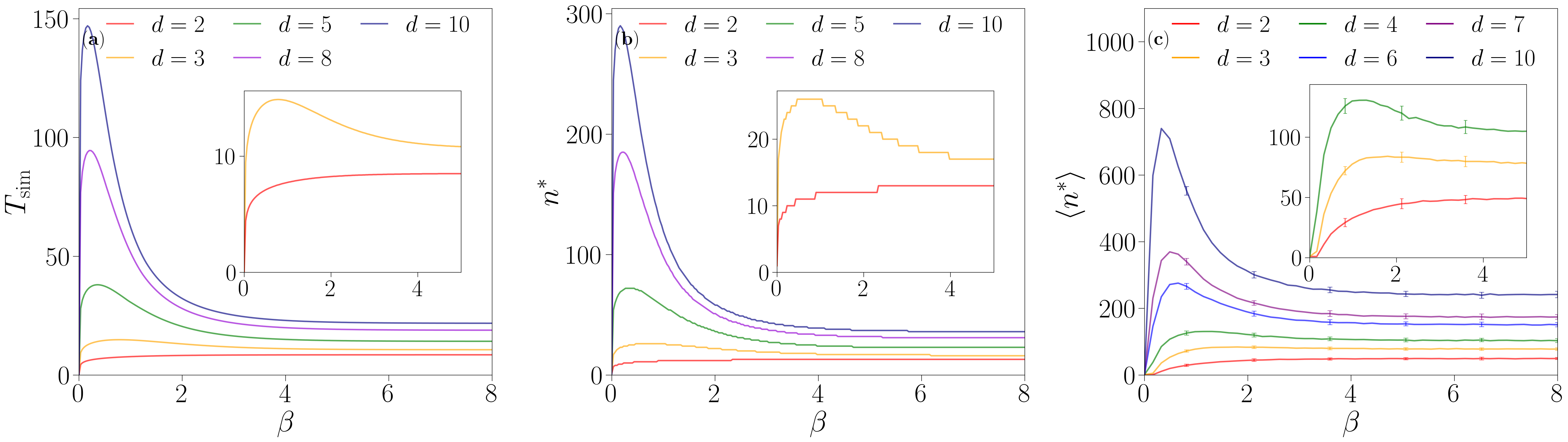} 
    \caption{(a) Simulation time $T_{\text{sim}}$ in the SL limit as a function of the inverse temperature $\beta$ for systems of dimension $d=2,3,5,8,10$, initialized in the maximally mixed state. The inset highlights the behavior of the $d=2$ and $d=3$ cases for $\beta \in[0,4.5]$. The coupling strength is $J=10$ and the interaction time is $\tau = 0.01$, efficiently capturing the SL limit.
    (b) Total number of collisions $n^*$ required to reach thermalization in $J\tau1$ regime, shown as a function of the inverse temperature $\beta$ for the same system dimensions. The inset also highlights the $d=2$ and $d=3$ system sizes for the same range of temperatures. The interaction product is fixed at $J\tau = \pi/4$ with $J = 10^{-3}$, capturing the $J\tau1$ regime. (c) Total number of collisions $\langle n^*\rangle$ for the randomized model. Every point is averaged over $m=100$ iterations. Before each collision every strength $J_{ij}$ from the interaction Hamiltonian is selected from a uniform random distribution $U(10^{-3},\pi\times10^{-3})$. The collision time is set to $\tau=100$. 
    In panels (a) and (b) the thermalization threshold is set to $\epsilon=10^{-4}$, while for the panel (c) it is set to $\epsilon=0.05$. The energy splitting for the three panels is $\omega=1$.}
    \label{fig:Tsimhigher_order_systems}
\end{figure*}

Our study so far has been mostly limited to considering a three-level system, and we have discussed thermal state preparation within this system. 
Next, we turn our attention to larger systems with higher energy levels. Our main question is: Does the Mpemba effect for thermal state preparation persist in higher-dimensional systems? The answer is positive: In Figs. \ref{fig:Tsimhigher_order_systems}(a) and (b), we focus on the Stroboscopic-Lindblad and $J\tau1$ regimes, respectively, and show that the Mpemba-like effect becomes {\it more pronounced} as the size of the system increases. This suggests that for intermediate temperatures in these larger systems, the dynamics is once again governed by the second slowest eigenvalue ($\lambda_2$ or $\xi_2$), which approaches zero for the SL limit and one in the $J\tau1$ regime as the dimension increases, thereby lengthening the total simulation time for thermal state preparation at nonzero temperature.

In Appendix \ref{sec:Generalization}, we give a mathematical justification to this phenomenon by obtaining the full spectrum of the Liouvillian $\mathcal{L}$ and the stochastic matrix $\Lambda$ for an arbitrary dimension $d$. There, we show that the second eigenvalue $\lambda_2^{(d)}$ ($\xi_2^{(d)}$) for the SL ($J\tau1$) limit, for a $d$-dimension system, is
\begin{equation}
    \label{eq:second_eigenvalue_recall}
    \begin{aligned}
        &\lambda_2^{(d)} = -\Gamma \left[1-2\theta\cos\left( \frac{\pi}{d}\right)\right],\\
        &\xi_2^{(d)} = \lambda_+ +2\theta \lambda_- \cos\left( \frac{\pi}{d}\right).
    \end{aligned}
\end{equation}
It is easy to show that as $d$ increases, in the intermediate temperature regime, $\lambda_2$ ($\xi_2$) become closer to zero (one), explaining why the Mpemba effect becomes more pronounced as we increase the dimensionality of the system. It is useful to recall that for $d=2$, the dynamics is governed by a single eigenvalue,  $\lambda_2^{(qb)}=-\Gamma$ ($\xi_2^{(qb)} = \lambda_+$), which is independent of temperature. This explains why, for a two-level system, preparing a low-temperature state from a completely mixed state always takes longer as the target temperature decreases.
 


\subsection{Mpemba effect with general and randomized interaction Hamiltonians}


\label{sec:random}
We explore thermalization dynamics 
and thermal state preparation when the system-ancilla interaction is more general, allowing for interactions that do not conserve energy. 
Concrete questions are: (i) What is the fixed point of the dynamics, that is, does the system thermalize to the temperature of the ancilla?
(ii) Does the nonmonotonic behavior of simulation time with temperature, in the form of a Mpemba effect, persist?

To address these questions, we first consider an extended version of the energy-conserving interaction Hamiltonian from Eq. (\ref{eq:interaction-Hamiltonian-qudit-qubit}) by including an additional non-conserving energy term of the form $J_{k+1,k}' \ket{k+1,\uparrow}\bra{k,\downarrow}$, yielding the more general interaction Hamiltonian,
\begin{align}
\label{eq:not_energy_preserving}
\hat{H}_I =& \sum_{k=0}^{d-2} \Big[ J_{k+1,k}
    \ket{k+1, \downarrow}\bra{k,\uparrow} \nonumber \\
    &+ J_{k+1,k}' \ket{k+1,\uparrow}\bra{k,\downarrow}
    + \text{H.c.} \Big].
\end{align}
%
This Hamiltonian allows transitions that do not conserve energy in the system. For example, in the two-level case ($d=2$), this Hamiltonian induces the same transitions as the Heisenberg-type model analyzed in Ref. \citenum{Segal2024}, in the particular case $J_{zz} = 0$. There, the term $J_{k+1,k}'$ corresponds to transitions associated with $J_{xx}-J_{yy}$, explicitly violating the energy conservation condition in the system, Eq. (\ref{eq: energy-preserving-condition}).

\begin{figure*}[tbp]
    \centering
    \includegraphics[width=\linewidth]{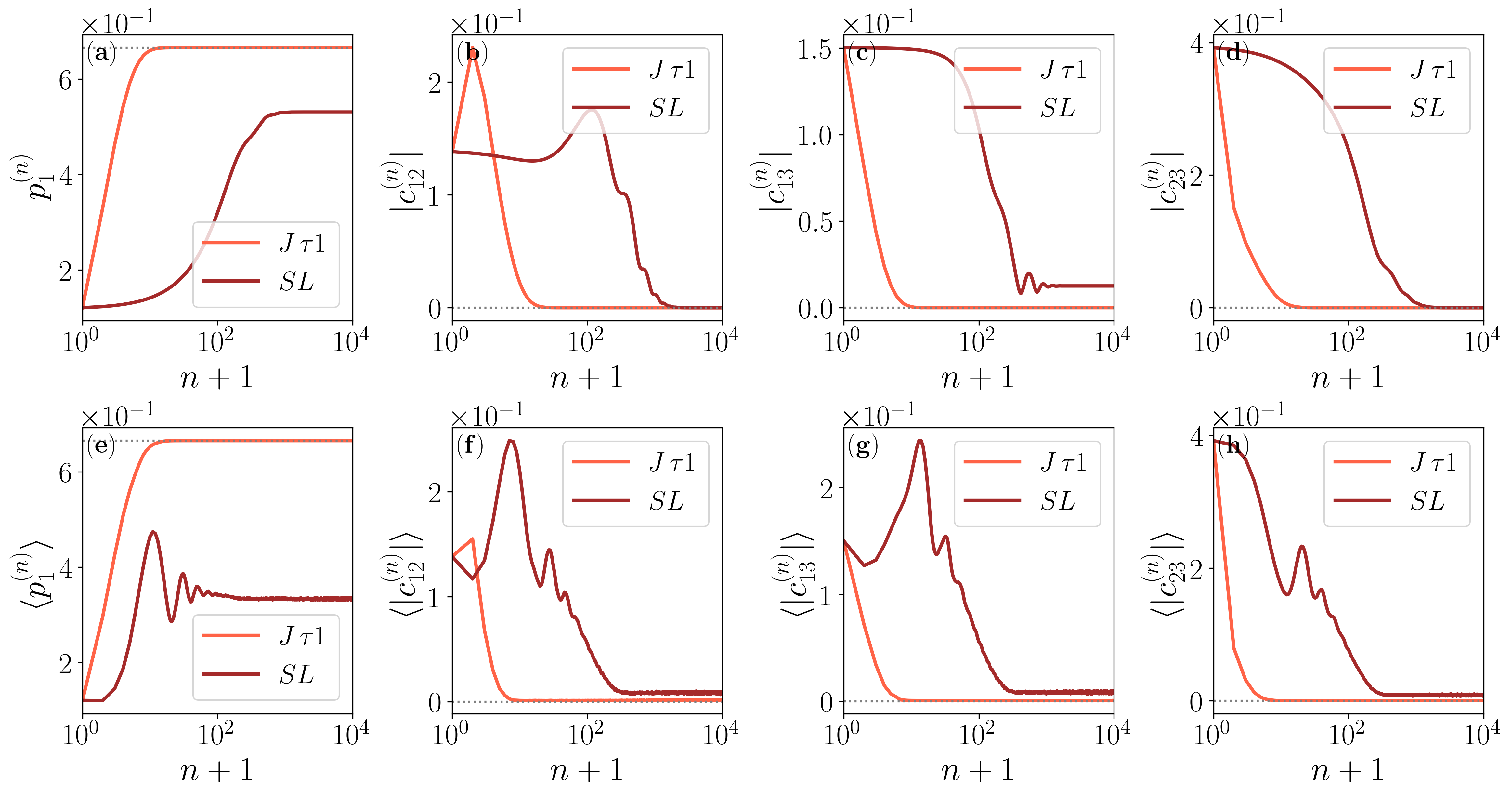} 
    \caption{(a)-(d) Dynamics of a $d=3$ system under the non-energy-conserving interaction Hamiltonian in Eq. (\ref{eq:not_energy_preserving}). Results shown in dark tones correspond to the SL limit with parameters $J =10$, $J'=5$, and $\tau = 0.01$, while lighter tones show the $J\tau1$ regime with $J = 10^{-3}$, $J' = \frac{1}{2}J$, and $\tau=10^3$ under the same Hamiltonian. (e)-(h) Corresponding averaged dynamics over $m = 100$ realizations of the fully random interaction Hamiltonian from Eq. (\ref{eq: random_Interaction_Hamiltonian}). For the SL limit, couplings $J_{ij}$ are sampled from a uniform distribution $U(5,20)$, with $\tau =0.01$. For the $J\tau1$ regime the couplings are sampled from $J_{ij}\in U(10^{-3},\pi\times10^{-3})$ with $\tau = 10^3$. The energy splitting is set for all the cases to $\omega = 1$, and the temperature of the ancilla is $\beta=1$. Error bars for the plots (e)-(h) are omitted as their magnitude was negligible.
    }
    \label{fig:multiplot-p1-and-coherences_random_Hamiltonian}
    \end{figure*}

For a three-level system, the total system-ancilla Hamiltonian takes the explicit matrix form
\begin{equation}
\label{eq:non-energy-conserving-total-hamiltonian-for-qutrit}
    \hat{H}_{tot}=
\begin{pmatrix}
-\frac{3\omega}{2} & 0 & 0 & J' & 0 & 0 \\
0 & -\frac{\omega}{2} & J & 0 & 0 & 0 \\
0 & J & -\frac{\omega}{2} & 0 & 0 & J'\\
J' & 0 & 0 & \frac{\omega}{2} &  J & 0 \\
0 & 0 & 0 & J & \frac{\omega}{2} & 0 \\
0 & 0 & J' & 0 & 0 & \frac{3\omega}{2}
\end{pmatrix}.
\end{equation}
As illustrated in Fig. \ref{fig:multiplot-p1-and-coherences_random_Hamiltonian}(a)-(d), in the SL regime, under this non-conserving energy interaction, the system does not reach the thermal state associated with the temperature of the ancilla. Instead, the system relaxes to a state where the nearest-neighbor coherence terms ($c_{12}, c_{23}$) vanish as shown in panels (b) and (d). On the other hand, the long-range coherence corresponding to the ground state and the second excited state  ($c_{13}$) remains nonzero, see panel (c). In Appendix \ref{sec: Dynamics_under_non_energy_conserving_interactions_SL_limit}, we show that in this limit the $c_{13}$ coherence couples to the populations, which explains the persistence of this term from a mathematical perspective. However, we lack a physical explanation for this behavior.
In contrast, in the $J\tau1$ regime, despite the fact that the interaction is not energy conserving, the system does reach the thermal state at the temperature of the ancillas. This demonstrates that the $J\tau1$ regime is robust for thermal state preparation, consistent with Ref. \cite{Segal2024}.

We further consider the scenario of a fully randomized interaction Hamiltonian that includes all possible off diagonal terms; therefore, the energy-conserving condition is not satisfied. This interaction Hamiltonian takes the form
\begin{equation}
\label{eq: random_Interaction_Hamiltonian}
    \hat{H}_I= \sum_{1\leq i<j\leq d} J_{ij} \left(\ket{i}\bra{j}+\ket{j}\bra{i} \right),
\end{equation}
where for every collision, the couplings $J_{ij}$ are sampled from a random uniform distribution. In Fig. \ref{fig:multiplot-p1-and-coherences_random_Hamiltonian} (e)-(h), we observe that the $J\tau1$ limit still reaches the target state, aligning our simulations with the results obtained in Ref. \citenum{Hagan}. On the other hand, for the SL limit, the steady state approximates a nonequilibrium steady state. 
%

Given that the $J\tau1$ regime provides thermalization for general interaction model, and in randomized interactions, we check again if the Mpemba effect still occurs under these conditions. The answer is  affirmative. As shown in Fig. \ref{fig:Tsimhigher_order_systems}(c), the Mpemba effect not only persists but becomes increasingly pronounced in larger system dimensions as observed in the previous section. Interestingly, in the randomized case, the Mpemba effect does not manifest itself clearly for the $d=3$; it appears more strongly for $d=4$ and persists then for higher dimensional systems with $d>3$.


\section{Summary and Open Problems}
\label{sec:summ}

The repeated interaction framework is commonly used to generate Lindbladian dynamics. In this work, however, we investigated the dynamics generated by the RI method beyond that. We focused on two complementary regimes: (i) Enacting frequent collisions, each at strong coupling. This regime indeed results in rate equations for the population and coherences of the Lindblad form.
(ii) Performing long collisions but with weak interaction, a regime that we refer to as the $J\tau1$ limit.
In both cases, we focused on a three-level system as a case study, 
derived the corresponding equations of motion, and solved them in the zero-temperature limit. 
Equipped with the general equations of motion for a three-level system, we studied the simulation time (or minimal number of collisions) necessary for thermal state preparation, focusing on achieving a finite-temperature state from a maximally mixed state.
We derived an exact expression at zero temperature and an approximate expression at arbitrary temperatures, which was shown to be quantitatively correct. This allowed us to demonstrate the existence of a Mpemba-like effect in the thermalization time as a function of target temperature. 

Concretely, our main results are the following:

(i) We proved that operating in the $J\tau1$ limit is preferable over the SL regime, as it leads to optimized performance (fewest simulation steps) when choosing to work at $J\tau=\pi/2, 3\pi/2, 5\pi/2...$.
Although this result was derived in the zero-temperature limit, simulations showed that these values for $J\tau$ continue to minimize the required number of collisions for thermal state preparation, $n^{*}$ even at higher temperatures.

(ii) We derived an accurate analytical expression for the simulation time (or number of RI steps) required to achieve thermal state preparation within a certain accuracy (\ref{eq:TsimMpembaJtau1}).
This general result holds for any value of the interaction strength and duration, and for general temperatures. We also studied the SL limit of the general result and provided closed-form approximate results for the associated simulation time, Eq. (\ref{eq:Mpemba_Tsim_lambda2}).

In both the general and the SL regimes, we observed a Mpemba-like effect in the simulation time, which varies nonmonotonically with the target temperature. We traced this behavior to the temperature dependence of the Liouvillian spectrum. Specifically, we demonstrated that the  slowest nonzero eigenvalue of the rate matrix, whether for continuous or discrete dynamics, became monotonically slower with increasing temperature. This leads to slower thermalization in the high-temperature regime.

Based on simulations, we also showed that the Mpemba effect persists for systems of high dimensions and that it remained robust even in the presence of random interactions. We proved analytically that as the dimensionality of the system increases, the thermalization process slows down.
As a result, the number of collisions required for thermal state preparation from a maximally mixed state increases, an effect that is most dramatically manifested at intermediate temperatures (Fig. \ref{fig:Tsimhigher_order_systems}). This important observation calls for developing new strategies for thermal state perpetration of high-dimensional systems. For example, one could try devising more complex RI models and protocols and using other types of resources (different interactions, coherences, etc.). 
Open questions include:
Could one devise a faster thermalization process by performing a gradual cooling? Or, the other way around, would a fast cooling to the ground state, then elevating temperature gradually allow for less resources?

Another open problem more broadly concerns the topic of resource estimation for state preparation.
In the context of quantum algorithms, in addition to the number of gates, (translating to run-time), energetic considerations should be taken into account \cite{Alexia}. Eventually, a trade-off between runtime and cost should guide the protocol of choice. 

Our work demonstrates that by choosing the interaction strength and its duration, the RI protocol can be optimized to minimize the number of collisions required for thermal state preparation. It is remarkable to note the rich thermalization dynamics captured within the simple RI scheme.
Future work will focus on RI dynamics with more complex interaction forms and when non-Markovian effects are taken into account. 
Implementing the RI scheme as a near-term quantum algorithm for thermal state preparation remains a key challenge. However, since our study identifies optimal interaction strengths and time intervals for this purpose, deploying the RI method on current quantum hardware presents an intriguing direction for exploration.

\begin{acknowledgements}
The work of C. R. E. was supported by the QUARMEN Erasmus Mundus Program and the University of Toronto.
D.S. acknowledges the NSERC Discovery Grant and the Canada Research Chairs Program.
The work of C. R. E. and A. P.  was supported by the Department of Physics at the University of Toronto and by the research project: ``Quantum Software Consortium: Exploring Distributed Quantum Solutions for Canada" (QSC). QSC is financed under the National Sciences and Engineering Research Council of Canada (NSERC) Alliance Consortia Quantum Grants \#ALLRP587590-23. We acknowledge fruitful discussions with Matthew Hagan and Matthew Pocrnic. 
\end{acknowledgements}

\clearpage
\begin{widetext}

\appendix

\section{Solution of the RI dynamics at zero temperature}
\label{AppendixT0}

We provide here a detailed solution for the equations of motion
(\ref{eq:populations-for-pA1-with-lambdaM})
 in the zero temperature limit, $p_A=1$, see Sec. \ref{subsec:EOM_Jtau1}. 
We recall the set of equations that we want to solve
\begin{equation}
\label{eq:populations-for-pA1-with-lambda}
    \begin{aligned}
        & p_1^{(n+1)}=p_1^{(n)}+\lambda_- p_2^{(n)},\\
        & p_2^{(n+1)} =\lambda_+p_2^{(n)}+\lambda_-p_3^{(n)},\\
        & p_3^{(n+1)}=\lambda_+p_3^{(n)},
    \end{aligned}
\end{equation}
with $\lambda =\cos(2J\tau)$ and $\lambda_{\pm} =\frac{1}{2}\left(1\pm\lambda\right )\leq 1$. 

We aim to express these recursive relations explicitly in terms of their initial values. We start with the last equation, which is decoupled from the rest. It has a straightforward solution,
\begin{equation}
\label{eq:p3npA1}
    p_3^{(n)} = (\lambda_+)^np_3^{(0)}.
\end{equation}
Substituting this result into the equation for $p_2^{(n+1)}$ we find
\begin{equation}
    p_2^{(n+1)} = \lambda_+p_2^{(n)}+\lambda_-(\lambda_+)^np_3^{(0)}.
\end{equation}
To solve this recurrence, we define the variable $X_n = (\lambda_+)^{1-n}\ p_2^{(n)}$, which allows us to rewrite the equation in a telescopic form
\begin{equation}
    X_{n+1} - X_n = \lambda_-p_3^{(0)}.
\end{equation}
Summing from $0$ to $n-1$ we obtain
\begin{equation}
    X_n-X_0 = n\lambda_-p_3^{(0)}.
\end{equation}
Solving for $p_2^{(n)}$, we therefore find
\begin{equation}
    p_2^{(n)} = (\lambda_+)^{(n-1)} \left[\lambda_+p_2^{(0)}+n\lambda_-p_3^{(0)}\right].
\end{equation}
Finally, substituting the expression for $p_2^{(n)}$ and the solution for $p_{3}^{(n)}$ into the equation for $p_1^{(n)}$, we obtain the following
\begin{equation}
\label{eq:p1-pA1-important-resultA}
    p_1^{(n)} = 1-(\lambda_+)^n \left[p_2^{(0)}+p_3^{(0)}\left(1+n\frac{\lambda_-}{\lambda_+}\right)\right].
\end{equation}

We now turn to the evolution of the coherences under the same limit of zero temperature for the ancilla, $p_A=1$. Substituting this condition into Eq. (\ref{eq:psi11andpsi13first}), we find that the coefficients simplify to
\begin{equation}
\label{eq: constants-of-the-coherencesApp}
\begin{aligned}
    &\psi_{11} = \psi_{22} = \mu,\quad
        \psi_{13} =\lambda_-, \quad
         \psi_{33} = \lambda_+, \quad
        \psi_{31} = 0,
\end{aligned}    
\end{equation}
where we define $\mu = \cos(J\tau)$. Substituting these into the recurrence relation for the coherences, Eq. (\ref{eq: equation for the coherences}) yields
\begin{equation}
    \label{eq: equation-coherences-pA1-}
    \begin{aligned}
        &c_{12}^{(n+1)} = e^{i\tau\omega}\left(c_{12}^{(n)}\mu+c_{23}^{(n)}\lambda_-\right),\\
        &c_{13}^{(n+1)} = e^{2i\tau \omega}c_{13}^{(n)}\mu,\\
        & c_{23}^{(n+1)} = e^{i\tau \omega}c_{23}^{(n)}\lambda_+.
    \end{aligned}
\end{equation}
We now solve these recurrence relations explicitly. Starting with $c_{13}^{(n)}$, we get
\begin{equation}
    \label{eq:c13-markovian-map}
    c_{13}^{(n)} = e^{2i\tau\omega n} \mu^n c_{13}^{(0)}.
\end{equation}
Similarly, the equation for $c_{23}^{(n)}$ is
\begin{equation}
\label{eq:c23-markovian map}
    c_{23}^{(n)}=e^{i\tau\omega n} (\lambda_+)^nc_{23}^{(0)}.
\end{equation}
For the coupled term $c_{12}^{(n)}$, we have to solve
\begin{equation}
    c_{12}^{(n+1)}=e^{i\tau\omega} \left[c_{12}^{(n)}\mu+\lambda_-e^{i\tau\omega n}(\lambda_+)^nc_{23}^{(0)}\right].
\end{equation}
Introducing the transformed variable $C_{12}^{(n)}=c_{12}^{(n)}e^{-i\tau\omega n}$, 
the recurrence relation then simplifies to an inhomogeneous linear difference equation
\begin{equation}
\label{eq: recurrence equation c12}
    C_{12}^{(n+1)}-\mu C_{12}^{(n)} = \lambda_- c_{23}^{(0)}(\lambda_+)^n.
\end{equation}
The solution to this equation consists of a homogeneous part and a particular solution $C_{12}^{(n)}=C_{12,h}^{(n)}+C_{12,p}^{(n)}$. We propose
\begin{equation}
    C_{12}^{(n)}=A\mu^n+B(\lambda_+)^n,
\end{equation}
where $A$ and $B$ are constants to be determined. Substituting into Eq. (\ref{eq: recurrence equation c12}) we find
\begin{equation}
    B = \frac{\lambda_-c_{23}^{(0)}}{\lambda_+-\mu}.
\end{equation}
To fix $A$, we use the initial condition $C_{12}^{(0)} =c_{12}^{(0)}$, giving
\begin{equation}
    A = c_{12}^{(0)}-\frac{\lambda_-c_{23}^{(0)}}{\lambda_+-\mu}.
\end{equation}
Therefore, the full solution for $c_{12}^{(n)}$ reads
\begin{equation}
    \label{eq:final-expression-Jtau1-c12}
    \begin{aligned}
        c_{12}^{(n)}=e^{i\tau\omega n} \bigg[\left(c_{12}^{(0)}-\frac{\lambda_-c_{23}^{(0)}}{\lambda_+-\mu}\right)\mu^n
        + \frac{\lambda_-c_{23}^{(0)}}{\lambda_+-\mu}(\lambda_+)^n
        \Bigg].
    \end{aligned}
\end{equation}

\section{Solving Eq. (\ref{eq:trace-distance-simplification}) and the Lambert $W$ function}
\label{sec: Lambert function}

\subsection{Equation (\ref{eq:trace-distance-simplification}): $n^*$ at zero temperature }

At zero temperature, the trace distance 
(\ref{eq:trace-distance-simplification})
reduces to  $D(\rho_S^{(n^*)},\rho_S^*) = 
        1-p_1^{(n^*)} \leq \epsilon$.
Substituting the solution for the ground state population of the system from Eq. (\ref{eq:p1-pA1-important-resultM}), and saturating the inequality, we solve for $n^*$,
\begin{equation}
    \label{eq: condition-nstar-qutrit-qubitApp}
    (\lambda_+)^{n^*}\left[p_2^{(0)}+p_3^{(0)}\left(1+n^*\frac{\lambda_-}{\lambda_+}\right)\right] = \epsilon.
\end{equation}
or
\begin{equation}
\label{eq: condition-nstar-qutrit-qubit-rewrittenApp}
p_2^{(0)}+p_3^{(0)}+p_3^{(0)}\frac{\lambda_-}{\lambda_+}n^* = \epsilon e^{-n^*\ln(\lambda_+)}.
\end{equation}
Defining 
        $A = p_2^{(0)}+p_3^{(0)}$,
        $B = p_3^{(0)}\frac{\lambda_-}{\lambda_+}$,
        $\alpha=-\ln(\lambda_+)$,
Eq. (\ref{eq: condition-nstar-qutrit-qubit-rewritten}) becomes
$   A+Bn^* = \epsilon e^{\alpha n^*}$.
Setting
$    y = A+Bn^*$, we obtain
$   y = \epsilon \exp{\left[\alpha\left(\frac{y-A}{B}\right)\right]}$,
which simplifies to
\begin{equation}
    y e^{-\frac{\alpha y}{B}} = \epsilon e^{-\frac{\alpha A}{B}}.
\end{equation}
Finally defining
$   w = -\frac{\alpha y}{B}$,
the equation takes the standard Lambert form
\begin{equation}
    w e^w = -\frac{\alpha \epsilon}{B}\exp{\left(-\frac{\alpha A}{B}\right)}.
\end{equation}
The solution of this equation is
\begin{equation}
\label{eq:nstar-valueApp}
n^* = -\frac{\lambda_+}{\lambda_-}\left(\frac{p_2^{(0)}+p_3^{(0)}}{p_3^{(0)}}\right)
+\frac{1}{\ln(\lambda_+)} \, W_{-1}\left[
(\ln \lambda_+)\frac{\lambda_+ \epsilon}{\lambda_- p_3^{(0)}}
\cdot (\lambda_+)^{\frac{\lambda_+\left(p_2^{(0)}+p_3^{(0)}\right)}{\lambda_- p_3^{(0)}}}
\right],
\end{equation}
which is Eq. (\ref{eq:nstar-value}) in the main text with
 $W_{-1}(z)$ as the lower $k=-1$ branch of the Lambert function, $w=W_{-1}(z)$ solving
$we^{w}=z$. 

\subsection{The Lambert function}

The Lambert function $ W_k(z)$, defined for an integer branch index $k$ and a complex argument $z$, is the multivalued inverse of the function $w \mapsto we^w$. That is, it satisfies the equation
\begin{equation}
    w(z)e^{w(z)}=z.
\end{equation}
Each choice of $k$ defines a different branch of the function. In our application, since the total simulation time must be a real value, only the principal branch $ W_0 $ and the lower branch $ W_{-1} $ are relevant. These branches are the only ones that provide real values for specific ranges of $z$ \cite{corless1996lambert}.

As shown in Fig. \ref{fig:Lambert_function}, only the principal branch $ W_0(z) $ takes positive values. However, in the context of our analysis, the Lambert function appears in expressions where the argument is proportional to the parameter $\epsilon$, which we take to be small and positive. Therefore, the argument of the Lambert function lies near the origin and is typically negative (close to $0^-$). To obtain a physically meaningful (positive) real solution, we must evaluate the function on the $ W_{-1} $ branch.

\begin{figure}[h!]
    \centering
    \includegraphics[width=0.5\linewidth]{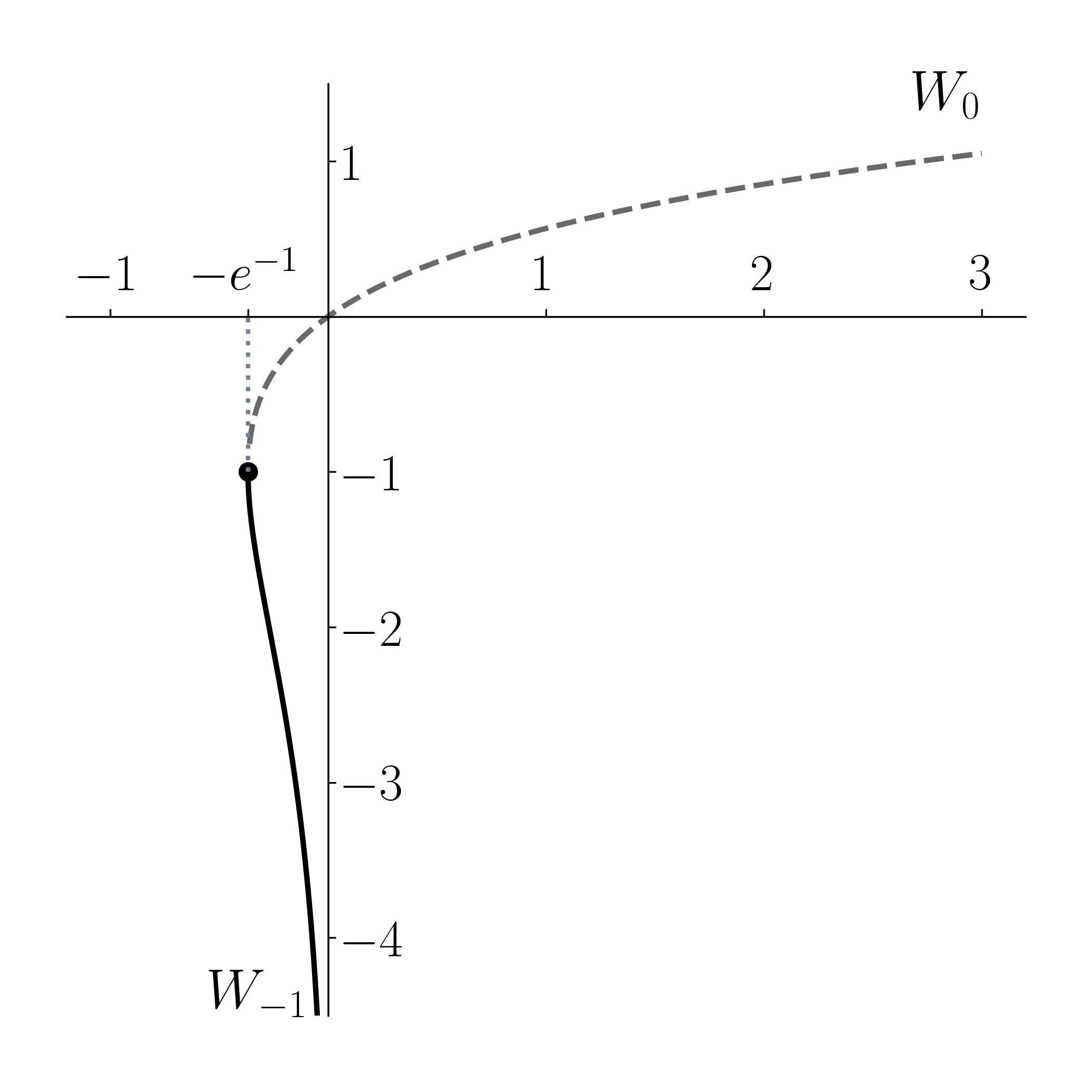} 
    \caption{Real branches of the Lambert-$W$ function. The dashed grey curve is the principal branch $W_0(z)$, defined for $z\geq -e^{-1}$ and increasing from $W_0(e^{-1})=-1$ to positive values as $z\xrightarrow[]{}\infty$. The solid black curve is the lower real branch $W_{-1}(z)$, defined on $-e^{-1}\leq z<0$; it starts at the shared branch point $(-e^{-1},-1)$ (black dot) and diverges to $-\infty$ as $z \xrightarrow[]{}0^-$. The dotted vertical line corresponds to the $z=-e^{-1}$ point.}
    \label{fig:Lambert_function}
\end{figure}

Moreover, to ensure that the closed-form expression of Eqs. (\ref{eq:nstar-value}) and (\ref{eq: Total_sim_time-qutrit-qubit}) give us real-valued results, we need to ensure that the argument $z$ of the Lambert function satisfies the condition
\begin{equation}
\label{eq: condition Lambert function}
    z \geq -\frac{1}{e}.
\end{equation}

This constraint, translate into two nontrivial conditions on the parameter $\epsilon$, depending on the regime under consideration.

\subsection{Constraint on \texorpdfstring{$\epsilon$}{epsilon} in the \texorpdfstring{$J\tau 1$}{J tau 1} Regime}

In the $J\tau 1$ regime, the condition $z \geq -1/e$ leads to the following upper bound on $\epsilon$
\begin{equation}
\label{eq: condition-for-epsilon-lambert}
    \epsilon \leq -\frac{1}{\ln(\lambda_+)} \cdot \frac{\lambda_- p_3^{(0)}}{\lambda_+ e} \cdot (\lambda_+)^{\frac{\lambda_- p_3^{(0)}}{\lambda_+ \left(p_2^{(0)} + p_3^{(0)}\right)}}.
\end{equation}
We remind that $\epsilon$ represents the convergence threshold, specifying how close the system state $\rho_S^{(n^*)}$ must be to the target thermal state. Therefore, to ensure physical consistency, we must enforce:
\begin{equation}
    0 < \epsilon \leq -\frac{1}{\ln(\lambda_+)} \cdot \frac{\lambda_- p_3^{(0)}}{\lambda_+ e} \cdot (\lambda_+)^{\frac{\lambda_- p_3^{(0)}}{\lambda_+ \left(p_2^{(0)} + p_3^{(0)}\right)}}.
\end{equation}

\subsection{Constraint on \texorpdfstring{$\epsilon$}{epsilon} in the Stroboscopic-Lindblad Limit}

\noindent In the Stroboscopic-Lindblad limit, the argument of the Lambert function takes the form
\begin{equation}
    z =  -\frac{1}{p_3(0)}e^{-\frac{p_2(0)+p_3(0)}{p_3(0)}}\epsilon.
\end{equation}
Imposing the condition $ z \geq -1/e $, we obtain
\begin{equation}
    \epsilon \leq {p_3(0)}\; e^{\frac{p_2(0)}{p_3(0)}}.
\end{equation}
Since $ \epsilon $ is defined to be a small positive quantity, this upper bound is always satisfiable and does not impose significant additional constraints in this regime.

\section{Dynamics and thermalization of systems with $d>3$}
\label{sec:Generalization}

In this Appendix, we generalize our derivations to systems with more than three levels. We begin with examples $d=4$, and $d=5$, with the objective of clarifying and bringing intuition on trends, then providing a general solution.

\subsection{General equations for a $d=4$ system}

For a four-level system, the system Hamiltonian is given by Eq. (\ref{eq: free-qudit-systemHamiltonian}) with $s=3/2$. 
The corresponding total Hamiltonian, incorporating the flip-flop interaction from Eq. (\ref{eq:interaction-Hamiltonian-qudit-qubit}) adapted to this case, is given by
\begin{equation}
    \hat{H}_{\text{tot}} = \begin{pmatrix}
-2\omega & 0 & 0 & 0 & 0 & 0 & 0 & 0 \\
0 & -\omega & J & 0 & 0 & 0 & 0 & 0 \\
0 & J & -\omega & 0 & 0 & 0 & 0 & 0 \\
0 & 0 & 0 & 0 & J & 0 & 0 & 0 \\
0 & 0 & 0 & J & 0 & 0 & 0 & 0 \\
0 & 0 & 0 & 0 & 0 & \omega & J & 0 \\
0 & 0 & 0 & 0 & 0 & J & \omega & 0 \\
0 & 0 & 0 & 0 & 0 & 0 & 0 & 2\omega
\end{pmatrix}.
\end{equation}
The collision unitary becomes 
\begin{equation}
    \hat{U}(\tau) =
\begin{pmatrix}
e^{2i \tau \omega} & 0 & 0 & 0 & 0 & 0 & 0 & 0 \\
0 & e^{i \tau \omega} \cos(J \tau) & -i e^{i \tau \omega} \sin(J \tau) & 0 & 0 & 0 & 0 & 0 \\
0 & -i e^{i \tau \omega} \sin(J \tau) & e^{i \tau \omega} \cos(J \tau) & 0 & 0 & 0 & 0 & 0 \\
0 & 0 & 0 & \cos(J \tau) & -i \sin(J \tau) & 0 & 0 & 0 \\
0 & 0 & 0 & -i \sin(J \tau) & \cos(J \tau) & 0 & 0 & 0 \\
0 & 0 & 0 & 0 & 0 & e^{-i \tau \omega} \cos(J \tau) & -i e^{-i \tau \omega} \sin(J \tau) & 0 \\
0 & 0 & 0 & 0 & 0 & -i e^{-i \tau \omega} \sin(J \tau) & e^{-i \tau \omega} \cos(J \tau) & 0 \\
0 & 0 & 0 & 0 & 0 & 0 & 0 & e^{-2i \tau \omega}
\end{pmatrix}
\end{equation}
Evolving the state using Eq. (\ref{eq: CPTPmap}), the resulting equations for the populations are 
\begin{equation}
\label{eq:ququart-qubit-populations}
\begin{aligned}
p_1^{(n+1)} &= \frac{1}{2} \left\{
p_1^{(n)} \left[1 + p_A + (1 - p_A) \cos(2J\tau)\right] +
p_2^{(n)} \, p_A \left[1 - \cos(2J\tau)\right]
\right\}, \\
p_2^{(n+1)} &= \frac{1}{2} \left\{
p_1^{(n)} \left[1 - p_A - (1 - p_A) \cos(2J\tau)\right] +p_2^{(n)} \left[1+ \cos(2J\tau)\right] +
p_3^{(n)} \, p_A \left[1 - \cos(2J\tau)\right]
\right\}, \\
p_3^{(n+1)} &= \frac{1}{2} \left\{
p_2^{(n)} \left[1 - p_A -(1 - p_A) \cos(2J\tau)\right] +
p_3^{(n)} \left[1+\cos(2J\tau)\right] +
p_4^{(n)} \, p_A \left[1 - \cos(2J\tau)\right]
\right\} ,\\
p_4^{(n+1)} &= \frac{1}{2} \left\{
p_3^{(n)} \left[1 - p_A - (1 - p_A) \cos(2J\tau)\right]+p_4^{(n)} \left[2 -p_A(1- \cos(2J\tau))\right]
\right\}.
\end{aligned}
\end{equation}
The steady state of the system is the canonical thermal state, $p_n^*$. As such, we rewrite Eq. (\ref{eq:ququart-qubit-populations}) in matrix form as
\begin{equation}
\label{eq:stochastic_matrix_ququart}
    \Delta\mathbf{p}^{(n+1)} = \begin{pmatrix}
        \eta_{11} & \eta_{12} &0 &0\\
        \eta_{21} & \eta_{22} & \eta_{12}&0\\
        0& \eta_{21} & \eta_{22}&\eta_{12}\\
        0&0&\eta_{21}& \eta_{33}
    \end{pmatrix}
    \Delta \mathbf{p}^{(n)},
\end{equation}
where the coefficients $\eta_{ij}$ are the same as in Eq. (\ref{eq:qutrit-qubit-eta-values}) and $\Delta\mathbf{p} =\mathbf{p}-\mathbf{p^*}$.

In the zero-temperature limit, $p_A = 1$, and Eq. (\ref{eq:stochastic_matrix_ququart}) becomes
\begin{equation}
\label{eq:stochastic_matrix_ququart_pA1}
    \Delta\mathbf{p}^{(n+1)} = \begin{pmatrix}
        1 & \lambda_- &0 &0\\
        0 & \lambda_+ & \lambda_-&0\\
        0& 0 & \lambda_+&\lambda_-\\
        0&0&0& \lambda_+
    \end{pmatrix}
    \Delta \mathbf{p}^{(n)}.
\end{equation}
Similarly to how we derived the equation of motion for the three-level system, we can derive the recursive relation for this case: 

\noindent First, for the decoupled term $p_4^{(n)}$ we  directly obtain
\begin{equation}
    p_4^{(n)} = (\lambda_+)^n p_4^{(0)}.
\end{equation}
Second, for $p_3^{(n)}$ we use the exact same trick as for the qutrit case, and find
\begin{equation}
    p_3^{(n)} = (\lambda_+)^{n-1}\left[\lambda_+p_3^{(0)} + n \lambda_-p_4^{(0)}\right].
\end{equation}
Next, for $p_2^{(n)}$ we have to solve $p_2^{(n+1)} = \lambda_+ p_2^{(n)}+ \lambda_- p_3^{(n)}$,
where substituting the above equation for $p_3^{(n)}$, we obtain
\begin{equation}
    p_2^{(n)} = (\lambda_+)^n\left[ p_2^{(0)}+n\frac{\lambda_-}{\lambda_+}p_3^{(0)}+\frac{n(n-1)}{2}\left(\frac{\lambda_-}{\lambda_+}\right)^2 p_4^{(0)}\right].
\end{equation}
Finally, the equation for $p_1^{(n)}$ by the normalization condition is
\begin{equation}
    p_1^{(n)} = 1-p_2^{(n)}-p_3^{(n)}-p_4^{(n)}.
\end{equation}

We also obtain the recurrence equations for the coherences. Reminding that $\lambda_+ = \cos^2(J\tau)$, $\lambda_- = \sin^2(J\tau)$, and $\mu =\cos(J\tau)$, the resulting equations are
\begin{equation}
   \begin{aligned}
   \label{eq: coherences_ququart}
        &c_{12}^{(n+1)} = e^{i\tau \omega}\left[c_{12}^{(n)}\left(\lambda_++p_A(\mu-\lambda_+) \right)+c_{23}^{(n)}p_A \lambda_-\right],\\
        &c_{13}^{(n)} = e^{2i\tau\omega}\left[c_{13}^{(n)}\left( \lambda_+ + p_A(\mu - \lambda_+)\right)+c_{24}^{(n)}p_A \lambda_- \right],\\
        & c_{14}^{(n+1)} = e^{3i\tau\omega}\mu c_{14}^{(n)}\\
        & c_{23}^{(n+1)} = e^{i\tau \omega} \left[c_{12}^{(n)}\lambda_-(1-p_A)+c_{23}^{(n)}\lambda_+ + c_{34}^{(n)}p_A \lambda_-\right]\\
        & c_{24}^{(n+1)} = e^{2i\tau\omega} \left[c_{13}^{(n)}\lambda_-(1-p_A)+c_{24}^{(n)}\left((1-p_A)\mu +p_A \lambda_+ \right) \right] \\
        & c_{34}^{(n+1)} = e^{i\tau\omega}\left[c_{23}^{(n)} \lambda_-(1-p_A)+c_{34}^{(n)}\left((1-p_A)\mu +p_A \lambda_+ \right) \right]
   \end{aligned}
\end{equation}

In the zero-temperature limit, we can solve the recursive equations. In this limit ($p_A = 1$) Eq. (\ref{eq: coherences_ququart}) reads

\begin{equation}
   \begin{aligned}
   \label{eq: coherences_ququart_zero_temperature}
        &c_{12}^{(n+1)} = e^{i\tau \omega}\left[c_{12}^{(n)}\mu+c_{23}^{(n)} \lambda_-\right],\\
        & c_{13}^{(n+1)} = e^{2i\tau \omega} \left[c_{13}^{(n)}\mu +c_{24}^{(n)}\lambda_- \right],\\
        & c_{23}^{(n+1)} = e^{i\tau \omega} \left[c_{23}^{(n)}\lambda_+ +c_{34}^{(n)} \lambda_-\right],\\
        & c_{14}^{(n)} = e^{3i\tau \omega} c_{14}^{(n)} \mu,\\
        &c_{24}^{(n)} = e^{2i\tau \omega} c_{24}^{(n)} \lambda_+,\\
        & c_{34}^{(n)} = e^{i\tau \omega} c_{34}^{(n)} \lambda_+.
   \end{aligned}
\end{equation}

\noindent The recurrence relation of the three last equations can be directly solved
\begin{equation}
     \label{eq: ququart: c12,c24,c23 at pA=1}
     \begin{aligned}
         &c_{14}^{(n)} = e^{3i\tau\omega n} \mu^n c_{14}^{(0)},\\
         &c_{24}^{(n)} = e^{2i\tau \omega n} (\lambda_+)^n c_{24}^{(0)},\\
         & c_{34}^{(n)} = e^{i\tau \omega n} (\lambda_+)^nc_{34}^{(0)}.
     \end{aligned}
\end{equation}

\noindent For the other three we have a similar scenario as we had for the qutrit case. By doing a similar change of variables as the one explained in Appendix \ref{AppendixT0}, we obtain
\begin{equation}
     \label{eq: ququart: c12,c24,c23 at pA=1}
     \begin{aligned}
         &c_{23}^{(n)} = e^{i\tau\omega n} (\lambda_+)^n\left[c_{23}^{(0)}+n\frac{\lambda_-}{\lambda_+}c_{34}^{(0)}\right],\\
         &c_{13}^{(n)} = e^{2i\tau \omega n} \mu^n\left[c_{13}^{(0)}+\frac{\lambda_-}{\mu} \frac{1-\mu^n}{1-\mu}c_{24}^{(0)} \right] ,\\
         & c_{12}^{(n)} = e^{i\tau \omega n} \left\{\mu^n c_{12}^{(0)}+ (\lambda_+)^n \frac{\lambda_-}{\lambda_+}\left[ \frac{1-\mu^n}{1-\mu} c_{23}^{(0)} + \frac{\lambda_-}{\lambda_+} \left(n-\frac{1-\mu^n}{1-\mu} \right)\right] \right\}.
     \end{aligned}
\end{equation}



\subsubsection{$n^*$ in the $J\tau1$ limit at $p_A = 1$.}

Once we have the rate equations as a function of the initial conditions, we can attempt to calculate the total number of collisions $n^*$ needed to reach a state $\epsilon$ close to the target thermal state by initializing the system from the maximally mixed state. 
Adapting Eq. (\ref{eq:trace-distance-simplification}) for this case, we saturate the inequality, and obtain the following transcendental equation from which one can  obtain $n^*$ numerically,
\begin{equation}
    \left(A_0 + A_1n^* +A_2\frac{n^*(n^*-1)}{2} \right)e^{\ln\lambda_+ n^*} = \epsilon.
\end{equation}
Here, $A_0 = p_2^{(0)}+p_3^{(0)}+p_4^{(0)}$, $A_1 = \frac{\lambda_-}{\lambda_+}\left( p_3^{(0)}+p_4^{(0)}\right)$, and $A_2 = \frac{\lambda_-^2}{\lambda_+^2}p_4^{(0)}$. When $p_4^{(0)} = 0$, we  recover Eq. (\ref{eq: condition-nstar-qutrit-qubit-rewritten}) for a qutrit. It is important to mention that this equation does not allow for a closed analytical expression; it can only be solved numerically.

\subsubsection{Optimal thermal state preparation at $J\tau = \pi/2$, $p_A=1$.}

Next, we are interested in studying the optimal limit, where we set $J\tau = \pi/2$. After one collision, we obtain from Eq. (\ref{eq:ququart-qubit-populations})
\begin{equation}
    \begin{aligned}
        &p_1^{(1)} = p_1^{(0)}+p_2^{(0)},\\
        &p_2^{(1)}=p_3^{(0)},\\
        & p_3^{(1)} = p_4^{(0)},\\
        &p_4^{(1)} = 0.
    \end{aligned}
\end{equation}
The second collision leads to
\begin{equation}
    \begin{aligned}
        &p_1^{(2)} = p_1^{(1)}+p_2^{(1)}=p_1^{(0)}+p_2^{(0)}+p_3^{(0)},\\
        &p_2^{(2)}=p_3^{(1)}=p_4^{(0)},\\
        & p_3^{(2)} = p_4^{(1)}=0,\\
        &p_4^{(2)} = 0.
    \end{aligned}
\end{equation}
And after one more collision 
\begin{equation}
    \begin{aligned}
        &p_1^{(3)} = p_1^{(2)}+p_2^{(2)}=p_1^{(1)}+p_2^{(1)}+p_3^{(1)}=p_1^{(0)}+p_2^{(0)}+p_3^{(0)}+p_4^{(0)}=1,\\
        &p_2^{(3)}=p_3^{(2)}=p_4^{(1)}=0,\\
        & p_3^{(3)} = p_4^{(2)}=0,\\
        &p_4^{(3)} = 0,
    \end{aligned}
\end{equation}
which corresponds to our zero temperature thermal state.

We also check that coherences die after three RI steps under these conditions (zero temperature and $J\tau=\pi/2)$. After one collision, we have
\begin{equation}
   \begin{aligned}
        & c_{34}^{(1)} = c_{24}^{(1)} = c_{14}^{(1)} = 0,\\
        & c_{23}^{(1)} =e^{i\tau \omega} c_{34}^{(0)}, \quad c_{12}^{(1)} = e^{i\tau\omega} c_{23}^{(0)}, \quad c_{13}^{(1)} = e^{2i\omega \tau} c_{24}^{(0)}.
   \end{aligned}
\end{equation}
After a second collision,
\begin{equation}
    c_{23}^{(2)} = e^{i\tau \omega} c_{34}^{(1)} = 0, \quad  c_{12}^{(2)} = e^{i\tau\omega} c_{23}^{(1)} = e^{2i\tau \omega} c_{34}^{(0)}, \quad c_{13}^{(2)} = e^{2i\omega \tau} c_{24}^{(1)} = 0.
\end{equation}
And finally after one more collision the last term vanishes
\begin{equation}
    c_{12}^{(3)} = e^{i\omega \tau} c_{23}^{(2)} = 0.
\end{equation}
Therefore, after $n^*=3=d-1$ collisions, both the populations and the coherences thermalize.
\subsubsection{Eigenvalues of the stochastic matrix}

Working in this $J\tau1$ limit, the next question that we are interested in understanding is why the Mpemba effect is more pronounced as we increase the dimensionality of the system. Recall that the Mpemba effect manifests itself due to the second largest eigenvalue, $\xi_2$, getting closer to one. This effect should become more pronounced as we increase the dimension.

From Eq. (\ref{eq:stochastic_matrix_ququart}), we obtain that the eigenvalues of the stochastic matrix as
\begin{equation}
    \label{eq:eigenvalues_stochastic_matrix4}
    \xi_1 = 1, \quad \xi_2 = \lambda_+ + \sqrt{2}\theta \lambda_-, \quad \xi_3 = \lambda_+, \quad\xi_4 = \lambda_+ - \sqrt{2}\theta \lambda_-.
\end{equation}
In Fig. \ref{fig:eigenvalues_multidimensional_case}(a) we show that these eigenvalues satisfy the properties of the eigenvalues of a stochastic matrix. Furthermore, we observe that $\xi_2^{(d=4)}>\xi_2^{(d=3)}$, 
see Eq. (\ref{eq:eigenvalues_stochastic_matrix}) with
$\xi_2^{(d=3)} = \lambda_+ + \theta\lambda_-$.
This suggests that
 as the dimension of the system increases, the Mpemba effect becomes more pronounced.

\subsubsection{Equations of motion in the SL limit and $T_{\text{sim}}$ estimation at $p_A = 1$}

Expanding Eq. (\ref{eq:stochastic_matrix_ququart})
to second order in $J\tau$, as we did for the qutrit case, and imposing the SL limit, 
we end up obtaining the following differential equation,
\begin{equation}
\label{eq:diff_eq_Lindblad_ququart}
    \frac{d}{dt} \begin{pmatrix}
        p_1(t)\\
        p_2(t)\\
        p_3(t)\\
        p_4(t)
    \end{pmatrix} = 
    \Gamma
    \begin{pmatrix}
        -(1-p_A) & p_A & 0&0\\
        1-p_A & -1 & p_A&0\\
        0 & 1-p_A & -1&p_A\\
        0&0&1-p_A&-p_A
    \end{pmatrix}
    \begin{pmatrix}
        p_1(t)\\
        p_2(t)\\
        p_3(t)\\
        p_4(t)
    \end{pmatrix}.
\end{equation}
It can be readily solved at zero temperature. 
Setting $p_A=1$, we have
\begin{equation}
    \frac{d}{dt} \begin{pmatrix}
        p_1(t)\\
        p_2(t)\\
        p_3(t)\\
        p_4(t)
    \end{pmatrix} = 
    \Gamma
    \begin{pmatrix}
        0 & 1 & 0&0\\
        0 & -1 & 1&0\\
        0 & 0 & -1&1\\
        0&0&0&-1
    \end{pmatrix}
    \begin{pmatrix}
        p_1(t)\\
        p_2(t)\\
        p_3(t)\\
        p_4(t)
    \end{pmatrix},
\end{equation}
which we solve by a direct exponentiation,
\begin{equation}
    \begin{pmatrix}
        p_1(t)\\
        p_2(t)\\
        p_3(t)\\
        p_4(t)
        \end{pmatrix}=
\begin{pmatrix}
1 & 1 - e^{-t \Gamma} & 1 - e^{-t \Gamma}(1 + t \Gamma) & 1 + \frac{1}{2} e^{-t \Gamma} \left( -2 - t \Gamma (2 + t \Gamma) \right) \\
0 & e^{-t \Gamma} & e^{-t \Gamma} t \Gamma & \frac{1}{2} e^{-t \Gamma} t^2 \Gamma^2 \\
0 & 0 & e^{-t \Gamma} & e^{-t \Gamma} t \Gamma \\
0 & 0 & 0 & e^{-t \Gamma}
\end{pmatrix}
\begin{pmatrix}
        p_1(0)\\
        p_2(0)\\
        p_3(0)\\
        p_4(0)
        \end{pmatrix}.
\end{equation}
Simplifying, the explicit solution for the populations are
\begin{equation}
    \begin{aligned}
        &p_1(t) = 1-e^{-\Gamma t}\left[p_2(0)+p_3(0)(1+\Gamma t)+\frac{p_4(0)}{2}(2+\Gamma t(2+\Gamma t))\right],\\
        &p_2(t) = e^{-\Gamma t}\left[p_2(0)+\Gamma tp_3(0)+\frac{p_4(0)}{2}\Gamma^2t^2\right],\\
        & p_3(t) = e^{-\Gamma t}\left[p_3(0)+p_4(0)\Gamma t\right],\\
        &p_4(t) = e^{-\Gamma t} p_4(0),
    \end{aligned}
\end{equation}
where we have new factors proportional to $\Gamma^2 t^2$, in comparison with the qutrit case.

To obtain $T_{\text{sim}}$, 
we need to compute the trace-distance condition, and saturate the inequality. The transcendental equation that we get is
\begin{equation}
\label{eq:Tsim4}
    e^{-\Gamma T_{\text{sim}}}\left[B_0+B_1 T_{\text{sim}}+B_2T_{\text{sim}}^2 \right]=\epsilon,
\end{equation}
where $B_0 = p_2(0)+p_3(0)+p_4(0)$, $B_1 = \Gamma\left[p_3(0)+p_4(0)\right]$, and $B_2 = \frac{1}{2}\Gamma^2 p_4(0)$. Again, due to the prefactor $ T_{\text{sim}}^2$, this equation can only be solved using numerical tools.

\subsubsection{Eigenvalues of the Liouvillian}

Focusing now on explaining the Mpemba effect for this system in the SL regime, we derive the eigenvalues by diagonalizing the Liouvillian
of Eq. (\ref{eq:diff_eq_Lindblad_ququart}),
\begin{equation}
    \lambda_1 =0, \quad \lambda_2 = -\Gamma(1-\sqrt{2}\theta),\quad\lambda_3 = -\Gamma, \quad \lambda_4 = -\Gamma(1+\sqrt{2}\theta).
\end{equation}
These eigenvalues are depicted in Fig. \ref{fig:eigenvalues_multidimensional_case}(b).

Recall that for the qutrit case we obtained $\lambda_2 = -\Gamma(1-\theta)$, see Eq. (\ref{eq:eigenvalues_Liouvillian}).
These results align with what we have been observing numerically: Since $\lambda_2$ for the ququart is smaller in magnitude than for the qutrit, we observe a more pronounced Mpemba effect in the former. 

\subsection{General equations for a $d=5$ system}

In the case of a ququint, the system Hamiltonian is Eq. (\ref{eq: free-qudit-systemHamiltonian}) with $s=2$. 
The total Hamiltonian using the interaction Hamiltonian of Eq. (\ref{eq:interaction-Hamiltonian-qudit-qubit}) adapted for this case is
\begin{equation}
    \hat{H}_{\text{tot}} = \begin{pmatrix}
-\frac{5\omega}{2} & 0 & 0 & 0 & 0 & 0 & 0 & 0 & 0 & 0 \\
0 & -\frac{3\omega}{2} & J & 0 & 0 & 0 & 0 & 0 & 0 & 0 \\
0 & J & -\frac{3\omega}{2} & 0 & 0 & 0 & 0 & 0 & 0 & 0 \\
0 & 0 & 0 & -\frac{\omega}{2} & J & 0 & 0 & 0 & 0 & 0 \\
0 & 0 & 0 & J & -\frac{\omega}{2} & 0 & 0 & 0 & 0 & 0 \\
0 & 0 & 0 & 0 & 0 & \frac{\omega}{2} & J & 0 & 0 & 0 \\
0 & 0 & 0 & 0 & 0 & J & \frac{\omega}{2} & 0 & 0 & 0 \\
0 & 0 & 0 & 0 & 0 & 0 & 0 & \frac{3\omega}{2} & J & 0 \\
0 & 0 & 0 & 0 & 0 & 0 & 0 & J & \frac{3\omega}{2} & 0 \\
0 & 0 & 0 & 0 & 0 & 0 & 0 & 0 & 0 & \frac{5\omega}{2}
\end{pmatrix}.
\end{equation}
The collision unitary is given by 
\begin{equation}
\resizebox{\textwidth}{!}{$
\hat{U}(\tau)= 
\begin{pmatrix}
\exp\left(\frac{5i\tau\omega}{2}\right) & 0 & 0 & 0 & 0 & 0 & 0 & 0 & 0 & 0 \\
0 & \exp\left(\frac{3i\tau\omega}{2}\right)\cos(J\tau) & -i\exp\left(\frac{3i\tau\omega}{2}\right)\sin(J\tau) & 0 & 0 & 0 & 0 & 0 & 0 & 0 \\
0 & -i\exp\left(\frac{3i\tau\omega}{2}\right)\sin(J\tau) & \exp\left(\frac{3i\tau\omega}{2}\right)\cos(J\tau) & 0 & 0 & 0 & 0 & 0 & 0 & 0 \\
0 & 0 & 0 & \exp\left(\frac{i\tau\omega}{2}\right)\cos(J\tau) & -i\exp\left(\frac{i\tau\omega}{2}\right)\sin(J\tau) & 0 & 0 & 0 & 0 & 0 \\
0 & 0 & 0 & -i\exp\left(\frac{i\tau\omega}{2}\right)\sin(J\tau) & \exp\left(\frac{i\tau\omega}{2}\right)\cos(J\tau) & 0 & 0 & 0 & 0 & 0 \\
0 & 0 & 0 & 0 & 0 & \exp\left(-\frac{i\tau\omega}{2}\right)\cos(J\tau) & -i\exp\left(-\frac{i\tau\omega}{2}\right)\sin(J\tau) & 0 & 0 & 0 \\
0 & 0 & 0 & 0 & 0 & -i\exp\left(-\frac{i\tau\omega}{2}\right)\sin(J\tau) & \exp\left(-\frac{i\tau\omega}{2}\right)\cos(J\tau) & 0 & 0 & 0 \\
0 & 0 & 0 & 0 & 0 & 0 & 0 & \exp\left(-\frac{3i\tau\omega}{2}\right)\cos(J\tau) & -i\exp\left(-\frac{3i\tau\omega}{2}\right)\sin(J\tau) & 0 \\
0 & 0 & 0 & 0 & 0 & 0 & 0 & -i\exp\left(-\frac{3i\tau\omega}{2}\right)\sin(J\tau) & \exp\left(-\frac{3i\tau\omega}{2}\right)\cos(J\tau) & 0 \\
0 & 0 & 0 & 0 & 0 & 0 & 0 & 0 & 0 & \exp\left(-\frac{5i\tau\omega}{2}\right)
\end{pmatrix}
$}
\end{equation}
Evolving the state using Eq. (\ref{eq: CPTPmap}), the populations obey the recursive relation,  
\begin{equation}
\label{eq:recurrence_ququint}
\begin{aligned}
p_1^{(n+1)} &= \frac{1}{2} \left\{
p_1^{(n)} \left[1 + p_A + (1 - p_A) \cos(2J\tau)\right] +
p_2^{(n)} \, p_A \left[1 - \cos(2J\tau)\right]
\right\}, \\
p_2^{(n+1)} &= \frac{1}{2} \left\{
p_1^{(n)} \left[1 - p_A - (1 - p_A) \cos(2J\tau)\right] +p_2^{(n)} \left[1+ \cos(2J\tau)\right] +
p_3^{(n)} \, p_A \left[1 - \cos(2J\tau)\right]
\right\}, \\
p_3^{(n+1)} &= \frac{1}{2} \left\{
p_2^{(n)} \left[1 - p_A -(1 - p_A) \cos(2J\tau)\right] +
p_3^{(n)} \left[1+\cos(2J\tau)\right] +
p_4^{(n)} \, p_A \left[1 - \cos(2J\tau)\right]
\right\}, \\
p_4^{(n+1)} &= \frac{1}{2} \left\{
p_3^{(n)} \left[1 - p_A - (1 - p_A) \cos(2J\tau)\right] +
p_4^{(n)} \left[1 + \cos(2J\tau)\right] +
p_5^{(n)} \, p_A \left[1 - \cos(2J\tau)\right]
\right\}, \\
p_5^{(n+1)} &= \frac{1}{2} \left\{p_4^{(n)}\left[1-p_A-(1-p_A)\cos(2J\tau)\right]+p_5^{(n)}\left[2-p_A+p_A\cos(2J\tau)\right]
\right\}.
\end{aligned}
\end{equation}
This equation can be rewritten in matrix form as
\begin{equation}
\label{eq:stochastic_matrix_ququint}
    \Delta\mathbf{p}^{(n+1)} = \begin{pmatrix}
        \eta_{11} & \eta_{12} &0 &0&0\\
        \eta_{21} & \eta_{22} & \eta_{12}&0&0\\
        0& \eta_{21} & \eta_{22}&\eta_{12}&0\\
        0&0&\eta_{21}& \eta_{22} &\eta_{12}\\
        0&0&0&\eta_{21}& \eta_{33}
    \end{pmatrix}
    \Delta \mathbf{p}^{(n)}.
\end{equation}
At zero temperature, setting $p_A = 1$, we obtain
\begin{equation}
\label{eq:stochastic_matrix_ququint_pA1}
    \Delta\mathbf{p}^{(n+1)} = \begin{pmatrix}
        1 & \lambda_- &0 &0&0\\
        0 & \lambda_+ & \lambda_-&0&0\\
        0& 0 & \lambda_+&\lambda_-&0\\
        0&0&0& \lambda_+& \lambda_-\\
        0&0&0&0&\lambda_+
    \end{pmatrix}
    \Delta \mathbf{p}^{(n)}.
\end{equation}
The recursive relations for the populations at zero temperature are given by
\begin{equation}
    \label{eq:recursive_equations_ququint}
    \begin{aligned}
        & p_5^{(n)} = (\lambda_+)^n p_5^{(0)},\\
        &p_4^{(n)} = (\lambda_+)^n p_4^{(0)} + n \lambda_- (\lambda_+)^{n-1} p_5^{(0)},\\
        & p_3^{(n)} = (\lambda_+)^n p_3^{(0)}+n\lambda_-(\lambda_+)^{n-1}p_4^{(0)}+\frac{n(n-1)}{2} \lambda_-^2(\lambda_+)^{n-2} p_5^{(0)},\\
        &p_2^{(n)} = (\lambda_+)^n p_2^{(0)} + n \lambda_-(\lambda_+)^{n-1}p_3^{(0)} + \frac{n(n-1)}{2}\lambda_-(\lambda_+)^{n-2} p_4^{(0)} + \frac{n(n-1)(n-2)}{6}\lambda_-^3(\lambda_+)^{n-3} p_5^{(0)},\\
        & p_1^{(n)} = 1-p_2^{(n)}-p_3^{(n)}- p_4^{(n)}-p_5^{(n)}.
    \end{aligned}
\end{equation}

Reminding that $\lambda_+ = \cos^2(J\tau)$, $\lambda_- = \sin^2(J\tau)$, and $\mu =\cos(J\tau)$, the resulting equations for the coherences $c_{ij}$ are
\begin{equation}
   \begin{aligned}
   \label{eq: coherences_ququint}
        &c_{12}^{(n+1)} = e^{i\tau \omega}\left[c_{12}^{(n)}\left(\lambda_++p_A(\mu-\lambda_+) \right)+c_{23}^{(n)}p_A \lambda_-\right],\\
        &c_{13}^{(n+1)} = e^{2i\tau\omega}\left[c_{13}^{(n)}\left( \lambda_+ + p_A(\mu - \lambda_+)\right)+c_{24}^{(n)}p_A \lambda_- \right],\\
        & c_{14}^{(n+1)} = e^{3i\tau\omega}\left[c_{14}^{(n)}\left( \lambda_+ + p_A(\mu - \lambda_+)\right)+c_{25}^{(n)}p_A \lambda_- \right],\\
        &c_{15}^{(n+1)} = e^{4i\tau\omega}\mu c_{15}^{(n)},\\
        & c_{23}^{(n+1)} = e^{i\tau \omega} \left[c_{12}^{(n)}\lambda_-(1-p_A)+c_{23}^{(n)}\lambda_+ + c_{34}^{(n)}p_A \lambda_-\right]\\
        & c_{24}^{(n+1)} = e^{2i\tau \omega} \left[c_{13}^{(n)}\lambda_-(1-p_A)+c_{24}^{(n)}\lambda_+ + c_{35}^{(n)}p_A \lambda_-\right] \\
        & c_{25}^{(n+1)} = e^{3i\tau\omega}\left[c_{14}^{(n)} \lambda_-(1-p_A)+c_{25}^{(n)}\left((1-p_A)\mu +p_A \lambda_+ \right) \right]\\
        & c_{34}^{(n+1)} = e^{i\tau \omega} \left[c_{23}^{(n)}\lambda_-(1-p_A) + c_{34}^{(n)}\lambda_+ + c_{45}^{(n)}p_A \lambda_- \right]\\
        & c_{35}^{(n+1)} = e^{2i\tau \omega} \left[c_{24}^{(n)}\lambda_-(1-p_A) + c_{35}^{(n)}\left(p_A \lambda_+ +(1-p_A) \mu \right) \right]\\
        & c_{45}^{(n+1)} = e^{i\tau \omega} \left[c_{34}^{(n)}\lambda_- (1-p_A) + c_{45}^{(n)} \left(\lambda_++(1-p_A) \mu \right) \right]
   \end{aligned}
\end{equation}

In the zero-temperature limit, corresponding to $p_A = 1$, the recurrence relations simplify and become anlytically solvable. In this regime, Eq. (\ref{eq: coherences_ququint}) reduces to

\begin{equation}
   \begin{aligned}
   \label{eq: coherences_ququint_zero_temperature}
        &c_{12}^{(n+1)} = e^{i\tau \omega}\left[c_{12}^{(n)}\mu+c_{23}^{(n)} \lambda_-\right],\\
        & c_{13}^{(n+1)} = e^{2i\tau \omega} \left[c_{13}^{(n)}\mu +c_{24}^{(n)}\lambda_- \right],\\
        &c_{14}^{(n+1)} = e^{3i\tau \omega} \left[c_{14}^{(n)}\mu +c_{25}^{(n)}\lambda_- \right],\\
        & c_{23}^{(n+1)} = e^{i\tau \omega} \left[c_{23}^{(n)}\lambda_+ +c_{34}^{(n)} \lambda_-\right],\\
        & c_{24}^{(n+1)} = e^{2i\tau\omega} \left[\lambda_+ c_{24}^{(n)} + \lambda_- c_{35}^{(n)} \right]\\
        & c_{34}^{(n+1)} = e^{i\tau \omega} \left[\lambda_+ c_{34}^{(n)} + \lambda_- c_{45}^{(n)} \right]\\
        &c_{15}^{(n+1)} = e^{4i\tau \omega} c_{15}^{(n+1)} \mu,\\
        & c_{25}^{(n)} = e^{3i\tau \omega} c_{25}^{(n)} \lambda_+.\\
        & c_{35}^{(n+1)} = e^{2i\tau \omega} \lambda_+ c_{35}^{(n)},\\
        &c_{45}^{(n+1)} = e^{i\tau \omega} \lambda_+ c_{45}^{(n)}.
   \end{aligned}
\end{equation}
We begin by solving the recurrence relations for the last four terms, which evolve independently and decay exponentially,
\begin{equation}
     \label{eq: ququint: c15,c25,c35,c45 at pA=1}
     \begin{aligned}
         &c_{15}^{(n)} = e^{4i\tau\omega n} \mu^n c_{15}^{(0)},\\
         &c_{25}^{(n)} = e^{3i\tau \omega n} (\lambda_+)^n c_{25}^{(0)},\\
         & c_{35}^{(n)} = e^{2i\tau \omega n} (\lambda_+)^nc_{35}^{(0)},\\
         & c_{45}^{(n)} = e^{i\tau \omega n} (\lambda_+)^nc_{45}^{(0)},
     \end{aligned}
\end{equation}
Next, we consider the coherences $c_{34}^{(n)}$, $c_{24}^{(n)}$, and $c_{14}^{(n)}$, which are coupled to the previous Eq. (\ref{eq: ququint: c15,c25,c35,c45 at pA=1}). Their solutions read

\begin{equation}
     \label{eq: ququint: c34,c24,c14 at pA=1}
     \begin{aligned}
         &c_{34}^{(n)} = e^{i\tau\omega n} (\lambda_+)^n\left[c_{34}^{(0)}+n\frac{\lambda_-}{\lambda_+}c_{45}^{(0)}\right],\\
         &c_{24}^{(n)} = e^{2i\tau\omega n} (\lambda_+)^n\left[c_{24}^{(0)}+n\frac{\lambda_-}{\lambda_+}c_{35}^{(0)}\right],\\
         &c_{14}^{(n)} = e^{3i\tau \omega n} \mu^n\left[c_{14}^{(0)}+\frac{\lambda_-}{\mu} \frac{1-\mu^n}{1-\mu}c_{25}^{(0)} \right].\\
     \end{aligned}
\end{equation}
We now proceed to the next layer of the hierarchy, where $c_{23}^{(n)}$ and $c_{13}^{(n)}$ depend on the solutions from above
\begin{equation}
     \label{eq: ququint: c23,c13 at pA=1}
     \begin{aligned}
         &c_{23}^{(n)} = e^{i\tau\omega n} (\lambda_+)^n\left[c_{23}^{(0)}+n\frac{\lambda_-}{\lambda_+}c_{34}^{(0)}+\frac{n(n-1)}{2}\left(\frac{\lambda_-}{\lambda_+} \right)^2 c_{45}^{(0)}\right],\\
         &c_{13}^{(n)} = e^{2i\tau\omega n} \mu^n\left[c_{13}^{(0)}+\frac{c_{24}^{(0)}}{\mu}\frac{1-\mu^n}{1-\mu}+\frac{\lambda_-}{\lambda_+}\frac{c_{35}^{(0)}}{\mu}\frac{\mu-n\mu^n+(n-1)\mu^{n+1}}{(1-\mu)^2}\right].\\
     \end{aligned}
\end{equation}
Finally, the expression for $c_{12}^{(n)}$ is
\begin{equation}
     \label{eq: ququint: c12 at pA=1}
     \begin{aligned}
         c_{12}^{(n)} =& e^{i\tau\omega n} \mu^n\Bigg\{c_{12}^{(0)}+ \frac{\lambda_-}{\mu}\frac{1-\mu^n}{1-\mu} c_{23}^{(0)}+\left(  
         \frac{\lambda_-}{\lambda_+}\right)^2c_{34}^{(0)}\left[\frac{\mu -n\mu^n+(n-1)\mu^{n+1}}{\mu(1-\mu)^2}\right]\\
         &+\left(\frac{\lambda_-}{\lambda_+} \right)^3c_{45}^{(0)}\left[\frac{(n^2-3n+2)\mu^{n}-2(n-2)n\mu^{n-1}+(n-1)n\mu^{n-2}-2}{2\mu(1-\mu)^3} \right]\Bigg\}.\\
     \end{aligned}
\end{equation}

\subsubsection{$n^*$ in the $J\tau 1$ limit at $p_A = 1$}

Initializing our system from the maximally mixed state and adapting Eq. (\ref{eq:trace-distance-simplification}) for this case, we saturate the inequality, and obtain the following transcendental equation from which one can  obtain $n^*$ numerically,
\begin{equation}
    \left(C_0 + C_1 n^* + C_2\frac{ n^*(n^*-1)}{2}+C_3 \frac{n(n-1)(n-2)}{6} \right) e^{\ln \lambda_+ n^*} = \epsilon.
\end{equation}
Here, $C_0 = p_2^{(0)}+p_3^{(0)}+p_4^{(0)}+p_5^{(0)}$, $C_1 = \frac{\lambda_-}{\lambda_+}\left(p_3^{(0)}+p_4^{(0)}+p_5^{(0)} \right)$, $C_2 = \frac{\lambda_-^2}{\lambda_+^2}\left(p_4^{(0)}+p_5^{(0)} \right)$, and $C_3 = \frac{\lambda_-^3}{\lambda_+^3}p_5^{(0)}$.

\subsubsection{Optimal simulation time at $J\tau = \pi/2$, $p_A = 1$}

We focus here on the
low-temperature limit, setting $p_A = 1$. We also set the interaction at $J\tau = \pi/2$, which we now prove to be optimal: Under this interaction and at zero temperature, one can readily prove that the system thermalizes (cools) with $n^*=4$ collisions, one more than for the four-level system:
\begin{equation}
    \begin{aligned}
        &p_1^{(1)} = p_1^{(0)}+ p_2^{(0)}, \quad p_2^{(1)} = p_3^{(0)}, \quad
        p_3^{(1)} = p_4^{(0)}, \quad
        p_4^{(1)} = p_5^{(0)}, \quad p_5^{(1)} = 0;\\
        &p_1^{(2)} = p_1^{(0)}+p_2^{(0)}+p_3^{(0)}, \quad p_2^{(2)} = p_4^{(0)} , \quad p_3^{(2)} = p_5^{(0)}, \quad p_4^{(2)}  = 0;\\
        & p_1^{(3)}  = p_1^{(0)}+p_2^{(0)}+p_3^{(0)}+p_4^{(0)}, \quad p_2^{(3)} = p_5^{(0)} , \quad p_3^{(3)}  = 0;\\
        & p_1^{(4)} = p_1^{(0)}+p_2^{(0)}+p_3^{(0)}+p_4^{(0)}+p_5^{(0)}, \quad p_2^{(4)} = 0.
    \end{aligned}
\end{equation}
We now verify that the coherences also vanish after $n^* = 4 = d - 1$ collisions in this optimal regime. At this point, the recursion relations simplify such that only feed-forward terms remain active, while all self-decay terms vanish due to $\mu = \lambda_+ = 0$ and $\lambda_- = 1$.

After the first collision, the only non-zero coherences are those directly fed by initial values
\begin{equation}
\begin{aligned}
    & c_{15}^{(1)} = c_{35}^{(1)} = c_{45}^{(1)}=c_{25}^{(1)} = 0, \\
    & c_{34}^{(1)} = e^{i\tau\omega} c_{45}^{(0)}, \quad
      c_{24}^{(1)} = e^{2i\tau\omega} c_{35}^{(0)}, \quad
      c_{14}^{(1)} = e^{3i\tau\omega} c_{25}^{(0)}, \\
    & c_{13}^{(1)} = e^{2i\tau\omega} c_{24}^{(0)}, \quad
      c_{23}^{(1)} = e^{i\tau\omega} c_{34}^{(0)}, \quad
      c_{12}^{(1)} = e^{i\tau\omega} c_{23}^{(0)}.
\end{aligned}
\end{equation}

After the second collision, the coherences continue to shift downward
\begin{equation}
\begin{aligned}
    & c_{34}^{(2)} = e^{i\tau\omega} c_{45}^{(1)} = 0, \quad
      c_{24}^{(2)} = e^{2i\tau\omega} c_{35}^{(1)} = 0, \quad
      c_{14}^{(2)} = e^{3i\tau\omega} c_{25}^{(1)} = 0, \\
    & c_{23}^{(2)} = e^{i\tau\omega} c_{34}^{(1)} = e^{2i\tau\omega} c_{45}^{(0)}, \quad
      c_{13}^{(2)} = e^{2i\tau\omega} c_{24}^{(1)} = e^{4i\tau\omega} c_{35}^{(0)}, \\
    & c_{12}^{(2)} = e^{i\tau\omega} c_{23}^{(1)} = e^{2i\tau\omega} c_{34}^{(0)}.
\end{aligned}
\end{equation}

After the third collision, only one coherence remains:
\begin{equation}
\begin{aligned}
    & c_{23}^{(3)} = e^{i\tau\omega} c_{34}^{(2)} = 0, \quad
      c_{13}^{(3)} = e^{2i\tau\omega} c_{24}^{(2)} = 0, \quad  c_{12}^{(3)} = e^{i\tau\omega} c_{23}^{(2)} = e^{3i\tau\omega} c_{45}^{(0)},
\end{aligned}
\end{equation}
and it vanishes after one more collision,
\begin{equation}
    c_{12}^{(4)} = e^{i\tau\omega} c_{23}^{(3)} = 0.
\end{equation}

We therefore conclude that the system thermalizes after $n^*=d-1=4$ collisions.

\subsubsection{Eigenvalues of the stochastic matrix}

Working in this $J\tau1$ limit, once again we can approach from a mathematical standpoint the question why
 the Mpemba effect becomes more pronounced as we increase the dimension of the system. 
From Eq. (\ref{eq:stochastic_matrix_ququint}) we obtain the eigenvalues of the stochastic matrix,
\begin{equation}
\label{eq:eigenvalues_stochastic_matrix5}
    \xi_1 = 1, \quad \xi_2 = \lambda_+ + \theta \lambda_-\left( \frac{1+\sqrt{5}}{2}\right), \quad \xi_3 = \lambda_++\theta\lambda_-\left( \frac{-1+\sqrt{5}}{2}\right), \quad\xi_4 = \lambda_+-\theta\lambda_-\left( \frac{-1+\sqrt{5}}{2}\right), \quad \xi_5 = \lambda_+-\theta\lambda_-\left( \frac{1+\sqrt{5}}{2}\right)
\end{equation}
These eigenvalues are shown in Fig. \ref{fig:eigenvalues_multidimensional_case}(c). We immediately note that  $\xi_2^{(d=5)}>\xi_2^{(d=4)}>\xi_2^{(d=3)}$, indicating that as the dimension increases, more steps are needed for thermal state preparation. As a result, the Mpemba effect becomes more pronounced.

\subsubsection{Equations of motion in the SL limit and $T_{\text{sim}}$ estimation at $p_A=1$}

We repeat the analysis of the $d=5$ system in the SL regime. 
Expanding Eq. (\ref{eq:stochastic_matrix_ququint}) to the second order in $J\tau$, we end up obtaining the following differential equation
\begin{equation}
\label{eq:diff_eq_Lindblad_ququint}
    \frac{d}{dt} \begin{pmatrix}
        p_1(t)\\
        p_2(t)\\
        p_3(t)\\
        p_4(t)\\
        p_5(t)
    \end{pmatrix} = 
    \Gamma
    \begin{pmatrix}
        -(1-p_A) & p_A & 0&0&0\\
        1-p_A & -1 & p_A&0&0\\
        0 & 1-p_A & -1&p_A&0\\
        0&0&1-p_A&-1&p_A\\
        0&0&0&1-p_A&-p_A
    \end{pmatrix}
    \begin{pmatrix}
        p_1(t)\\
        p_2(t)\\
        p_3(t)\\
        p_4(t)\\
        p_5(t)
    \end{pmatrix}
\end{equation}
Setting $p_A = 1$, we have
\begin{equation}
    \frac{d}{dt} \begin{pmatrix}
        p_1(t)\\
        p_2(t)\\
        p_3(t)\\
        p_4(t)\\
        p_5(t)
    \end{pmatrix} = 
    \Gamma
    \begin{pmatrix}
        0 & 1 & 0&0&0\\
        0 & -1 & 1&0&0\\
        0 & 0 & -1&1&0\\
        0&0&0&-1&1\\
        0&0&0&0&-1
    \end{pmatrix}
    \begin{pmatrix}
        p_1(t)\\
        p_2(t)\\
        p_3(t)\\
        p_4(t)\\
        p_5(t)
    \end{pmatrix}
\end{equation}
which can be solved by direct exponentiation,
\begin{equation}
    \begin{pmatrix}
        p_1(t)\\
        p_2(t)\\
        p_3(t)\\
        p_4(t)\\
        p_5(t)
        \end{pmatrix}=
\begin{pmatrix}
1 & 1 - e^{-t\Gamma} & 1 - e^{-t\Gamma}(1 + t\Gamma) & 1 + \frac{1}{2} e^{-t\Gamma} \left(-2 - t\Gamma(2 + t\Gamma)\right) & 1 + \frac{1}{6} e^{-t\Gamma} \left(-6 - t\Gamma(6 + t\Gamma(3 + t\Gamma)) \right) \\
0 & e^{-t\Gamma} & e^{-t\Gamma} \, t\Gamma & \frac{1}{2} e^{-t\Gamma} \, t^2 \Gamma^2 & \frac{1}{6} e^{-t\Gamma} \, t^3 \Gamma^3 \\
0 & 0 & e^{-t\Gamma} & e^{-t\Gamma} \, t\Gamma & \frac{1}{2} e^{-t\Gamma} \, t^2 \Gamma^2 \\
0 & 0 & 0 & e^{-t\Gamma} & e^{-t\Gamma} \, t\Gamma \\
0 & 0 & 0 & 0 & e^{-t\Gamma}
\end{pmatrix}
\begin{pmatrix}
        p_1(0)\\
        p_2(0)\\
        p_3(0)\\
        p_4(0)\\
        p_5(0)
        \end{pmatrix}.
\end{equation}
Simplifying, the explicit solutions for the populations are
\begin{equation}
    \begin{aligned}
        &p_1(t) = 1-e^{-\Gamma t}\left[p_2(0)+p_3(0)(1+\Gamma t)+\frac{p_4(0)}{2}(2+\Gamma t(2+\Gamma t))+\frac{p_5(0)}{6}(6+\Gamma t(6+\Gamma t(3+\Gamma t)))\right],\\
        &p_2(t) = e^{-\Gamma t}\left[p_2(0)+\Gamma tp_3(0)+\frac{p_4(0)}{2}\Gamma^2t^2+\frac{p_5(0)}{6}\Gamma^3 t^3\right],\\
        & p_3(t) = e^{-\Gamma t}\left[p_3(0)+p_4(0)\Gamma t + \frac{1}{2}\Gamma^2 t^2\right],\\
        &p_4(t) = e^{-\Gamma t} [p_4(0)+\Gamma t p_5(0)],\\
        &p_5(t) = e^{-\Gamma t}p_5(0).
    \end{aligned}
\end{equation}
Saturating the trace distance condition, for this case we obtain
\begin{equation}
\label{eq:Tsim5}
    e^{-\Gamma T_{\text{sim}}}\left[D_0 +D_1T_{\text{sim}}+D_2T_{\text{sim}}^2+D_3T_{\text{sim}}^3\right]=\epsilon,
\end{equation}
where $D_0 = p_2(0)+p_3(0)+p_4(0)+p_5(0)$, $D_1 = \Gamma\left[p_3(0)+p_4(0)+p_5(0)\right]$, and $D_2 = \frac{1}{2}\Gamma^2 \left[p_4(0) + p_5(0) \right]$, and $D_3 = \frac{\Gamma^3}{6}p_5(0)$.
This equation can be solved numerically to extract the simulation time, $T_{sim}$.

\subsubsection{Eigenvalues of the Liouvillian}

We focus now on explaining the Mpemba effect for this $d=5$-level system. Using Eq. (\ref{eq:diff_eq_Lindblad_ququint}), we obtain its eigenvalues by diagonalization,
\begin{equation}
    \lambda_1 =0, \quad \lambda_2 = -\Gamma\left(1-\frac{1+\sqrt{5}}{2}\theta\right),\quad \lambda_3 =  -\Gamma\left(1-\frac{\sqrt{5}-1}{2}\theta\right), \quad \lambda_4 =  -\Gamma\left(1+\frac{\sqrt{5}-1}{2}\theta\right), \quad \lambda_5 =  -\Gamma\left(1+\frac{1+\sqrt{5}}{2}\theta\right),
\end{equation}
We plot these eigenvalues in Fig. \ref{fig:eigenvalues_multidimensional_case}(d), and confirm that
 $\lambda_2^{(d=5)}< \lambda_2^{(d=4)}<\lambda_2^{(d=3)}$. 

 \begin{figure}[h!]
     \centering
     \includegraphics[width=1\linewidth]{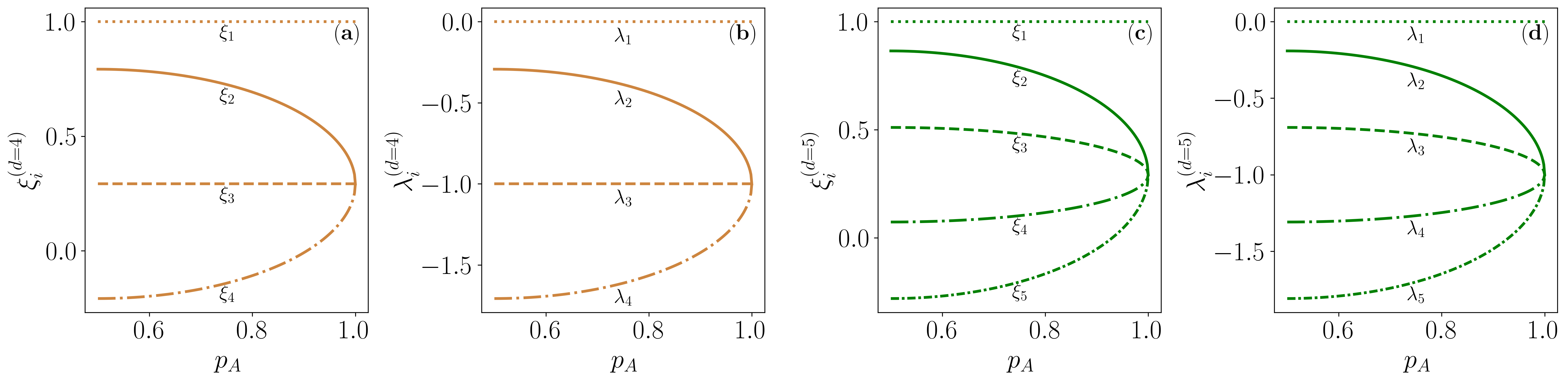} 
     \caption{(a), (c) Eigenvalues of the stochastic matrix for a $d=4$, and $d=5$ systems, respectively. For both, the product $J\tau$ is set to one. (b), (d) Eiganvalues of the Liouvillian for a $d=4$, and $d=5$ systems, respectively with $\Gamma=1$. 
     }
     \label{fig:eigenvalues_multidimensional_case}
 \end{figure}

\subsection{Generalization for a $d$-dimensional system.}
We addressed the thermal state preparation problem for $d=3$ in the main test, and for $d=4$ and $d=5$ in this Appendix. We now derive results for a $d-$dimensional system.

At zero temperature, $p_A=1$, we can generalize the recurrence equations for the populations. These can be expressed as
\begin{equation}
    \label{eq:general_equations_d_dim_system}
    \begin{aligned}
        &p_{d-k}^{(n)} = \sum_{j=0}^k \begin{pmatrix}
            n\\j
        \end{pmatrix} \lambda_-^j(\lambda_+)^{n-j}p_{d-k+j}^{(0)}, \quad \text{for } k=0,1,2, \ldots,d-2 \\
        & p_1^{(n)} = 1- \sum_{i=2} ^d p_i^{(n)}.
    \end{aligned}
\end{equation}
\subsubsection{$n^*$ in the $J\tau1$ limit at $p_A = 1$.}

Next, we obtain a compact expression for the total number of collisions needed, $n^*$, for a $ d$-dimensional system. From the trace distance condition, we have 
\begin{equation}
    D(\rho_S^{(n^*)},\rho_S^*) = \frac{1}{2}\sum_{i=1}^d \left|p_i^{(n^*)}-p_i^* \right| =\sum_{i=2}^dp_i^{(n^*)} = \sum_{k=0}^{d-2}p_{d-k}^{(n^*)} \leq \epsilon,
\end{equation}
where we have used that in the zero temperature limit, $p_1^*=1$, and $p_{i\neq 1} = 0$. Saturating the inequality and using Eq. (\ref{eq:general_equations_d_dim_system}) we obtain

\begin{equation}
    \sum_{k=0}^{d-2} \sum_{j=0}^{d-2} \begin{pmatrix}
        n^*\\
        j
    \end{pmatrix} \lambda_-^j \lambda_+^{n^*-j} p_{d-k+j}^{(0)} = \epsilon.
\end{equation}
This equation can be rewritten in a more compact form as 
\begin{equation}
    \label{eq:general_nstar_value}
    e^{\ln \lambda_+ n^*}\sum_{j=0}^{d-2} \begin{pmatrix}
        n^*\\
        j
    \end{pmatrix} \left(\frac{\lambda_-}{\lambda_+} \right)^j \left[\sum_{k=j}^{d-2}p_{d-k+j}^{(0)} \right] = \epsilon.
\end{equation}

\subsubsection{Thermal state preparation at $J\tau = \pi/2$, $p_A = 1$}

As we have seen before for the $d=3,4$ and $5$ cases, choosing $J\tau = \pi/2$ allows thermalization with only $d-1$ collisions. Next, using Eq. (\ref{eq:general_equations_d_dim_system}) we aim to generalize this result for the populations. The equation for the excited states leads to
\begin{equation}
    p_{d-k}^{(n)} = \sum_{j=0}^k\begin{pmatrix}
        n\\
        j
    \end{pmatrix} 1^j 0^{n-j} p_{d-k+j}^{(0)}.
\end{equation}
Here, the term $0^{n-j} = 0$ as long as $n-j>0$, being only equal to one when $n=j$. That term contributes to the sum if $n\leq k$. Concretely,
\begin{equation}
    p_{d-k}^{(n)} = \begin{cases}
        p_{d-k+n}^{(0)}, \quad n\leq k,\\
        0, \quad \quad \quad \ n>k.
    \end{cases}
\end{equation}
Setting now $n=d-1$, for every excited-state index $d-k$ with $k = 0,1,\ldots, d-2$, we have $n=d-1>k \implies p_{d-k}^{(d-1)} = 0$. In other words, $p_2^{(d-1)} = p_3^{(d-1)} = \ldots=p_d^{(d-1)} = 0$. On the other hand, the $p_1^{(d-1)}$ term, obeys by normalization
\begin{equation}
    p_1^{(d-1)} = 1- \sum_{i=2}^d p_i^{(d-1)} = 1.
\end{equation}
Therefore, after exactly $d-1$ collisions, the system thermalizes (cools to zero). In Fig. \ref{fig:Tsimhigher_order_systems_Jtau1_pi2} we observe this phenomenon using numerical simulations; focus on the low-temperature limit, $\beta \xrightarrow[]{}\infty$.

\begin{figure}[h]
    \centering
\includegraphics[width=0.5\linewidth]{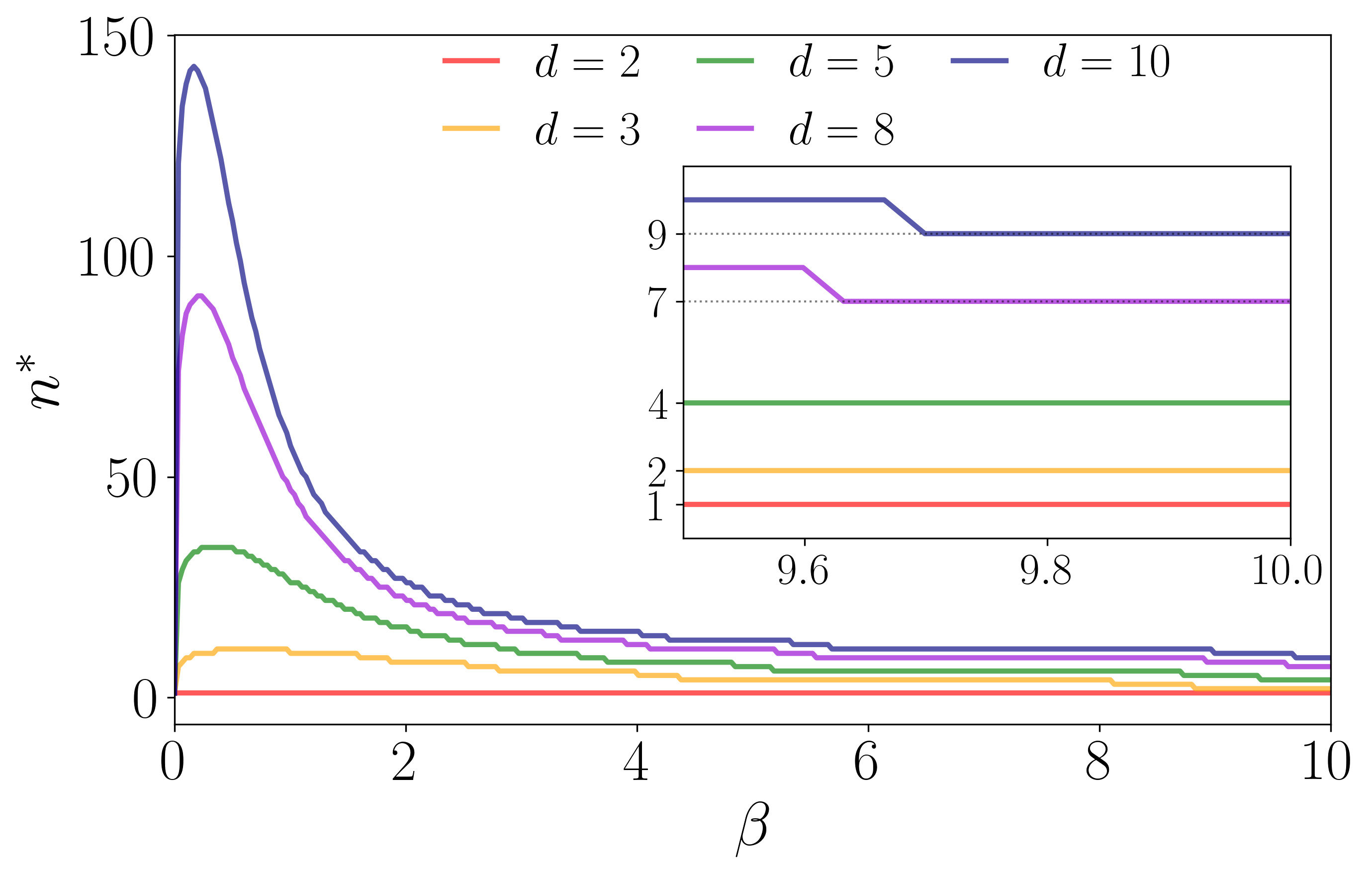} 
    \caption{Total number of collisions $n^*$ in the $J\tau1$ regime as a function of the inverse temperature $\beta$ for systems with $d=2,3,5,8,10$ levels, starting from a completely mixed state. 
    The inset zooms in on the low-temperature regime corresponding to $\beta \in [9.5,10]$. The interaction and collision time is always set to $J\tau = \pi/2$ with $J = 10^{-3}$, efficiently capturing the $J\tau1$ regime. The precision is set to $\epsilon = 10^{-4}$.}
\label{fig:Tsimhigher_order_systems_Jtau1_pi2}
\end{figure}

\subsubsection{General expression for $T_{\text{sim}}$}

The SL regime can be similarly generalized to an $d$-level system.
Given previous expressions for the $d=2,3,4,5$ system, see Eqs. (\ref{eq: trace-distance-qt-qb-fulleq}),
(\ref{eq:Tsim4}), (\ref{eq:Tsim5}).
we generalize these result for any $d$-dimensional system, 
\begin{equation}
    \label{eq:general_expression_Tsim}
    e^{-\Gamma T_{\text{sim}}}\left[\sum_{k=0}^{d-2} \frac{(\Gamma T_{\text{sim}})^k}{k!} \left( \sum_{i=k+2}^d p_i(0) \right) \right] = \epsilon.
\end{equation}
This equation can be solved numerically to obtain the total simulation time.

\subsubsection{Eigenvalues of a $d$-dimensional Liouvillian in the SL limit}

Before attempting to obtain the eigenvalues of the stochastic matrix, we first proceed to obtain the eigenvalues of the Liouvillian, since it has a more tractable expression. 

For a $d-$dimensional system, the Liouvillian, describing the dynamics of population only, takes the following tridiagional form
\begin{equation}
    \mathcal{L}_d = \Gamma \begin{pmatrix}
        -(1-p_A)&p_A&\\
        1-p_A&-1&p_A&\\
        &1-p_A&-1&p_A\\
        & & 1-p_A &-1 & p_A\\
        & & & \ddots&\ddots& \ddots\\
        & & & & 1-p_A&-1&p_A\\
        & & & & &1-p_A& -p_A
    \end{pmatrix},
\end{equation}
which can be expressed in a more general way as
\begin{equation}
\label{eq:tridigonal_matrix_known_eigenvalues}
    \mathcal{L}_d = \begin{pmatrix}
        -\alpha+b&c&\\
        a&b&c&\\
        &\ddots&\ddots& \ddots\\
        & &a&b&c\\
        & & &a& -\kappa +b
    \end{pmatrix},
\end{equation}
with the coefficients $a = \Gamma(1-p_A)$, $c = \Gamma p_A$, $b = -\Gamma$, $\alpha = -\Gamma p_A = -c$ and $\kappa = \Gamma(-1+p_A)=-a$.

This class of matrices has been studied extensively in Refs. \cite{Yueh2005,Willms2009}. In Ref. \cite{Yueh2005} Yueh established that the eigenvalues of this $d\times d$ matrix are
\begin{equation}
    \lambda = b +2\sqrt{ac} \cos(\gamma),
\end{equation}
where $\gamma$ is a solution of
\begin{equation}
\label{eq:gamma_parameter_angle_d_dim_Liouvillian}
    ac \sin((n+1)\gamma) + (\alpha+\kappa)\sqrt{ac} \sin(n\gamma) + \alpha \kappa \sin((n-1)\gamma) = 0, \hspace{0.5cm} \gamma \neq k\pi, k \in \mathbb{Z}.
\end{equation}
In Ref. \cite{Willms2009}, Willms solved the eigenvalues of this tridiagonal matrix for the specific case where $\alpha = -c$ and $\kappa = -a$, which, as can be verified, corresponds to our particular case. From Eq. (\ref{eq:gamma_parameter_angle_d_dim_Liouvillian}), using $\sin((n+1)\gamma)+\sin((n-1)\gamma) = 2\sin(n\gamma)\cos(\gamma)$, we obtain
\begin{equation}
    \sin(n\gamma) = 0 \quad \text{or} \quad 2\sqrt{ac}\cos(\gamma) -(a+c) = 0,
\end{equation}
which results in the eigenvalues
\begin{equation}
\label{eq:general_eigenvalues_expression_tridiagonal_matrix}
\begin{aligned}
    & \lambda_1 = b+a+c,\\
    &\lambda_m = b+2\sqrt{ac}\cos\left( \frac{(m-1)\pi}{d}\right), \quad \text{for} \; m=2, \ldots,d.
\end{aligned}
\end{equation}
Substituting the expressions for $a$, $b$, and $c$, we obtain that the full spectrum as given by
\begin{equation}
    \begin{aligned}
    &\lambda_1 = 0, \\
    &\lambda_m = \Gamma \left[-1+2\theta\cos\left( \frac{(m-1)\pi}{d}\right)\right], \quad \text{for} \; m=2, \ldots,d.\\
\end{aligned}
\end{equation}
where $\theta = \sqrt{p_A(1-p_A)}$. This closed-form solution correctly reproduces the eigenvalues obtained for the cases $d=2,3,4,5$ reported in this paper.

In particular, the second smallest eigenvalue (in magnitude), which controls the relaxation timescale, is  given by 
\begin{equation}
\label{eq: second_eigenvalue_d_dimensional_case_SL_limit}
    \lambda_2^{(d)} = -\Gamma \left[1-2\theta\cos\left( \frac{\pi}{d}\right)\right].
\end{equation}
We display a series of these eigevalues in Fig. \ref{fig:d-dimensional-eigenvalue-Liouvillian}(a). We observe that $\lambda_2^{(d)}$ becomes smaller in magnitude as $d$ increases. This trend explains the emergence of the Mpemba effect with increasing dimensionality.

\subsubsection{Eigenvalues of the $d$-dimensional stochastic matrix}

We construct next the stochastic matrix describing the recursive equations for a $d\times d$ system, 
\begin{equation}
    \Lambda_d = 
 \begin{pmatrix}
        \eta_{11}&\eta_{12}&\\
        \eta_{21}&\eta_{22}&\eta_{12}&\\
        &\eta_{21}&\eta_{22}&\eta_{12}\\
        & & \eta_{21} &\eta_{22} & \eta_{12}\\
        & & & \ddots&\ddots& \ddots\\
        & & & & \eta_{21}&\eta_{22}&\eta_{12}\\
        & & & & &\eta_{21}& \eta_{33}
    \end{pmatrix}.
\end{equation}
It generalizes Eq. (\ref{eq:stochastic_matrix_ququint}).
By rewriting this matrix in the same general form as Eq. (\ref{eq:tridigonal_matrix_known_eigenvalues}), with  $a = \eta_{21}$, $c = \eta_{12}$, $b= \eta_{22}$, $\alpha =-\eta_{12}=-c$, and $\beta = -\eta_{21} = -a$, we find that the eigenvalues follow the structure
\begin{equation}
\label{eq:general_eigenvalues_stochastic_matrix}
\begin{aligned}
    & \xi_1 = 1,\\
    &\xi_m =\eta_{22}+2\sqrt{\eta_{21} \eta_{12}}\cos\left( \frac{(m-1)\pi}{d}\right)=\lambda_++2\theta \lambda_-\cos\left( \frac{(m-1)\pi}{d}\right) , \quad \text{for} \; m =2, \ldots,d.\\
\end{aligned}
\end{equation}
This analytical expression reproduces results obtained for $d=3$, $4$, and $5$, presented in this work.

The second-largest eigenvalue $\xi_2^{(d)}$ that governs the convergence to the steady state is given by
\begin{equation}
    \label{eq: xi_2d_eigenvalue_stochastic_matrix}
    \xi_2^{(d)} = \lambda_+ +2\theta \lambda_- \cos\left( \frac{\pi}{d}\right).
\end{equation}
We plot this eigenvalue in Fig. \ref{fig:d-dimensional-eigenvalue-Liouvillian}(b). We can observe that, as the dimension of the system increases, $\xi_2^{(d)}$ approaches one. This corresponds to a more pronounced Mpemba effect observed with increasing dimensionality.

\begin{figure}[h!]
    \centering
    \includegraphics[width=0.9\linewidth]{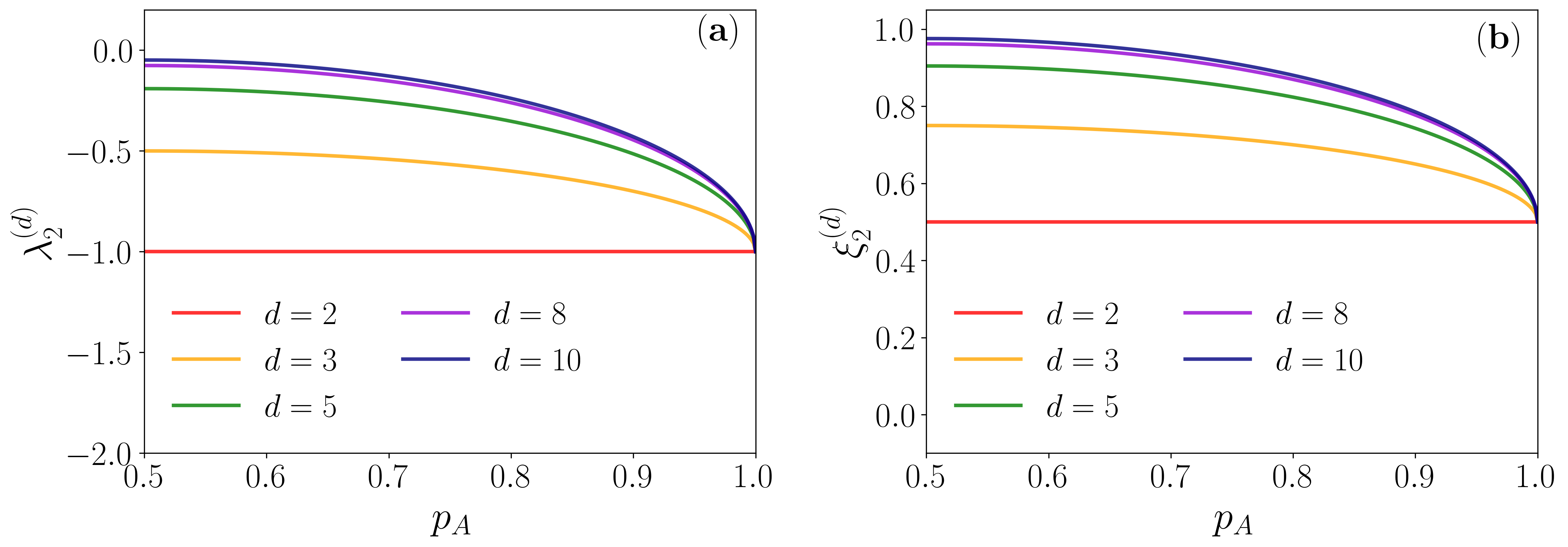} 
    \caption{(a) $\lambda_2^{(d)}$ for $d=2$ (red), $d=3$, yellow, $d=5$ (green), $d=8$ (purple), and $d=10$ (blue) computed using Eq. (\ref{eq: second_eigenvalue_d_dimensional_case_SL_limit}). The value of $\Gamma$ is set to one. (b) $\xi_2^{(d)}$ using Eq. (\ref{eq: xi_2d_eigenvalue_stochastic_matrix}). The product $J\tau = \pi/4$, which gives $\lambda_+=\lambda_- = \frac{1}{2}$. }
    \label{fig:d-dimensional-eigenvalue-Liouvillian}. 
\end{figure}

\section{Dynamics under the non-energy-conserving Hamiltonian at the SL limit.}
\label{sec: Dynamics_under_non_energy_conserving_interactions_SL_limit}

In this Appendix, we aim to study the dynamics in the stroboscopic-Lindblad under the interaction Hamiltonian Eq. (\ref{eq:not_energy_preserving}). Our goal is to understand the steady state shown in Fig. \ref{fig:multiplot-p1-and-coherences_random_Hamiltonian} with $|c_{13} \neq 0|$, which causes the system not to reach an equilibrium thermal state. 

Expanding the collision unitary until the second order in $J\tau$ with the total Hamiltonian of Eq. (\ref{eq:non-energy-conserving-total-hamiltonian-for-qutrit}) gives
\begin{equation}
\resizebox{\textwidth}{!}{$
\hat{U}(\tau) \approx \begin{pmatrix}
1 + \frac{1}{8} \tau \left( -4 {J'}^2 \tau + 3 \omega (4i - 3 \tau \omega) \right) & 0 & 0 & \frac{1}{2} J' \tau (-2i + \tau \omega) & -\frac{1}{2} J J' \tau^2 & 0 \\
0 & \frac{1}{8} \left( 8 - 4 J^2 \tau^2 + \tau \omega (4i - \tau \omega) \right) & \frac{1}{2} J \tau (-2i + \tau \omega) & 0 & 0 & -\frac{1}{2} J J' \tau^2 \\
0 & \frac{1}{2} J \tau (-2i + \tau \omega) & \frac{1}{8} \left( 8 + 4i \tau \omega - \tau^2 \left( 4(J^2 + {J'}^2) + \omega^2 \right) \right) & 0 & 0 & -\frac{1}{2} J' \tau (2i + \tau \omega) \\
\frac{1}{2} J' \tau (-2i + \tau \omega) & 0 & 0 & 1 - \frac{1}{8} \tau \left( 4(J^2 + {J'}^2)\tau + 4i \omega + \tau \omega^2 \right) & -\frac{1}{2} J \tau (2i + \tau \omega) & 0 \\
-\frac{1}{2} J J' \tau^2 & 0 & 0 & -\frac{1}{2} J \tau (2i + \tau \omega) & 1 - \frac{1}{8} \tau \left( 4 J^2 \tau + \omega (4i + \tau \omega) \right) & 0 \\
0 & -\frac{1}{2} J J' \tau^2 & -\frac{1}{2} J' \tau (2i + \tau \omega) & 0 & 0 & 1 - \frac{1}{8} \tau \left( 4 {J'}^2 \tau + 3 \omega (4i + 3 \tau \omega) \right)
\end{pmatrix}
$}
\end{equation}
With this, using the RI protocol, Eq. (\ref{eq: CPTPmap}), 
and expanding until second order in $J\tau$, 
equations for the populations are
\begin{equation}
\label{eq:populations-non-energy-conserving-Hamiltonian}
    \begin{aligned}
        &p_1^{(n+1)}=p_1^{(n)} + \left( -J_1^2 p_1^{(n)} + J_2^2 p_2^{(n)} + (J_1 - J_2)(J_1 + J_2)(p_1^{(n)} + p_2^{(n)})p_A \right) \tau^2 
    - J_1 J_2 \tau^2 \, \mathrm{Re}[c_{13}^{(n)}],\\
    &p_2^{(n+1)} = p_2^{(n)}+ \Big[ 
        J_2^2 (-p_2^{(n)} + p_3^{(n)} + p_1^{(n)} p_A - p_3^{(n)} p_A)+ J_1^2 (p_1^{(n)} - p_2^{(n)} - p_1^{(n)} p_A + p_3^{(n)} p_A) 
    \Big] \tau^2 
    + 2 J_1 J_2 \tau^2 \, \mathrm{Re}[c_{13}^{(n)}],\\
    &p_3^{(n+1)} = p_3^{(n)}+ \Big[ 
        J_1^2 \left( p_2^{(n)} - (p_2^{(n)} + p_3^{(n)})p_A \right) 
        + J_2^2 \left( -p_3^{(n)} + (p_2^{(n)} + p_3^{(n)})p_A \right)\Big] \tau^2 - J_1 J_2 \tau^2 \, \mathrm{Re}[c_{13}^{(n)}].
    \end{aligned}
\end{equation}
In our notation here compared to the main text we use, $J_1 \equiv J, \; J_2 \equiv J'$. Based on Eq. (\ref{eq:populations-non-energy-conserving-Hamiltonian}) we observe that the coherence term $c_{13}$ is coupled to the populations.

Taking now the usual SL limit, we define the following rates $J_1^2 \tau = \Gamma_1$, $J_2^2\tau = \Gamma_2$, and $J_1J_2\tau=\Gamma_{12}$. These rates are constructed constant as $\tau \to0$. With this, the differential equation for the populations becomes
\begin{equation}
    \label{eq:diff_eq_SL_limit_non_energy_conserving}
    \begin{aligned}
        &\dot{p}_1(t) = -\Gamma_1(1-p_A)p_1(t) + \Gamma_2(1-p_A)p_2(t) - \Gamma_{12} \text{Re}[c_{13}(t)],\\
        & \dot{p}_2(t) = p_1(t)\left[\Gamma_2p_A+\Gamma_1\left(1-p_A \right) \right]-p_2(t)\left(\Gamma_2+\Gamma_1\right)+p_3(t)\left[\Gamma_2(1-p_A)+\Gamma_1 p_A \right] +2 \Gamma_{12} \text{Re}[c_{13}(t)],\\
        & \dot{p}_3(t) = p_2(t)\left[ \Gamma_1 (1-p_A)+\Gamma_2 p_A\right]-p_3(t)\left[\Gamma_1 p_A + \Gamma_2(1-p_A) \right]- \Gamma_{12} \text{Re}[c_{13}(t)]
    \end{aligned}
\end{equation}
For the coherences, the recursive equations are 
\begin{equation}
\begin{aligned}
    &c_{12}^{(n+1)}=\frac{1}{2} \Big[
2 c_{23}^{(n)} \left( -J_2^2 (-1 + p_A) + J_1^2 p_A \right) \tau^2
+ c_{12}^{(n)} \Big( 
2 + \tau \Big(J_1^2 (-2 + p_A) \tau\\
&+ 2i \omega 
- \tau \left( J_2^2 (1 + p_A) + \omega^2 \right)
\Big)
\Big)
+ J_1 J_2 \tau^2 \left( 2 \overline{c}_{12}^{(n)} - \overline{c}_{23}^{(n)} \right)
\Big],\\
& c_{13}^{(n+1)} = c_{13}^{(n)} 
- \frac{1}{2} J_1 J_2 (p_1^{(n)} - 2p_2^{(n)} + p_3^{(n)}) \tau^2
- \frac{1}{2} c_{13}^{(n)}  \, \tau \left( -4i \omega + \tau (J_1^2 + J_2^2 + 4 \omega^2) \right),\\
& c_{23}^{(n+1)} =\; c_{23}^{(n)}+ c_{12}^{(n)} \left( -J_1^2 (-1 + p_A) + J_2^2 p_A \right) \tau^2 \\
&- \frac{1}{2} c_{23}^{(n)} \, \tau \left( 
    -J_2^2 (-2 + p_A) \tau 
    + J_1^2 (1 + p_A) \tau 
    + \omega (-2i + \tau \omega) 
\right) - \frac{1}{2} J_1 J_2 \tau^2 \left( \overline{c}_{12}^{(n)} - 2 \overline{c}_{23}^{(n)} \right).
\end{aligned}
\end{equation}
To obtain differential equations, we make the additional assumption as we did during this work, where $\omega\tau \xrightarrow[]{}0$ and $\omega^2\tau\to0$, which places us in the limit of $J\gg\omega$. Under this assumption, the differential equations are given by
\begin{equation}
    \label{eq: diff-eq-non-energy-preserving-Hamiltonian}
    \begin{aligned}
        &\dot{c}_{12}(t)= c_{23}(t)\left[\Gamma_2(1-p_A)+\Gamma_2 p_A \right]-c_{12}(t) \left[\Gamma_1(2-p_A)+\Gamma_2(1+p_A) \right] + \Gamma_{12} \left(2\overline{c}_{12}(t)-\overline{c}_{23}(t) \right),\\
        & \dot{c}_{13}(t) = \frac{1}{2}\Gamma_{12} \left(2p_2(t)-p_1(t)-p_3(t)\right)-\frac{1}{3}c_{13}(t)\left( \Gamma_1+\Gamma_2\right),\\
        &\dot{c}_{23}(t) = c_{12}(t)\left[ \Gamma_1(1-p_A)+\Gamma_2p_A\right] -\frac{1}{2}c_{23}(t)\left[\Gamma_2(2-p_A)+\Gamma_1(1+p_A)\right]-\frac{1}{2}\Gamma_{12} \left(\overline{c}_{12}(t)-2\overline{c}_{23}(t)\right).
    \end{aligned}
\end{equation}
From these differential equations, we observe that coherences corresponding to adjacent energy levels are coupled to each other, as we have already seen for the energy-conserving interaction Hamiltonian. The main difference appears when dealing with the coherence term $c_{13}$, whose evolution is coupled now to populations. This explains the phenomenon observed in Fig. \ref{fig:multiplot-p1-and-coherences_random_Hamiltonian}, where the $c_{13}$ term does not vanish, resulting in the following steady state
\begin{equation}
    c_{13}^* = \frac{3}{2}\frac{\Gamma_{12}}{\Gamma_1 + \Gamma_2} (2p_2^*-p_1^*-p_3^*).
\end{equation}

\end{widetext}


\end{document}